\documentclass[12pt]{article}
\usepackage{amsmath}
\usepackage{graphicx,psfrag,epsf}
\usepackage{amsthm}
\usepackage{xr-hyper}
\usepackage{xr}
\usepackage{pdfpages}
\allowdisplaybreaks[1]
\usepackage{enumerate}
\usepackage{natbib}
\usepackage{url} 

\newcommand{\blind}{1}

\usepackage[
top    = 1in,
bottom = 1in,
left   = 1in,
right  = 1in]{geometry}

\RequirePackage[OT1]{fontenc}

\usepackage{amssymb}
\usepackage{epsfig}
\usepackage{color}
\usepackage{enumerate,verbatim}
\usepackage{bm}
\usepackage{subfig}
\usepackage{xcolor}
\usepackage{float}
\usepackage{url}
\usepackage{multirow}

\newtheorem{thm}{Theorem}[section]

\newcommand{\smc}{\twelvesmc}

\newtheorem*{remark}{Remark}

\newtheorem{asmp}{Assumption}
\newcommand{\bd}{\begin{document}}
	\def\qed{\hfill$\diamondsuit$}
	\newcommand{\ed}{\end{document}}
\def\bse{\begin{eqnarray*}}
	\def\ese{\end{eqnarray*}}
\def\be{\begin{eqnarray}}
\def\ee{\end{eqnarray}}
\newcommand{\ben}{\begin{eqnarray}}\newcommand{\een}{\end{eqnarray}}
\newcommand{\bn}{\begin{enumerate}}
	\newcommand{\en}{\end{enumerate}}
\newcommand{\im}{\item}
\newcommand{\bc}{\begin{cases}}
	\newcommand{\ec}{\end{cases}}
\newcommand{\bt}{\begin{tabular}}
	\newcommand{\et}{\end{tabular}}
\newcommand{\bct}{\begin{center}}
	\newcommand{\ect}{\end{center}}
\def\wt{\widetilde}
\def\diag{\hbox{diag}}
\def\wh{\widehat}
\def\AIC{\hbox{AIC}}
\def\BIC{\hbox{BIC}}
\newcommand{\tr}{{\rm tr}}
\newcommand{\cid}{\buildrel d \over \longrightarrow}
\newcommand{\cas}{\buildrel a.s \over \longrightarrow}
\newcommand{\cw}{\buildrel w \over \longrightarrow}
\newcommand{\cip}{\buildrel p \over \longrightarrow}
\newcommand{\cil}{\buildrel L^2 \over \longrightarrow}
\newcommand{\bl}{{\pmb l}}

\newcommand{\bk}{{\pmb k}}
\newcommand{\by}{{\pmb y}}
\newcommand{\bda}{{\pmb a}}
\newcommand{\bdb}{{\pmb b}}
\newcommand{\bdc}{{\pmb c}}
\newcommand{\bdd}{{\pmb d}}
\newcommand{\bdf}{{\pmb f}}
\newcommand{\bdg}{{\pmb g}}
\newcommand{\bdx}{{\pmb x}}
\newcommand{\bdu}{{\pmb u}}
\newcommand{\bdv}{{\pmb v}}

\newcommand{\bone}{{\pmb 1}}
\newcommand{\bdepsilon}{{\pmb\epsilon}}
\newcommand{\bdmu}{{\pmb \mu}}
\newcommand{\bdpsi}{{\pmb \psi}}
\newcommand{\bdalpha}{{\pmb \alpha}}
\newcommand{\bdbeta}{{\pmb \beta}}
\newcommand{\bdtheta}{{\pmb \theta}}
\newcommand{\bdvartheta}{{\pmb \vartheta}}

\newcommand{\CA}{{\cal A}}
\newcommand{\CB}{{\cal B}}
\newcommand{\CC}{{\cal C}}
\newcommand{\CD}{{\cal D}}
\newcommand{\CE}{{\cal E}}
\newcommand{\CF}{{\cal F}}
\newcommand{\CG}{{\cal G}}
\newcommand{\CH}{{\cal H}}
\newcommand{\CJ}{{\cal J}}
\newcommand{\CK}{{\cal K}}
\newcommand{\CL}{{\cal L}}
\newcommand{\CM}{{\cal M}}
\newcommand{\CN}{{\cal N}}
\newcommand{\CO}{{\cal O}}
\newcommand{\CP}{{\cal P}}
\newcommand{\CQ}{{\cal Q}}
\newcommand{\CR}{{\cal R}}
\newcommand{\CS}{{\cal S}}
\newcommand{\CT}{{\cal T}}
\newcommand{\CV}{{\cal V}}
\newcommand{\CW}{{\cal W}}
\newcommand{\CX}{{\cal X}}
\newcommand{\CY}{{\cal Y}}
\newcommand{\CZ}{{\cal Z}}
\newcommand{\BA}{{\pmb A}}
\newcommand{\BB}{{\pmb B}}
\newcommand{\BS}{{\pmb S}}
\newcommand{\BG}{{\pmb G}}
\newcommand{\BI}{{\pmb I}}
\newcommand{\bs}{{\pmb s}}
\newcommand{\BR}{{\bold R}}
\newcommand{\BT}{{\pmb T}}
\newcommand{\BX}{{\pmb X}}
\newcommand{\BU}{{\pmb U}}
\newcommand{\BV}{{\pmb V}}
\newcommand{\BZ}{{\pmb Z}}
\newcommand{\BY}{{\pmb Y}}
\newcommand{\FC}{\field{C}}
\newcommand{\la}{{\langle}}
\newcommand{\ra}{{\rangle}}
\newcommand{\E}{{\rm E}}

\newcommand{\BLambda}{{\pmb \Lambda}}
\newcommand{\BSigma}{{\pmb \Sigma}}
\newcommand{\BUpsilon}{{\pmb \Upsilon}}

\newcommand{\el}{L^2[0,1]}
\def\II{I\negthinspace I}
\def\III{I\negthinspace I\negthinspace I}
\renewcommand{\baselinestretch}{1.2}
\newcommand{\bepsilon}{{\pmb varepsilon}}
\newcommand{\ep}{\vskip-.3cm\noindent\begin{flushright}
		${{\hfill\llap{$\sqcup\!\!\!\!\sqcap$}}}$
	\end{flushright}}
	
	\def\wt{\widetilde}
	\def\diag{\hbox{diag}}
	\def\wh{\widehat}
	\def\AIC{\hbox{AIC}}
	\def\BIC{\hbox{BIC}}
	\newcommand{\Appendix}
	{
		\def\thesection{Appendix~\Alph{section}}
		\def\thesubsection{A.\arabic{subsection}}
	}
	\def\diag{\hbox{diag}}
	\def\log{\hbox{log}}
	\def\bias{\hbox{bias}}
	\def\sd{\hbox{sd}}
	\def\Siuu{\boldSigma_{i,uu}}

	\def\dfrac#1#2{{\displaystyle{#1\over#2}}}
	\def\VS{{\vskip 3mm\noindent}}
	\def\boxit#1{\vbox{\hrule\hbox{\vrule\kern6pt
				\vbox{\kern6pt#1\kern6pt}\kern6pt\vrule}\hrule}}
	\def\refhg{\hangindent=20pt\hangafter=1}
	\def\refmark{\par\vskip 2mm\noindent\refhg}
	\def\naive{\hbox{naive}}
	\def\itemitem{\par\indent \hangindent2\parindent \textindent}
	\def\var{\hbox{var}}
	\def\cov{\hbox{cov}}
	\def\corr{\hbox{corr}}
	\def\trace{\hbox{trace}}
	\def\refhg{\hangindent=20pt\hangafter=1}
	\def\refmark{\par\vskip 2mm\noindent\refhg}
	\def\Normal{\hbox{Normal}}
	\def\povr{\buildrel p\over\longrightarrow}
	\def\ccdot{{\bullet}}
	\def\bse{\begin{eqnarray*}}
		\def\ese{\end{eqnarray*}}
	\def\be{\begin{eqnarray}}
	\def\ee{\end{eqnarray}}
	\def\bq{\begin{equation}}
	\def\eq{\end{equation}}
	\def\bse{\begin{eqnarray*}}
		\def\ese{\end{eqnarray*}}
	\def\pr{\hbox{pr}}
	\def\wh{\widehat}
	\def\trans{^{\rm T}}
	\def\myalpha{{\cal A}}
	\def\th{^{th}}
	\def\wi{{\hbox{\scriptsize WI}}}
	
	\def\boxit#1{\vbox{\hrule\hbox{\vrule\kern6pt
				\vbox{\kern6pt#1\kern6pt}\kern6pt\vrule}\hrule}}
	\def\licomment#1{\vskip 2mm\boxit{\vskip 2mm{\color{black}\bf#1} {\color{blue}\bf -- Yehua\vskip 2mm}}\vskip 2mm}
	\def\zhangcomment#1{\vskip 2mm\boxit{\vskip 2mm{\color{black}\bf#1} {\color{blue}\bf -- Haozhe\vskip 2mm}}\vskip 2mm}

\begin{document}

\def\spacingset#1{\renewcommand{\baselinestretch}%
{#1}\small\normalsize} \spacingset{1}


\if1\blind
{
  \title{\bf Unified Principal Component Analysis for Sparse and Dense Functional Data under Spatial Dependency}
  \author{Haozhe Zhang\\
   	Microsoft Corporation, Redmond, United States\\
    Email: haozhe.zhang@microsoft.com\\
    \hspace{.2cm}\\
    Yehua Li \\
    Department of Statistics, University of California, Riverside\\
    Email: yehuali@ucr.edu}

  \maketitle
} \fi

\if0\blind
{
  \bigskip
  \bigskip
  \bigskip
  \begin{center}
    {\LARGE\bf Unified Principal Component Analysis for Sparse and Dense Functional Data under Spatial Dependency}
\end{center}
  \medskip
} \fi

\bigskip
\begin{abstract}
We consider spatially dependent functional data collected under a geostatistics setting, where locations are sampled from a spatial point process. The functional response is the sum of a spatially dependent functional effect and a spatially independent functional nugget effect. Observations on each function are made on discrete time points and contaminated with measurement errors. Under the assumption of spatial stationarity and isotropy, we propose a tensor product spline estimator for the spatio-temporal covariance function. When a coregionalization covariance structure is further assumed, we propose a new functional principal component analysis method that borrows information from neighboring functions. The proposed method also generates nonparametric estimators for the spatial covariance functions, which can be used for functional kriging. Under a unified framework for sparse and dense functional data, infill and increasing domain asymptotic paradigms, we develop the asymptotic convergence rates for the proposed estimators. Advantages of the proposed approach are demonstrated through simulation studies and two real data applications representing sparse and dense functional data, respectively. 
\end{abstract}

\noindent%

\vfill

\newpage
\spacingset{1.45} 

\section{Introduction}\label{sec:introduction}
\subsection{Literature review}
Modern technology and data collection methods produce massive data with repeated measurements over time and space, thus give rise to functional data \citep{Ramsay-Silverman05, HorvathKokoszka2012, KokoszkaReimherr2017}. 
In many applications, functional data collected at different times or locations are naturally correlated. There have been a lot of recent theory and methodology developments for dependent functional data, including multi-level functional data \citep{crainiceanu2009generalized, XuLiNettleton2017}, functional time series \citep{hormann2010weakly, Aue2015jasa}, and spatially dependent functional data \citep{staicu2010fast, zhou2010reduced, gromenko2012estimation, Zhang2016jasa, Kokoszka2020, liang2020modeling}. There has also been some work on modeling spatio-temporal point process data using a functional data approach \citep{li2014functional}.

Functional data are commonly viewed as infinite dimensional random vectors in a Hilbert space, and dimension reduction is crucial for visualization, interpretation and inference on these data \citep{Hsing-Eubank15}. There has been a lot of methodological and theoretical developments on dimension reduction for independent data using the functional principal component analysis (FPCA) \citep{yao2005functional, hall2006bproperties, li2010uniform}. The functional principal component scores are also widely used as predictors in linear or nonlinear regression models to predict other variables of interest \citep{CaiHall06, Wong2019}. 

There has also been some work on FPCA on spatially dependent functional data. \cite{hormann2013consistency} provide some theoretical justification on spatial FPCA, assuming the functions are fully observed. In practice, however, functional data are often observed on discrete time points and the measurements are contaminated with errors. Based on the number of observations on each curve, functional data are traditionally classified as sparse functional data \citep{yao2005functional} and dense functional data \citep{hall2006bproperties}. For independent functional data, it is known that the convergence rates for various functional estimators (such as the mean, covariance and principal components) are different under different sampling schemes. 
\cite{Wang2018jrssb} show that nonparametric hypothesis tests have different properties under sparse and dense functional data, in terms of asymptotic null distribution and power. However, sparse and dense functional data are asymptotic concepts, which are not clearly defined in any practical contexts. A lot of recent research efforts were focused on developing unified estimation and inference strategies for all types of functional data \citep{li2010uniform, ZhangWang2016AOS, Wang2018jrssb}. No such results yet exist for spatially dependent functional data.

\subsection{Motivating data examples}\label{sec:data examples}
Our work is motivated by two real data examples from business applications, representing sparse and dense spatially dependent functional data, respectively.

\begin{figure}
	\centering
	\subfloat[]{\includegraphics[width = 0.48\linewidth, height=0.42\linewidth]{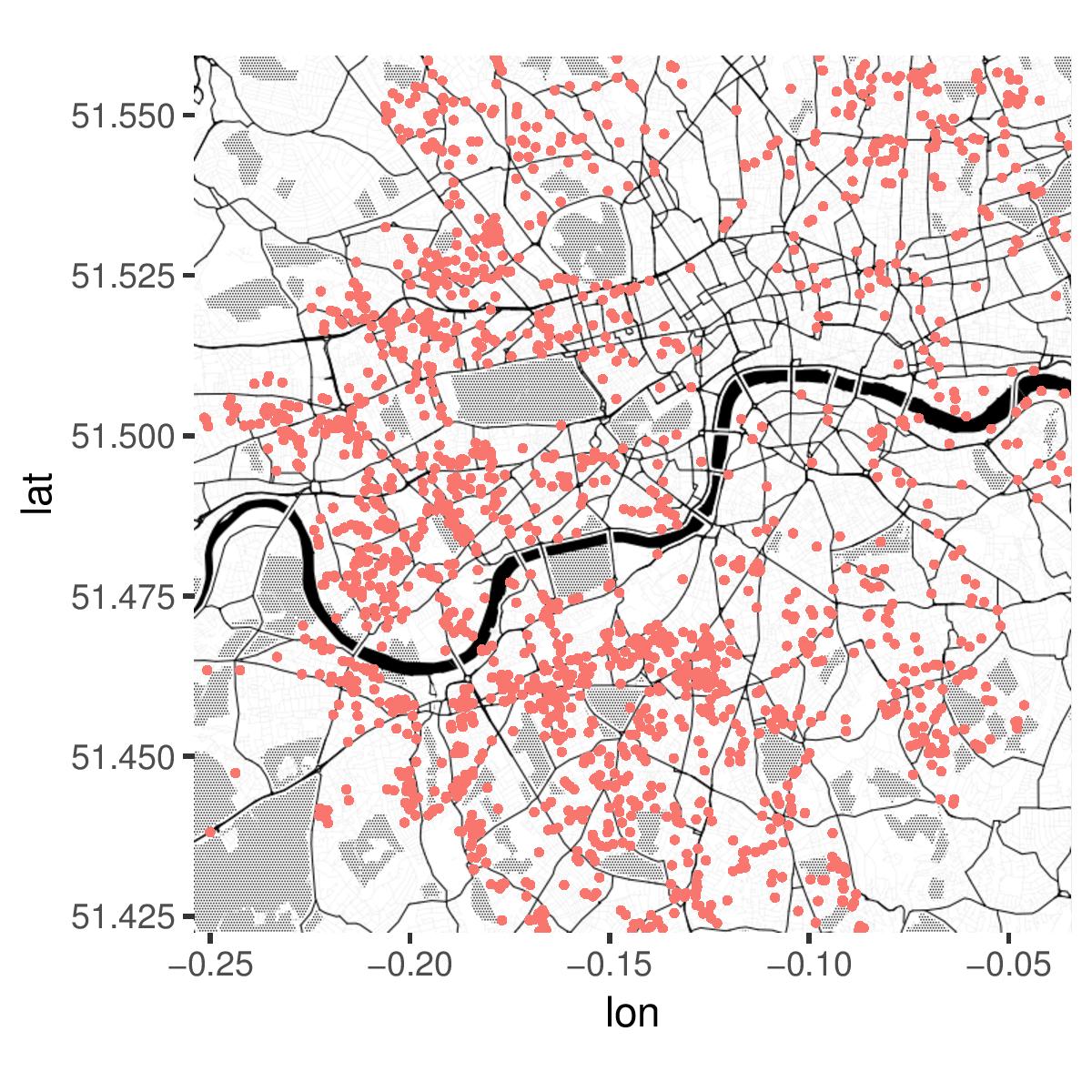}}
	\subfloat[]{\includegraphics[width = 0.48\linewidth, height=0.42\linewidth]{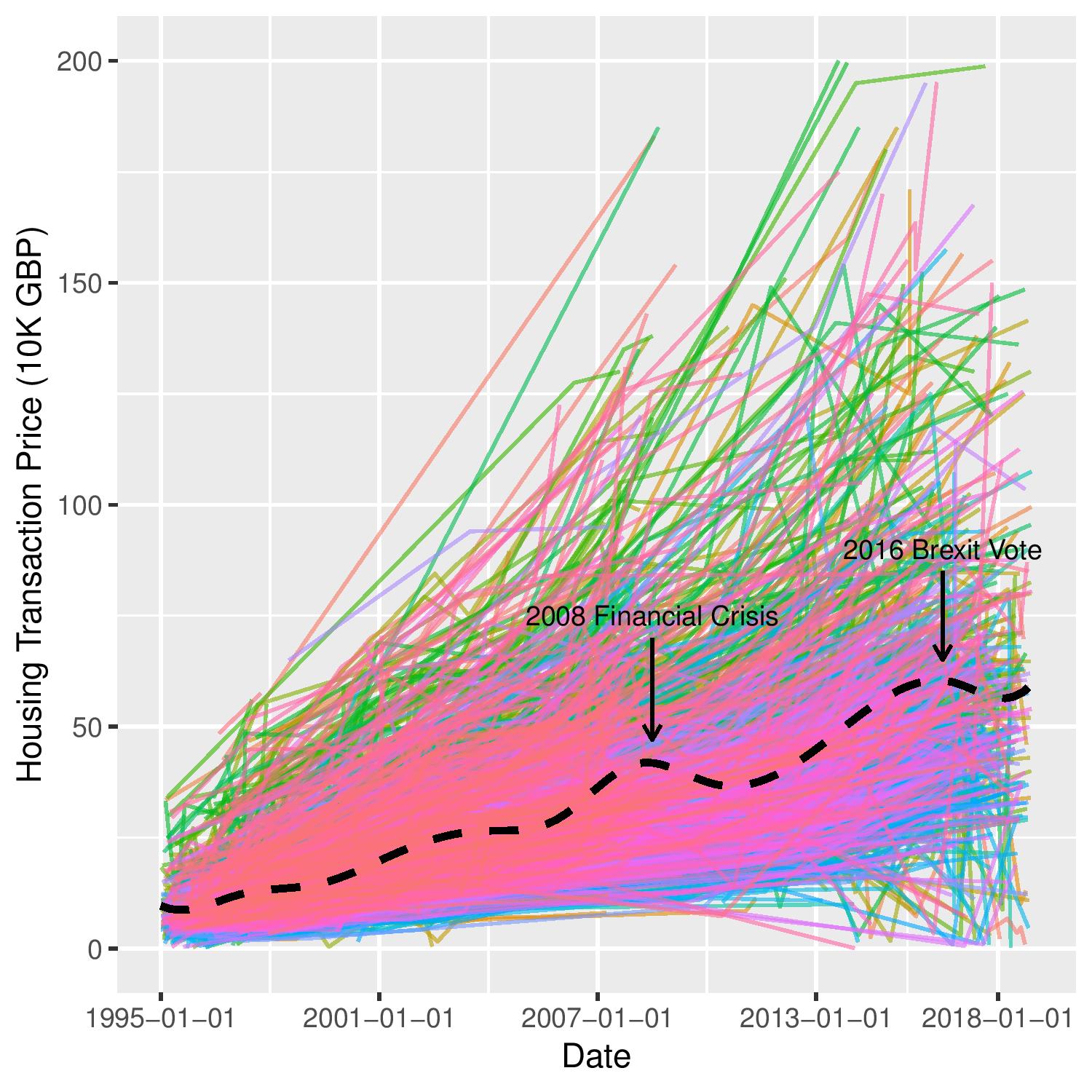}}
	\caption{London house price data. (a) Locations of houses in the Greater London Area; (b) trajectories of the house prices and the estimated mean function (dashed line).
	}
	\label{fig: london_data}
\end{figure}

{\bf Example 1: sparse functional data on London house price.}
The data are public records of home sales from the UK government website (\url{https://www.gov.uk/government}). The dataset includes all houses with at least 5 transactions between Jan 1, 1995 and Dec 31, 2018 in the Greater London Area. Each transaction record contains information on the price, date, and property address. Exact locations, including longitudes and latitudes, of the houses are obtained by searches of the property addresses on Google Map API. The house locations are shown in Panel (a) of  Figure \ref{fig: london_data}.

The value of a house changes continuously over time, the trajectory of which we model as functional data. However, the value is measured by the market only when a sale is made, and the number of sale transactions per house ranges between 5 and 12. The house price trajectories are shown in Panel (b) of Figure \ref{fig: london_data}. As we can see, the transaction times are sparse, irregular and house-specific. 

{\bf Example 2: dense functional data from Zillow Real Estate.}
Zillow (\url{https://www.zillow.com/research}) publishes real estate data for research purposes for all major cities in the US. Our variable of interest is the ``home price-to-rent ratio'', defined as the ratio of residential real estate price to the annual rent, which has attracted broad interests in economics and social sciences \citep{campbell2009moves, kishor2015factors}. It has strong relationships with market fundamentals, and has been widely used as an indicator for housing market bubbles. This variable is updated monthly for geographical units called ``neighborhoods'' defined by Zillow.

\begin{figure}[htb]
	\centering
	\subfloat[]{\includegraphics[width = 0.5\linewidth, height=0.45\linewidth]{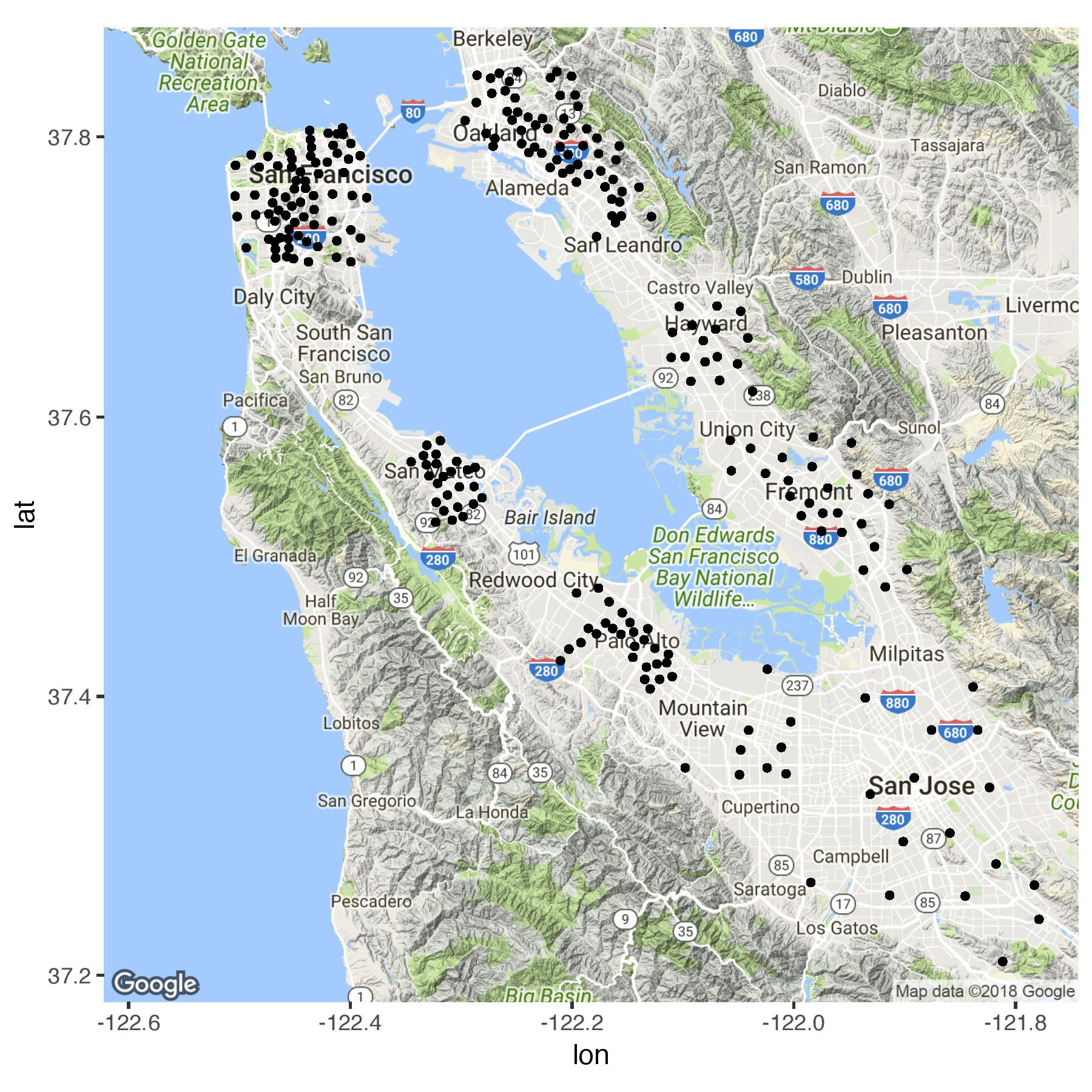}} 
	\subfloat[]{\includegraphics[width = 0.5\linewidth, height=0.45\linewidth]{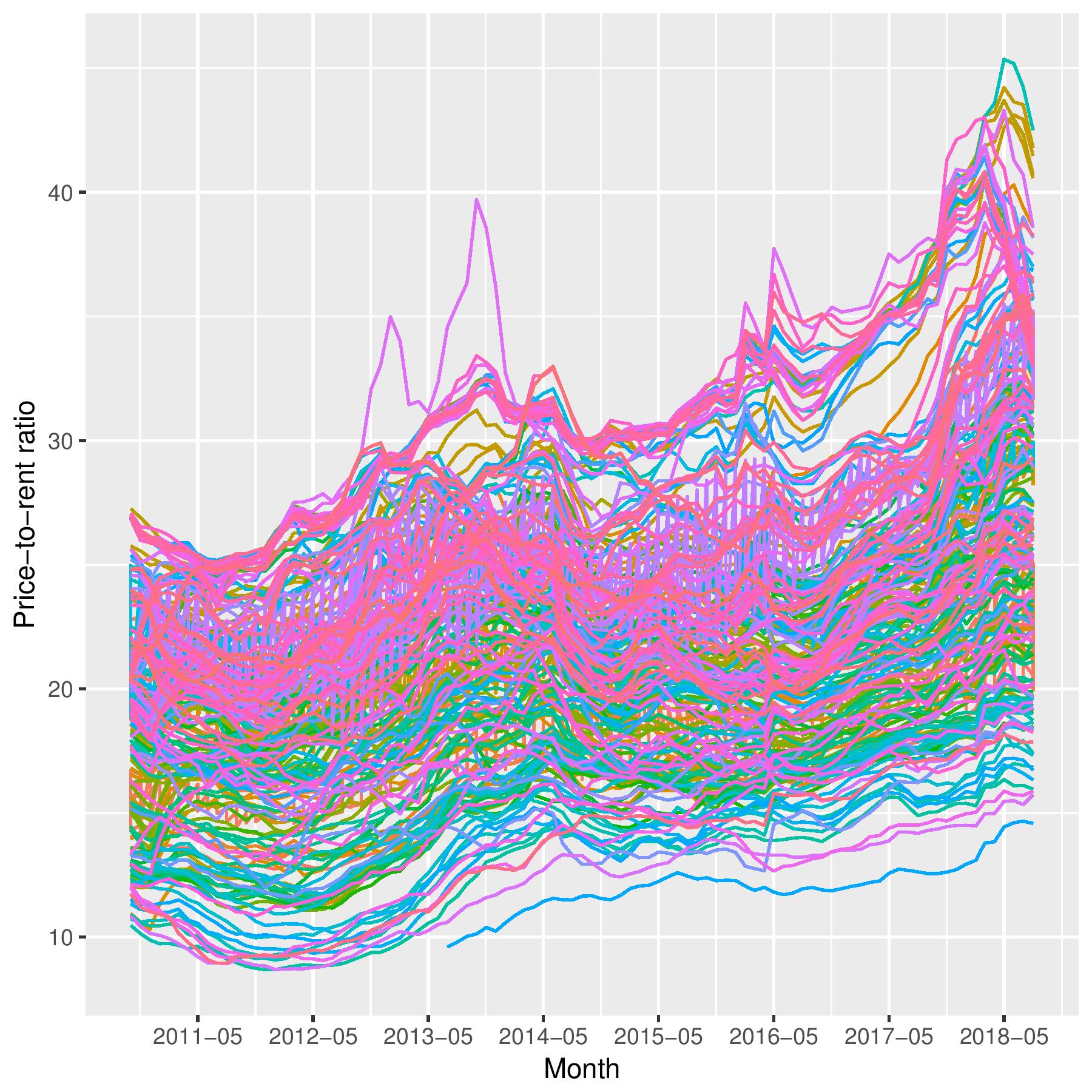}}
	\caption{(a) The locations of the 234 neighborhoods in the San Francisco Bay Area;  (b) trajectories of the home price-to-rent ratios, observed monthly from October 2010 to August 2018 in the 234 neighborhoods.}
	\label{fig: zillow_map_data}
\end{figure}

The dataset we analyze consists of monthly median price-to-rent ratios from $234$ neighborhoods in the San Francisco Bay Area from October 2010 to August 2018, with 95 observations on each curve at a missing rate of $1.48\%$. Figure~\ref{fig: zillow_map_data}  illustrates the geographic locations of these neighborhoods and their price-to-rent ratio trajectories. 




\subsection{Our contributions}

We propose a unified FPCA method that is applicable to both sparse and dense functional data collected under a geostatistics setting, where locations are sampled from a spatial point process. We assume that the trajectory of a random function is determined by two effects: a temporal process that is spatially correlated with neighboring functions and a location-specific random process independent from neighbors. The location-specific random process is also interpreted as the ``nugget'' effect following classic geostatistics literature\citep{Cressie1993}.
Observations on each function are made on discrete time points and contaminated with measurement errors. Under the assumption of spatial stationarity and isotropy, we propose a tensor product spline estimator for the spatio-temporal covariance function. If a coregionalization covariance structure \citep{Banerjee2003, gelfand2004nonstationary} is further assumed, we propose a new FPCA method that borrows information from neighboring functions. Byproducts of our approach also include nonparametric estimators for the spatial covariance functions of the principal component scores. Under a unified framework that combines both infill and increasing domain asymptotic paradigms, we develop unified asymptotic convergence rates for the proposed estimators which demonstrate a phase transition from sparse to dense functional data.

The rest of the paper is organized as follows. We introduce the model and framework in Section \ref{Section: Model}, propose our estimation procedure in Section \ref{Section: Method}, and investigate the theoretical properties of the proposed estimators in Section \ref{Section: theory}. We address some important implementation issues in Section \ref{Section: Implementation} and further extend our method for functional kriging in Section \ref{Section: Kriging}. Numerical performance of the proposed methods is illustrated by simulation studies in Section \ref{Section: Simulation}, where we also show existing methods ignoring the functional nugget effect can lead to biased results.
We analyze the two motivating data examples in Section \ref{Section: Data Analysis} and provide concluding remarks in Section \ref{Section: Discussion}. Technical proofs of the main theorems and additional figures from our numerical studies are collected in the online Supplementary Material. 

\section{Model and assumptions}  \label{Section: Model}
\subsection{Random field modeling for spatially dependent functional data}
Suppose random functions of time defined on a time domain $T$ are sampled from locations in a spatial domain $\CD_{n} \subseteq\mathbb{R}^{2}$. 
Let $Y_{ij}= Y(\bs_{i},t_{ij})$ be the discrete observation at time $t_{ij}$ on the random curve sampled at spatial location $\bs_i$, $ i=1,\ldots,N$, $j=1,\ldots,M_{i}$, and assume the following model
\begin{equation}\label{eq:model}
Y(\bs_{i},t_{ij})=X(\bs_{i},t_{ij})+U_{i}(t_{ij})+\epsilon_{ij},
\end{equation}
where $X(\cdot,\cdot)$ is a spatio-temporal process on $\CD_{n}\times T$ representing a spatially correlated functional effect, $\{U_{i}(\cdot)\}$ are zero-mean, independent temporal processes called the functional nugget effects, and $\{\epsilon_{ij}\}$ are the independent measurement errors with $E(\epsilon_{ij})=0$ and $\var(\epsilon_{ij})=\sigma^{2}_{\epsilon}$. 
The functional nugget effects $U_i(\cdot)$ characterize local variations that are not correlated with neighboring functions, with the covariance function denoted by $\Lambda (t_{1}, t_{2}) = \cov\{U(t_{1}), U(t_{2})\}$. The three model components $X(\cdot, \cdot)$, $U(\cdot)$ and $\epsilon$ are mutually independent.

Assuming that the spatial dependency is second-order stationary and isotropic, the general covariance function of $X(\bs,t)$ can be written as
\ben\label{eq:spt_cov}
R(\|\bs_{1}-\bs_{2}\|,t_{1},t_{2}) = \cov\{X(\bs_{1},t_{1}),X(\bs_{2},t_{2})\},
\een
for any $(\bs_1, t_1), (\bs_2, t_2) \in \CD_n\times T$. 
We consider $X_\bs(t)= X(\bs,t)$ as spatial replicates of a temporal process with a standard Karhunen-Lo\`eve expansion
\begin{equation} \label{Exp: FPC of X}
\hbox{$ X_\bs(t)=\mu(t)+\sum_{j=1}^{\infty}\xi_{j}(\bs)\psi_{j}(t)$},
\end{equation}
where  $\mu(t) = E\left\lbrace X_\bs(t)\right\rbrace$ is the mean function, $\psi_{j}(\cdot)'s$ are orthonormal functions known as the principal components, and the principal component score $\xi_{j}(\bs)= \int_T \{X(\bs,t) -\mu(t)\} \psi_j(t) dt$ is the loading of $X(\bs, t)$ on the $j$th principal component.
We assume $\{\xi_{j}(\bs)\}$ are zero-mean, second-order {stationary} and {isotropic} random fields, that are uncorrelated across different $j$.  Spatial dependence among the function data is induced by the dependence within each $\xi_{j}(\bs)$. Denote the spatial covariance function of $\xi_j(\bs)$ as $\CC_{j}(\|\bs_{1}-\bs_{2}\|) = \cov\{\xi_{j}(\bs_{1}),\xi_{j}(\bs_{2})\}$, for any $ \bs_1, \bs_2 \in \CD_n$, then the covariance function for $X(\bs,t)$ can be written as 
\begin{align}\label{Exp: cross-variance function}
R(\|\bs_{1}-\bs_{2}\|,t_{1},t_{2})& = \hbox{$ \cov\left\lbrace\sum_{j=1}^{\infty}\xi_{j}(\bs_{1})\psi_{j}(t_{1}),\sum_{j=1}^{\infty}\xi_{j}(\bs_{2})\psi_{j}(t_{2})\right\rbrace$}  \\
& = \hbox{$ \sum_{j=1}^{\infty}\CC_{j}(\|\bs_{1}-\bs_{2}\|) \psi_{j}(t_{1})\psi_{j}(t_{2}) $} .\nonumber
\end{align}
%
Denote $\varpi_j= \CC_j(0)$ as the marginal variance for $\xi_j(\bs)$, and assume the principal components are ordered according to their magnitudes such that $\varpi_1\ge \varpi_2\ge \cdots > 0$. It is easy to see that $\varpi_j$'s and $\psi_j(\cdot)$'s are the eigenvalues and eigenfunctions of the covariance function $R(0, \cdot, \cdot)$, which reveals an important connection between our model and classic models for independent functional data. 
The functional nugget effect $U_{i}(\cdot)$, on the other hand, may have an entirely different covariance structure with different eigenvalues and eigenfunctions.

In many applications, including the two real data examples in Section \ref{sec:introduction}, we are interested in the temporal processes defined on some spatially distributed entities, e.g. houses. These entities may not exist on all locations, and the random field framework is a tool of choice to describe the spatial dependence. Model (\ref{Exp: FPC of X}) is also analogous to recent developments in factor models for high dimensional multivariate time series \citep{Fan2018} in the sense that $\xi_j(\bs)$ can be considered as latent factors that govern the dynamics of the temporal process $X_\bs(t)$ and provides reduced rank representations of these temporal processes. In some applications, the latent factors $\xi_j(\bs)$ are of interest and can used as predictors in a second stage regression analysis \citep{Wong2019}. Similar FPC expansion as (\ref{Exp: FPC of X}) was also promoted by \cite{HorvathKokoszka2012} for spatially dependent functional data, who argued that, even if stationarity in space is mildly violated, the mean and eigenfunctions still provide meaningful marginal summary statistics for the data.
By allowing different orders of FPC score to have different spatial covariances, covariance function (\ref{Exp: cross-variance function}) is a ``coregionalization'' model \citep{Banerjee2003, gelfand2004nonstationary}, which is the sum of many separable spatio-temporal covariance functions, and it reduces to a separable structure if $\CC_j(\cdot)= \varpi_j  \rho(\cdot)$ for all $j$.

\subsection{Sampling scheme for spatial locations and observation times}
As illustrated by the two examples in Section \ref{sec:introduction}, the spatial locations $\{\bs_i\}$ are often irregular and random, and can be best described by a spatial point process $\CN_{s}(\cdot )$. The simplest spatial point process is the inhomogeneous Poisson process, where given the total number the locations are independent and identically distributed. A point process can be used to describe more complicated location patterns, such as clustered or regular patterns \citep{Cressie1993}. The correlation between locations are described by the higher-order intensity functions. 

For any location $\bs$, let $d\bs$ be a small neighborhood around $\bs$, and denote $|d\bs|$ as the area of $d\bs$ and $\CN_s(d \bs)$ as the number of locations sampled in $d \bs$.
The $k$-th order intensity function of $\CN_s(\cdot)$ is defined as \citep{Cressie1993}
\begin{equation}
\lambda_{s,k}(\bs_{1},\ldots, \bs_{k}) =\lim\limits_{\tiny \begin{array}{c}
	|d\bs_{r}| \rightarrow 0,\\
	r=1,\ldots, k
	\end{array}} \frac{E\left\lbrace\CN_{s} (d\bs_{1})\ldots\CN_{s} (d\bs_{k})\right\rbrace}{|d\bs_{1}|\ldots|d\bs_{k}|},  
\end{equation}
and we assume $\CN_s$ has up to the $4$th order intensity function well defined.
The collection of observation time points on $Y(\bs, \cdot)$ is a realization of a temporal point process $\CN_{t}(dt|\bs)$. 
Assume that temporal point processes at different locations are independent and identically distributed. 
Denote the first and second intensity functions of $\CN_t(\cdot | \bs)$ as 
\begin{equation}
\lambda_{t,1}(t) =\lim\limits_{
	|dt| \rightarrow 0}\frac{E\CN_{t} (dt|\bs)}{|dt|}, \ \ 
\lambda_{t,2}(t_{1}, t_{2}) =\lim\limits_{
	|dt_{1}|, |dt_{2}|\rightarrow 0}\frac{E\left\lbrace\CN_{t} (dt_{1}|\bs)\CN_{t} (dt_{2}|\bs)\right\rbrace}{|dt_{1}\|dt_{2}|},
\end{equation}
which are independent of $\CN_s(d \bs)$.
This setting also implies that the number of repeated measures on $Y(\bs_i, \cdot)$ is a random variable $M_i :=\int_T \CN_t( dt |\bs_{i}) dt$. We can also define the joint point process for sampling locations and times as $\CN(d\bs, dt)=\CN_s(d\bs) \CN_t( dt | \bs)$. 

As further discussed in Section \ref{Section: theory}, we do not require $\CN_{s}(\cdot )$ or $\CN_t(\cdot | \bs)$ to be stationary, but rather need the intensity functions of these point processes to be bounded from zero so that we have a positive chance to sample from any location and time.  By allowing the intensity functions, $\lambda_{s,k}(t)$ and $\lambda_{t,k}(t)$, to diverge to infinity, we also allow the ``infill'' paradigm: the number of sampled locations in unit space and the number of measurements in unit time are allowed to diverge to infinity.

\section{Estimation method}\label{Section: Method}
We now propose nonparametric estimators for various model components described in Section \ref{Section: Model}, where the core issue is estimating the spatio-temporal covariance function $R(\cdot, \cdot, \cdot)$ in (\ref{eq:spt_cov}). We then use the estimated covariance function to further derive estimators for the principal components $\psi_j(\cdot)$ and spatial covariance functions $\CC_j(\cdot)$, which are of fundamental importance to dimension reduction and understanding the spatial dependence. We will also estimate the covariance function $\Lambda(\cdot, \cdot)$ for the functional nugget effect and the variance of the measurement error $\sigma_\epsilon^2$, which will be further used in the functional kriging.

\subsection{Estimation of the spatio-temporal covariance function} \label{subsetction: estimating covariance function}

For ease of exposition, we assume $\mu(t) \equiv 0$ for Sections \ref{Section: Method} and \ref{Section: theory}. In practice, one can estimate $\mu(t)$ using the smoothing method described in Section \ref{Section: Implementation}, center the response as $\wt Y(\bs_i, t_{ij})= Y(\bs_i,t_{ij})-\wh \mu(t_{ij})$, and then the rest of our methods and theory still apply.

We will only estimate $R(u,\cdot, \cdot)$ up to a pre-determined spatial distance $\Delta>0$. As pointed out by many authors \citep{hall1994nonparametric, li2007nonparametric}, spatial dependency usually decays to zero beyond certain distance; the spatial covariance estimator at a large spatial lag tends to be highly variable, consisting of more noise than signal. To determine $\Delta$, one needs to get a rough estimate for the range of spatial dependency based on a pilot study, for example using the nonparametric method in \cite{li2007nonparametric} based on a more stringent separable spatio-temporal covariance structure.
We consider $R(u, t_{1}, t_{2})$ as a function over a 3-dimensional domain $H:=[0, \Delta]\times T\times T$, and propose to estimate it using 3-dimensional tensor product B-splines. 
For independent functional data, many nonparametric smoothing methods have been proposed to estimate the covariance function, including kernel methods \citep{yao2005functional, li2010uniform} 
and penalized splines \citep{Xiao2013jrssb}. 
In this paper, we focus on tensor product regression spline methods for their computational merits \citep{huang2004identification}, but our methods and theory can be naturally extended to other smoothers.

Without loss of generality, assume $T=[0,1]$. 
Let $\BB_{T}(t) =\{ B_{1, K_{t}}^{p_{t}}(t), B_{2, K_{t}}^{p_{t}}(t), \ldots,$ $B_{K_{t}+p_{t}, K_{t}}^{p_{t}}(t)\}\trans$
be a vector of normalized B-spline functions \citep{de2001practical,huang2004identification} of order $p_t$, defined on time domain $T$ with equally spaced interior knots $\kappa_{j} = j/(K_t+1)$, $j=1,\ldots, K_t$, and denote the corresponding spline space as $\CS_{K_t}^{p_t}[0,1]$. Similarly, let 
$\BB_{S}(u) =\{ B_{1, K_{s}}^{p_{s}}(u), B_{2, K_{s}}^{p_{s}}(u)$, $\ldots$, $B_{K_{s}+p_{s}, K_{s}}^{p_{s}}(u)\}\trans$ be a vector of B-spline basis functions on $[0,\Delta]$ with equally spaced interior knots, where the order $p_s$ and number of knots $K_s$ can be different from $p_t$ and $K_t$ allowing different amount of smoothing in spatial and temporal directions. The assumption of knots being equally spaced is for ease of theoretical derivations, but can be relaxed in practice. Denote the spline space spanned by $\BB_S(u)$ as $\CS_{K_{s}}^{p_{s}}[0,\Delta]$.
%
Then the 3-dimensional tensor product spline space is defined as $\CS_{[3]}\equiv \CS_{K_{s}}^{p_{s}}[0,\Delta] \otimes \CS_{K_{t}}^{p_{t}}[0,1] \otimes \CS_{K_{t}}^{p_{t}}[0,1]$, which is spanned by basis functions  $B_{j_{1}j_{2}j_{3}}(u,t_{1},t_{2})=B_{j_{1}, K_{s}}^{p_{s}}(u) B_{j_{2}, K_{t}}^{p_{t}}(t_{1}) B_{j_{3}, K_{t}}^{p_{t}}(t_{2}) $. Pool the tensor product spline basis functions into a vector 
$
\BB_{[3]}(u,t_{1},t_{2})= \BB_S(u)\otimes \BB_T(t_1) \otimes \BB_T (t_2)
%
$, 
where $\otimes$ is the Kronecker product.

Define
$\CN_{s,2}(d\bs_{1},d\bs_{2}) :=  \CN_{s}(d\bs_{1})\CN_{s}(d\bs_{2})I(\bs_{1}\neq \bs_{2})$, and the tensor product spline estimator of the spatio-temporal covariance function is 
\begin{eqnarray} \label{Exp: estimator of R}
&&\hskip-0.24in \wh{R}(\cdot,\cdot,\cdot) = \underset{g(\cdot,\cdot,\cdot) \in \CS_{[3]}}{\hbox{argmin}}
\int_{\CD_{n}} \int_{\CD_{n}} \int_{T} \int_{T}
\left\lbrace Y(\bs_{1},t_{1})Y(\bs_{2},t_{2})-g(\|\bs_{1}-\bs_{2}\|,t_{1},t_{2})\right\rbrace^{2} \nonumber\\
&&\hskip 1.1in \times I(\|\bs_{1}-\bs_{2}\|\leq \Delta)\CN_{t}(dt_{1}|\bs_{1})\CN_{t}(dt_{2}|\bs_{2})\CN_{s,2}(d\bs_{1},d\bs_{2}),
\end{eqnarray}
where $I(\cdot)$ is the indicator function.
The estimator above can be equivalently written as $\wh{R}(u, t_{1}, t_{2})= \BB_{[3]}\trans(u,t_{1},t_{2})\wh{\bdbeta}$,
where $\wh\bdbeta$ minimizes 
%
\begin{equation} \label{Exp: estimated beta}
{\small
	\CL(\bdbeta)= 
	\sum_{i=1}^N \sum_{ \tiny \begin{array}{c} \hbox{ $i'\neq i$} \\ \hbox{$\|\bs_i-\bs_{i'}\|\leq \Delta$} \end{array}}  \sum_{j=1}^{M_i}  \sum_{j'=1}^{M_{i'}}  \left\lbrace Y_{ij} Y_{i'j'}
	-\BB_{[3]}\trans\left(\|\bs_{i}-\bs_{i'}\|,t_{ij},t_{i'j'}\right)\bdbeta\right\rbrace^{2}. }
%
%
\end{equation}
The numbers of knots $K_{s}$ and $K_{t}$ decide the amount of smoothing and can be selected by data-driven methods described in Section \ref{Section: Implementation}.


%
%
%
%
%
%
%
%
%
%
%
%

\subsection{Estimation of the functional principal components} \label{Subsection:Estimation of the functional principal components }
When the coregionalization structure in (\ref{Exp: cross-variance function}) is assumed, define
\begin{eqnarray}
\Omega(t_{1},t_{2}) := \int_0^\Delta R(u,t_{1},t_{2})\CW(u)du 
=\sum_{j=1}^{\infty}\omega_{j}\psi_{j}(t_{1})\psi_{j}(t_{2}),
\end{eqnarray}
where $\CW(\cdot) \in L^{2}$ is a non-negative and bounded weight function and $\omega_{j} = \int_0^\Delta \CC_{j}(u)\CW(u)du$. 
For all numerical studies in this paper, we use a simple weight function $\CW(u)\equiv 1$ for $u\in [0,\Delta]$ and 0 otherwise.
It is easy to see that the FPCs $\psi_j(t)$ are eigenfunctions of $\Omega (\cdot, \cdot)$. An estimator of $\Omega$ is obtained as
\begin{equation}\label{eq:Omega}
\wh{\Omega}(t_{1},t_{2}) = \int_0^\Delta \wh{R}(u,t_{1},t_{2})\CW(u)du,
\end{equation}
and the estimated eigenvalues and eigenfunctions of $\Omega(\cdot, \cdot)$, denoted as $\{\wh \omega_j, \wh\psi_j(t)\}$, are obtained by solving the eigen-decomposition problem
\begin{equation}
\int_{T}\wh{\Omega}(t_{1},t_{2})\wh{\psi}_{j}(t_{1})dt_{1}=\wh{\omega}_{j}\wh{\psi}_{j}(t_{2}), \quad j = 1, 2,\ldots, \label{Exp: Estimation of the functional principal components}
\end{equation}
subject to the orthonormal constraints $\int_T \wh\psi_j(t) \wh \psi_{j'}(t) dt =I(j=j')$.

From the right hand side of (\ref{eq:Omega}), it is easy to see that all B-splines in the spatial direction are integrated out, and $\wh{\Omega}(\cdot, \cdot)$ is contained in a bivariate tensor product spline space $\CS_{[2]}$ spanned by the basis $\BB_{[2]}(t_{1}, t_{2}) := \BB_{T}(t_{1})\otimes\BB_{T}(t_{2})$. Hence, the functional eigen-decomposition problem in (\ref{Exp: Estimation of the functional principal components}) can be translated into a multivariate problem. Notice that our estimator $\wh{\Omega}$ is inherently symmetric. We can arrange the coefficient vector into a symmetric matrix $\widehat{\BS}$, so that $\wh{\Omega}(t_{1}, t_{2}) = \BB\trans_{T}(t_{1})\widehat{\BS}\BB_{T}(t_{2})$. Define an inner product matrix $\CJ= \int_{T}\BB_{T}(t)\BB\trans_{T}(t)dt$, then the eigen-decomposition problem in (\ref{Exp: Estimation of the functional principal components}) is equivalent to the multivariate generalized eigenvalue decomposition
\bse
\wh{\bm{\phi}}_{j}\trans\CJ \widehat{\BS} \CJ \wh{\bm{\phi}}_{j} = \wh{\omega}_{j}, \quad
\text{ subject to } \quad \wh{\bm{\phi}}_{j'}\trans\CJ \wh{\bm{\phi}}_{j} = I(j=j'),
\ese
and $\wh{\psi}_{j}(t) =\BB_T\trans(t)\wh{\bm{\phi}}_{j}$, $j=1,2 ,\ldots$. 
%
%
%
%
%
%
%
%
%
%
%
%
%
%
%

\subsection{Estimation of the spatial covariance and correlation functions}

By the orthogonality of $\psi_j(t)$'s and  (\ref{Exp: cross-variance function}),
$
\CC_{j}(u)=\int_{T}\int_{T}R(u,t_{1},t_{2})\psi_{j}(t_{1})\psi_{j}(t_{2})dt_{1}dt_{2}
$, 
which motivates the following estimator of the spatial covariance function
\begin{equation}\label{eq:C_j_hat}
\wh{\CC}_{j}(u)=\int_{T}\int_{T}\wh{R}(u,t_{1},t_{2})\wh{\psi}_{j}(t_{1})\wh{\psi}_{j}(t_{2})dt_{1}dt_{2}.
\end{equation}
We then estimate the variance of the $j$th FPC by $\wh \varpi_j=\wh{\CC}_{j}(0)$ and estimate the spatial correlation function $\rho_j(u)=\CC_j(u)/\CC(0)$ by
$
\wh{\rho}_{j} (u)= \wh{\CC}_{j}(u)/\wh{\CC}_{j}(0).
$

\subsection{Covariance estimation for the functional nugget effect} 
\label{section: functional nugget effect}
Define $\Gamma(t_{1}, t_{2}) = R(0,t_{1}, t_{2})$ $+\Lambda(t_{1}, t_{2}) $.
By independence between $X(\bs_i,t)$ and the functional nugget effect $U_i(t)$, it is easy to see $\cov\left\{Y(\bs, t_{1}), Y(\bs, t_{2})\right\}= \Gamma(t_{1}, t_{2}) $ for $t_{1} \neq t_{2}$, which motivates another spline estimator 
%
\begin{equation} \label{eq:Gamma_hat}
\wh{\Gamma} (\cdot,\cdot) =\underset{g(\cdot,\cdot) \in \CS_{[2]}^\Gamma}{\hbox{argmin}} \int_{\CD_{n}}\int_{T}\int_{T}
\left\lbrace Y(\bs,t_{1})Y(\bs,t_{2})-g(t_{1},t_{2})\right\rbrace^{2} I(t_{1} \neq t_{2}) \CN_{t}(dt_{1}|\bs)\CN_{t}(dt_{2}|\bs) \CN_{s}(d\bs).
\end{equation}
Here, $\CS_{[2]}^\Gamma$ is a functional space of bivariate tensor product splines of order $p_{\Gamma}$ defined on $K_{\Gamma}$ interior knots. This spline space can be defined on a different set of temporal knots than those used to estimate $R(\cdot, \cdot,\cdot)$, thus allowing a different amount of smoothing. 
A natural covariance estimator for the functional nugget effect is 
\begin{equation}\label{eq:Lambda_hat}
\wh{\Lambda}(t_{1}, t_{2})= \wh{\Gamma}(t_{1}, t_{2}) - \wh{R}(0,t_{1}, t_{2}),
\end{equation}
where $\wh{R}(0,t_{1}, t_{2})$ is the estimator defined in (\ref{Exp: estimator of R}) evaluated at $u=0$.

\subsection{Variance estimation for the measurement errors}
The variance function of the response is $\sigma_{Y}^{2}(t) = \var \{Y(\bs, t)\} = R(0, t, t) + \Lambda(t,t) +\sigma_{\epsilon}^{2} = \Gamma (t,t) +\sigma_{\epsilon}^{2}$. We estimate $\sigma_{Y}^{2}(t) $ by the following spline estimator, 
\begin{equation} \label{eq: sigma_Y_hat}
\wh{\sigma}_{Y}^{2}(\cdot) = \underset{g(\cdot) \in \CS^\epsilon_{[1]}}{\hbox{argmin}} \int_{\CD_{n}}\int_{T} \left\lbrace Y^{2}(\bs,t)-g(t)\right\rbrace^{2} \CN_{t}(dt|\bs)\CN_{s}(d\bs),
\end{equation}
where $\CS^\epsilon_{[1]}$ is a univariate spline space of order $p_{\epsilon}$ defined on $K_{\epsilon}$ interior knots.
The following variance estimator is similar in spirit with those proposed by \cite{yao2005functional} 
\begin{equation}
\wh{\sigma}_{\epsilon}^{2} = \frac{1}{|T|}\int_{T}\{\wh{\sigma}_{Y}^{2}(t) - \wh{\Gamma}(t, t) \}dt.
\end{equation}
Both  $\wh\sigma_\epsilon^2$ and $\wh\Lambda$ are important quantities we will later use for functional kriging.

\begin{remark}
	Our estimation procedure involves integration of (multivariate) spline functions, when calculating $\wh{\Omega}(\cdot, \cdot)$, $\wh{\psi}_{j}(\cdot)$, $\wh{\CC}_{j}(\cdot)$ and $\wh{\sigma}_{\epsilon}$. 
	In our R code that supplements this paper, we compute the exact values of these integrals, using close-form expressions for integrals and the Gram matrix of B-splines \citep{de2001practical}. 
\end{remark}

\section{Theoretical properties} \label{Section: theory}
One important theoretical challenge in our problem is that there is only one copy of the spatio-temporal random field and all data are correlated. Under such a setting, it is well-known that infill asymptotics may lead to inconsistent estimation of spatial covariance \citep{zhang2005towards}. We therefore adopt a theoretical framework that combines both the infill and increasing domain asymptotic paradigms. \cite{Lu2014} proposed a different way to combine the increasing domain and infill paradigms, which does not rely on point process modeling of the sampling locations, but their message was in line with ours that we need to combine the two asymptotic paradigms for good statistical properties and flexible modeling of the data.

For any function $f(\cdot)$ (univariate or multivariate) defined on a compact support, denote $\|f\|_{L^2}$ and $\|f\|_{\infty}$ as its $L^2$ and $L^\infty$ norms. 
For any positive sequences $\{a_n\}$ and $\{b_n\}$, we write $a_n\lesssim b_n$ if $a_n/b_n$ is bounded above by a constant, and    $a_n \asymp b_n$ if $C_1\le a_n/b_n \le C_2$ for all $n$ and some $C_1, C_2>0$.
For any subset $E \subset \mathbb{R}^{2}$, let $\CF_{X}(E)$ be the $\sigma$-algebra generated by $\{X(\bm{s}, t): (\bm{s}, t) \in E \times T\}$. Suppose the spatial dependence of the functional data can be described by the $\alpha$-mixing coefficients \citep{rosenblatt1956remarks}:
\begin{eqnarray} \label{Exp: strong mixing coefficient}
\alpha_{X}(h)&=&\sup\limits_{\substack{E_1, E_2  \subset \mathbb{R}^{2}\\
		dist(E_{1}, E_{2})\geq h}}\sup\limits_{\substack{A_{1} \in \CF_{X}(E_{1}), \\A_{2}\in \CF_{X}(E_{2})}}|P(A_{1}\cap A_{2})-P(A_{1})P(A_{2})|,
\end{eqnarray}
where $dist(E_{1}, E_{2})$ denotes the minimal Euclidean distance between $E_{1}$ and $E_{2}$. We make the following assumptions for our theoretical investigation. 

\begin{asmp}\label{asmp:increase} 
	While the time domain $T$ is fixed, consider a sequence of spatial domains $\{\CD_n\}$ with the same shape such that, as $n\to\infty$, $C_{1} n \leq |\CD_{n}| \leq C_{2} n$, and $C_{1}\surd{n} \leq |\partial \CD_{n}| \leq C_{2}\surd{n} $, for some $C_{1}, C_{2}> 0$. Here, $|\CD_n|$ and $|\partial \CD_n|$ are the area and perimeter of $\CD_n$. 
\end{asmp}

\begin{asmp} \label{Assumption: 4th moment} Assume $X(\bs,t)$ is strictly stationary in $\bs$ and, for some $\nu>4$, $\sup\limits_{ t\in T}E|X(\bs, t)|^{\nu} < \infty$ and $\sup\limits_{t \in T} E | U(t) | ^\nu<\infty$.
\end{asmp}

\begin{asmp} \label{Assumption: strong mixing}
	The $\alpha$-mixing coefficient (\ref{Exp: strong mixing coefficient}) is well defined for $X(\bs,t)$, and there exist constants $\delta_1 > 2 \nu/ (\nu-4)$ and $C>0$ such that $\alpha_{X}(h)  \le C h^{-\delta_1}$ for all $h\ge 0$ 
	\citep{guyon1995random}.
\end{asmp}


\begin{asmp}\label{Assumption:lambda_s}
	Suppose $\CN_{s}(d\bs)$ is also $\alpha$-mixing with the coefficient, denoted as $\alpha_\CN (h)$, similarly defined as (\ref{Exp: strong mixing coefficient}), and assume $\alpha_{\CN}(h) \le C \exp(-\delta_2 h)$ for some $C>0$ and $\delta_2>0$. There exists a sequence of positive numbers $\{L_{n}\}$, that is either constant or monotonically increasing to infinity with $n$, and constants $C_2> C_1> 0$ such that  
	$C_1 L_{n}^k \le  \lambda_{s,k}(\bs_{1},\ldots, \bs_{k})\le L_{n}^k C_2$ for $k=1,\ldots,4$ and all $\bs_{1},\ldots,\bs_{4} \in \CD_{n}$. 
\end{asmp}

\begin{asmp}\label{Assumption:lambda_t}		
	Let $M_{n}$ be a sequence of positive constants depending on $n$, such that there exist some $C_1, C_2>0$ such that $ C_1 M_{n}^k \le \lambda_{t,k}(t_{1},\ldots, t_{k}) \le C_2 M_{n}^{k}$ for all $t_1, t_2\in T$ and $k=1,2$.
\end{asmp}

\begin{asmp}\label{Assumption: knots} As $n\to \infty$, both $K_s$ and $K_t\to \infty$, and $K_sK_t^{2} = o\left\lbrace {n}/{\log^2(n)}\right\rbrace$.
\end{asmp}

\begin{asmp}\label{Assumption:R} Restricting $R(\cdot,\cdot,\cdot)$ on the compact 3-dimensional domain $H=[0,\Delta]\times T\times T$, for order $\bm{r}=(r_{1}, r_{2}, r_{3})$ and $a>0$, define the H\"older class of functions on $H$ as $C_{3}^{\bm{r},a}(H):=\{f:\underset{\bm{x}_{1}\bm{x}_{2}\in H }{sup}|f^{(\ell_1, \ell_2, \ell_3)}(\bm{x}_{1})-f^{(\ell_1, \ell_2, \ell_3)}(\bm{x}_{2})|/\|\bm{x}_{1}-\bm{x}_{2}\|^{a}<\infty,  0\le \ell_i \le r_i, i=1,2,3\}$. Assume that $R \in C_{3}^{\bm{p},a}$, where $\bm{p}=(p_{s}, p_{t}, p_{t})$ is the order of the 3-dimensional tensor product spline function and $a >0$.
\end{asmp}

\begin{asmp}\label{Assumption:Gamma_Lambda}
	Define a class of bivariate H\"older continuous functions on $T^{2}$ as $C_{2}^{\bm{r},a}(T^{2}):=\{f:\underset{\bm{x}_{1}\bm{x}_{2}\in T^{2} }{sup}|f^{(\ell_{1},\ell_{2})}(\bm{x}_{1})-f^{(\ell_{1}, \ell_{2})}(\bm{x}_{2})|/\|\bm{x}_{1}-\bm{x}_{2}\|^{a}<\infty, \bm{r}=(r_{1}, r_{2}), 0\leq \ell_{1} \leq r_{1}, 0\leq \ell_{2} \leq r_{2} \}$. Assume that $\Gamma(\cdot,\cdot)$ and $\Lambda(\cdot, \cdot) \in C_{2}^{(p_{t}, p_{t}),a}\left(T^{2}\right)$, where $a>0$.
\end{asmp}


Assumption \ref{asmp:increase} describes a typical increasing domain asymptotic framework \citep{guan2004nonparametric}. A rectangular or circular spatial domain $\CD_n$ with the same shape but increasing area would satisfy Assumption \ref{asmp:increase}. Assumption \ref{Assumption: 4th moment} is a standard moment condition in functional data analysis \citep{li2010uniform}. Assumption \ref{Assumption: strong mixing} allows the spatial dependency in $X(\bs, t)$ to decay in a slow polynomial rate. In Assumption \ref{Assumption:lambda_s}, we assume that the sampling spatial point process is also weakly dependent and there is a positive chance to sample any four points in $\CD_n$. A homogenous Poisson process would satisfy Assumption \ref{Assumption:lambda_s}. By allowing $L_n\to \infty$, our framework also accommodates the infill paradigm, meaning we allow $\lambda_{s,k}(\cdot)$ and hence the expected number of sampling points on any unit space to diverge to infinity. It is also worth pointing out that the expected number of repeated measures on $Y(\bs_i,\cdot)$ is $\int_T \lambda_{t,1}(t) dt \asymp M_n$ under Assumption \ref{Assumption:lambda_t}. When $M_n$ are bounded by a constant, the data are spatially correlated sparse functional data; on the other hand, if $M_n\to \infty$ fast enough as a function of $n$, the data are dense functional data. In all of our theoretical results below, we allow $M_n$ to be of any rate relative to $n$, thus admit all types of functional data in a unified framework. Assumption \ref{Assumption: knots} is a standard assumption on the number of knots and sets a range for the tuning parameters. Assumptions \ref{Assumption:R} and \ref{Assumption:Gamma_Lambda} govern the smoothness of the functions that we estimate.

%
%
%
%
%
%
%
%


The following theorem provides the asymptotic convergence rate for the tensor-product spline estimator of the spatio-temporal covariance function.

\begin{thm} \label{thm:3dim covariance} Under the model framework in Section \ref{Section: Model} and Assumptions \ref{asmp:increase} -- \ref{Assumption:R}, 
	\bse	
	\| \wh{R} -R\|_{L^{2}}   =O_{p}\left[ | \CD_{n}| ^{-1/2} \{ \sqrt{K_{s}} +\sqrt{ K_{s}K_{t} / ( M_{n}L_{n})}+\sqrt{ K_{s}K_{t}^{2} / (M_{n}^{2}L_{n}^2  )} \} +K_{s}^{-p_{s}}+K_{t}^{-p_{t}}\right].
	\ese
\end{thm}

\begin{remark}[Effect of Infill]
Theorem \ref{thm:3dim covariance} implies that the most dominating factor in achieving consistent covariance estimation is the domain size $|\CD_n|$. The infill factor $L_n$ only plays a secondary role in the convergence rate: letting $L_n \to \infty$ but holding $|\CD_n|$ fixed will result in an inconsistence covariance estimator, which is in agreement with the results of \cite{zhang2005towards} and \cite{hormann2013consistency}. Intuitively, increasing the sampling locations in a unit spatial domain will result in increasingly correlated data but not more information that is equivalent to independent samples. The factor $L_n M_n$ measures the number of spatio-temporal measurements in a unit spatial neighborhood. In an ideal case $L_n\to\infty$ in a fast enough rate so that we can choose $ K_s \lesssim K_t \lesssim L_nM_n$, the dominant terms in $ \| \wh{R} -R\|_{L^{2}}$ are of order  $O_p( K_s ^{1/2} |\CD_n|^{-1/2} +K_s^{-p_s})$.
\end{remark}

\begin{remark}[Phase Transition from Sparse to Dense Functional Data]
	For simplicity, the following discussion is restricted to a standard increasing domain framework where $|\CD_n| \to \infty$ and $L_n$ is a fixed constant.
	For sparse functional data where $M_{n}$ is a bounded constant, assume $K_s=K_t\equiv K$ and $p_s=p_t\equiv p$ for simplicity, then the result in Theorem \ref{thm:3dim covariance} can be simplified to $\|\wh R-R\|_{L^2}= O_p( K^{3/2} |\CD_n|^{-1/2} +K^{-p})$. Since $|\CD_n| \asymp \E(N)$ is proportional to the sample size (i.e. the number of functions) under this setting, such a rate is the classic convergence rate for a 3-dimensional nonparametric regression using splines \citep{stone1994use}.
	For dense functional data with $M_n\gtrsim n^{1/(2 p_t)}$ and choosing $K_t \asymp M_n$, we have $\|\wh R-R\|_{L^2}= O_p( K_s ^{1/2} |\CD_n|^{-1/2} +K_s^{-p_s})$, which is the nonparametric convergence rate for estimating a stationary, isotropic spatial covariance function \citep{li2007nonparametric}. This result suggests $M_n\asymp n^{1/(2 p_t)}$ is a transition point \citep{li2010uniform, ZhangWang2016AOS, Wang2018jrssb}, where estimating the 3-dim spatio-temporal covariance function is as efficient as estimating a 1-dim spatial covariance, and further increasing the number of repeated measures on each curve would not improve the convergence rate of $\wh R$.
\end{remark}

The bivariate function $\Omega(\cdot, \cdot)$ in (\ref{eq:Omega}) is of fundamental importance to our FPCA methodology, where we borrow spatial information up to a distance $\Delta > 0$. The following theorem provides the convergence rate of $\wh \Omega$.

\begin{thm} \label{thm:2dim covariance} 
	Under the assumptions in Theorem \ref{thm:3dim covariance} and the coregionalization structure in (\ref{Exp: cross-variance function}),
	$
	%
	\| \wh{\Omega}-\Omega\|_{L^{2}} = O_{p}\left[ | \CD_{n}|^{-1/2} \{ 1+  \sqrt{K_{t} / (M_{n}L_n)} \}+K_{s}^{-p_{s}}+K_{t}^{-p_{t}}\right].
	$
	%
\end{thm}
\begin{remark}
	By integrating over the spatial dimension of $\wh R$, we apply another step of smoothing and therefore obtain a faster convergence rate for $\wh \Omega$ than $\wh R$. 
	By undersmoothing in the spatial direction letting $K_s\gtrsim n^{1/(2p_s)}$, the $O_p(K_s^{-p_s})$ nuisance of estimating spatial covariance becomes negligible, then the rate in Theorem \ref{thm:2dim covariance} is comparable to the classic covariance estimation convergence rate \citep{li2010uniform} for independent functional data using kernel smoothing. The convergence rate above becomes a typical bivariate spline smoothing rate $O_p( K_t/ |\CD_n |^{1/2} +K_t^{-p_t})$ when the data are sparse (the total number of measurements in a unit area $ L_nM_n$ is bounded); and the root-$n$ convergence rate, $\| \wh{\Omega}-\Omega\|_{L^{2}} = O_{p}(| \CD_{n}|^{-1/2})$, is attainable, if the data are dense enough with $L_n M_n\gtrsim n^{1/(2p_t)}$ and if we choose $K_t \asymp L_n M_n$.
\end{remark}

The convergence rate for $\wh \psi_j(t)$ is a direct result from the perturbation theory in \cite{hall2006aproperties} and is provided in the following theorem.

\begin{thm} \label{thm:functional principal components} 
	Under the assumptions in Theorem \ref{thm:2dim covariance} and suppose all eigenvalues of $\Omega(\cdot, \cdot)$ are distinct,
		\bse	
	\| \wh{\psi}_{j}-\psi_{j}\|_{L^{2}} = O_{p}\left[ | \CD_{n}|^{-1/2}\{ 1+\sqrt{ K_{t}/ (M_{n}L_n)} \} +K_{s}^{-p_{s}}+K_{t}^{-p_{t}}\right],
	\ese
	for $j=1,2,\ldots, J$, up to any fixed order $J$.
\end{thm}

\begin{remark}
	Results in Theorem \ref{thm:functional principal components} are comparable to those in \cite{hall2006bproperties} and \cite{li2010uniform} for independent functional data. For sparse functional data where $L_nM_n$ is bounded by a constant, by adopting an undersmoothing strategy in the spatial direction (i.e. $K_s \gtrsim n^{1/(2p_s)}$), we get $\| \wh{\psi}_{j}-\psi_{j}\|_{L^{2}} = O_p\{ (K_t / |\CD_n|)^{1/2} +K_t^{-p_t}\}$. This is a 1-dim spline smoothing convergence rate, even though $\wh \psi_j(t)$ is a byproduct of a 2-dim nonparametric estimator $\wh \Omega(\cdot, \cdot)$ that converges in a slower 2-dim rate. For dense functional data ($L_n M_n\gtrsim n^{1/(2p_t)}$), by choosing $K_t \asymp L_n M_n$, we get $\| \wh{\psi}_{j}-\psi_{j}\|_{L^{2}} = O_{p} (| \CD_{n}|^{-1/2})$, which is a root-$n$ rate.
\end{remark}

%
%
%
%
%
%
%
%

Restricting $\CC_j(u)$ and $\widehat{\CC}_{j}$ on $[0,\Delta]$, the following theorem provides convergence rates for the estimated spatial covariance functions.

\begin{thm} \label{thm: spatial correlation}	
	Under the assumptions of Theorem \ref{thm:functional principal components},
	\bse	
	\| \widehat{\CC}_{j}- \CC_{j}\|_{L^{2}} = O_{p}\left[ | \CD_{n}|^{-1/2} \{ \sqrt{K_{s}}+\sqrt{K_{t}/( M_{n}L_n)}\}  +K_{s}^{-p_{s}}+K_{t}^{-p_{t}}\right],
	\ese	
	for $j=1,2,\ldots, J$ up to any fixed order $J$.
\end{thm}

\begin{remark}
	Suppose the covariance function $R$ is smoother in the temporal directions than the spatial direction, i.e. $p_t\ge  p_s$, by choosing $K_s^{p_s/p_t} \lesssim K_t \lesssim K_s$, the convergence rate in Theorem \ref{thm: spatial correlation} becomes $O_{p}\left\{ (K_{s}/| \CD_{n}|)^{1/2}+K_{s}^{-p_{s}}\right\}$, which is comparable to the results in \cite{li2007nonparametric} developed for 1-dimensional spatial domain, multivariate response and under a rather stringent separable covariance assumption.
\end{remark}

With the additional smoothness conditions in Assumption \ref{Assumption:Gamma_Lambda}, 
we have the following results on the covariance estimator $\wh \Lambda$ for the functional nugget effect and the variance estimator $\wh\sigma_\epsilon^2$ for the measurement errors.
\begin{thm} \label{thm: nugget effect}
	Under Assumptions \ref{asmp:increase}--\ref{Assumption:Gamma_Lambda} and assume $K_\Gamma \asymp K_t$ and $p_{\Gamma}=p_{t}$, 
	$	
	\|\widehat{\Lambda} - \Lambda \|_{L^{2}} =  O_{p}\left [ | \CD_{n}|^{-1/2} \{ \sqrt{K_{s}}+\sqrt {K_{s}K_{t}/ (M_{n}L_n)}+\sqrt{K_{t}^{2}/ (M_{n}^{2}L_n)}+\sqrt{K_{s}K_{t}^{2}/ (M_{n}^{2}L_n^2)}\}+K_{s}^{-p_{s}}+K_{t}^{-p_{t}} \right].
	$	
\end{thm}

\begin{thm} \label{thm: measurement error}
	Under Assumptions \ref{asmp:increase} -- \ref{Assumption:Gamma_Lambda} and further assume $K_\Gamma \asymp K_\epsilon \asymp K_t$ and $p_\Gamma =p_{\epsilon}=p_{t}$, 
	$
	\wh{\sigma}_{\epsilon}^{2} - \sigma_{\epsilon}^{2} = O_{p}\left[ | \CD_{n}|^{-1/2} \{ 1+\sqrt{ K_{t}/  (M_{n}L_n )} \} +K_{t}^{-p_{t}}\right].
	$
\end{thm}

\begin{remark}
	%
	The convergence rate of $\wh\sigma_\epsilon^2$ in Theorem \ref{thm: measurement error} is comparable to Theorem 3.4 of \cite{li2010uniform} for independent functional data. 
	Both $\wh\Lambda$ and $\wh\sigma_\epsilon^2$ are important quantities we will later use for functional kriging.
\end{remark}

\section{Implementation} \label{Section: Implementation}

\subsection{Positive semi-definite adjustment for spatial covariance functions}
The spatial covariance functions $\left\lbrace\CC_{j}(u): j=1, \cdots, J\right\rbrace$ are required by definition to be positive semi-definite in $\mathbb{R}^{2}$, meaning 
$\int\int \CC_{j} (\|\bs_{1}-\bs_{2}\|)a(\bs_{1})a(\bs_{2})d\bs_{1}d\bs_{2} \geq 0$, 
for any integrable functions $a(\cdot)$ defined on $\mathbb{R}^{2}$. The spline estimators $\wh C_j (u)$ defined in (\ref{eq:C_j_hat}), even though consistent, are not guaranteed to be positive semidefinite. Nevertheless, this violation can be easily corrected using a correction procedure similar to that used in \cite{hall1994nonparametric}.

By Bochner's theorem \cite[p.~141]{schabenberger2017statistical},  $\CC_j(u)$ is positive semidefinite if $\CC^{+}_{j}(\theta) \geq 0$ for all $\theta$, where 
$\CC^{+}_{j}(\theta) = \int_{0}^{\infty}\CC_{j}(u)J_{0}(\theta u)udu$ is the Hankel transformation of $\CC_{j}(\cdot)$ and $J_{0}(\cdot)$ is the Bessel function of the first kind with order $0$.
This motivates us to take a nonnegative truncation on the Hankel transformation of $\wh{\CC}_{j}(\cdot)$, i.e., 
$
\wh{\CC}^{+}_{j}(\theta) = \max\left\{\int_{0}^{\infty}\wh{\CC}_{j}(u)J_{0}(\theta u)udu, 0\right\}
$.
In practice, $\CC_{j}(u)$ decays to zero beyond the range of spatial dependence and $\wh \CC_j(u)$ is unstable for a large $u$. We therefore multiply $\wh{\CC}_{j}$ by a weight function $w(u) \leq 1$ when taking the Hankel transformation,
\begin{equation}\label{eq:hankel}
\wh{\CC}^{+}_{j}(\theta) = \max\left\{\int_{0}^{\infty}\wh{\CC}_{j}(u) J_{0}(\theta u) w(u) u d u, 0 \right\}.
\end{equation}
Possible choices of $w(\cdot)$ suggested by \cite{hall1994nonparametric} are $w_{1}(u) = I(|u|\leq D)$ for a threshold $D>0$; and $w_{2}(u) = 1$ if $|u|<D_{1}$, $(D_{2}-|u|)/(D_{2}-D_{1})$ for $D_{1} \leq |u| \leq D_{2}$ and $0$ if $|u|>D_{2}$.
Then the adjusted covariance estimators are the inverse Hankel transformations
\begin{equation}
\widetilde{\CC}_{j}(u) = \int_{0}^{\infty} \wh{\CC}_{j}^{+}(\theta) J_{0}(\theta u)\theta d\theta.
\end{equation}
And the correlation functions are adjusted as $\widetilde{\rho}_j(u) = \widetilde{\CC}_{j}(u)/\widetilde{\CC}_{j}(0)$ and an adjusted estimator for the spatio-temporal covariance function $R(\cdot, \cdot, \cdot)$ can be constructed as 
\begin{equation}
\wt{R}(u,t_1,t_2) =\sum_{j=1}^{J}\wt{\CC}_{j}(u)\wh{\psi}_{j}(t_1)\wh{\psi}_{j}(t_2),
\end{equation}
where $J$ is a large enough number such that the first $J$ principal components capture most of the variation in the data. For the choice of the weight function in (\ref{eq:hankel}), we use $w_1(u)=I(|u|\le D)$ and set $D=\Delta$ in all of our numerical studies, which leads to satisfactory results.

\subsection{Choosing the number of B-spline knots}\label{subsection: knot selection}
The amount of smoothing in our spline covariance estimator $\wh R$ is governed by the numbers of knots $K_{s}$ and $K_{t}$. Following \cite{huang2004identification}, we choose these tuning parameters by minimizing the following Bayesian Information Criterion (\BIC) 
\begin{equation}
\BIC(K_{s}, K_{t}) = \wt{N}\log \{\CL(\wh{\bdbeta})\} + df \times \log(\wt{N}),
\end{equation}
where $\CL(\cdot)$ is the square loss function defined in (\ref{Exp: estimated beta}), the degree of freedom $df=(K_{s}+p_{s})(K_{t}+p_{t})^{2}$ is the total number of tensor product B-spline basis functions, and
$\wt{N}=\int_{\CD_{n}}\int_{\CD_{n}}\int_{T}\int_{T}I(\|\bs_{1}-\bs_{2}\|\leq \Delta) \CN_{t}(dt_{1}|\bs_{1})\CN_{t}(dt_{2}|\bs_{2})
\CN_{s,2}(d\bs_{1},d\bs_{2})$ is the total sample size for estimating $R(\cdot, \cdot, \cdot)$.
Similar BIC criteria are used to choose the number of knots in $\wh \Gamma(\cdot, \cdot)$ and $\wh \sigma_Y^2(\cdot)$.
%
%
%
%
%
%
%

\subsection{Estimation of the mean function} \label{subsection: Estimation of the mean function}
Up to this point, we assume $\mu(t)\equiv 0$. In practice, we first estimate $\mu(t) $ by 
\begin{equation} \label{Exp: mean function estimation}
\wh{\mu}(\cdot) = \underset{g(\cdot) \in \CS_{K_{m}}^{p_{m}}[0,1]}{\hbox{argmin}}
\int_{\CD_{n}}\int_{T} \left\lbrace Y(\bs,t)-g(t)\right\rbrace^{2}\CN_{t}(dt|\bs)\CN_{s}(d\bs),
\end{equation}
where $\CS_{K_{m}}^{p_{m}}[0,1]$ is a spline space with order $p_{m}$ and $K_{m}$ interior knots, and then proceed with the methods described in Section \ref{Section: Method} using the centered response $\wt Y(\bs_i, t_{ij})= Y(\bs_i,t_{ij})-\wh \mu(t_{ij})$. For fully observed functional data with simple parametric spatial covariance  and no measurement error, \cite{KokoszkaReimherr2017} proposed a method to improve estimation efficiency for the mean function taking into account the spatial dependence. However, it is not yet clear how to extend this method to the discretely observed functional data with non-separable covariance structures in our paper, especially with the complication of functional nugget effect and measurement error.

\section{Kriging of spatially dependent functional data} \label{Section: Kriging}
Spatial prediction or kriging is a major interest in spatial statistics \citep{stein2012interpolation} and there has been some recent work on kriging for spatially dependent functional data.
The FPCA-then-kriging two-step procedure \citep{nerini2010cokriging, menafoglio2016universal} is to first perform the classic FPCA \citep{yao2005functional} ignoring any spatial dependence and then perform co-kriging on the estimated FPC scores by fitting parametric spatial covariance models such as those in the Mat\'ern family. There are several issues with this procedure: first, it does not consider functional nugget effect and, as shown in our simulation studies, may suffer from large estimation biases; 
second, the estimated FPC scores are contaminated with estimation errors, which bring a lot of nuisance into spatial covariance estimation; third, the spatial covariance models are limited to a few parametric families which may be mis-specified. The trace kriging method \citep{giraldo2011ordinary, menafoglio2013} does not depend on dimension reduction (e.g. FPCA) and requires fully observed functional data without measurement error nor nugget effect. 

%

We now propose a new functional kriging method under our model. Let $\bs_{0} \in \CD_{n}$ be a new location where no data are observed, and our goal is to predict the unobserved functional data $X(\bs_{0},t)$ using information from neighboring locations. Under our framework, $X(\bs_{0},t)=\mu(t)+\sum_{j=1}^{\infty}\xi_{j}(\bs_{0})\psi_{j}(t)$. In practice, the infinite principal component expansion of $X(\bs_0,t)$ needs to be truncated at a finite order $J$, which can be determined by a simple ``percentage of variation explained'' method \citep{yao2005functional}. We then predict $X(\bs_{0},t)$ by $ \wh{X}(\bs_{0},t)=\wh{\mu}(t)+\sum_{j=1}^{J}\wh{\xi}_{j}(\bs_{0})\wh{\psi}_{j}(t)$, where $\wh\xi_j(\bs_0)$ is the Best Linear Unbiased Predictor (BLUP) of $\xi_{j}(\bs_{0})$ using data collected from locations close to $\bs_0$. 

Let $\CN(\bs_0, \Delta)$ be the collection of sampled locations within a distance $\Delta$ from $\bs_0$, and $\BY_{\bs_0, \Delta}= \{ Y(\bs_i, t_{ij}), \bs_i\in \CN(\bs_0, \Delta)\}\trans$ be the vector of observed data from the neighboring locations. Similarly, let $\BX_{\bs_0, \Delta}=  \{ X(\bs_i, t_{ij}), \bs_i\in \CN(\bs_0, \Delta)\}\trans$ and $\BU_{\bs_0, \Delta}= \{ U_i ( t_{ij}), \bs_i\in \CN(\bs_0, \Delta)\}\trans$ be the latent random vectors in $\BY_{\bs_0, \Delta}$.  Suppose $\BR_{\bs_0,\Delta}=\cov( \BX_{\bs_0, \Delta})$ is the covariance matrix interpolated from the spatio-temporal covariance function $R(\cdot, \cdot, \cdot)$, $\BLambda_{\bs_0,\Delta}=\cov( \BU_{\bs_0, \Delta})$ is a block diagonal matrix representing the covariance of the functional nugget effect, then $\BSigma_{\bs_0, \Delta}=\cov(\BY_{\bs_0, \Delta})= \BR_{\bs_0,\Delta}+ \BLambda_{\bs_0,\Delta}+\sigma_\epsilon^2 \BI$ is the covariance matrix of the observed data within the neighborhood $\CN(\bs_0, \Delta)$. Define $\BUpsilon_{\bs_0, j}=\cov\{ \xi_j(\bs_0), \BY_{\bs_0, \Delta}\}= \{ \CC_j(\| \bs_i-\bs_0\|) \psi_j(t_{i\ell}), \bs_i\in \CN(\bs_0, \Delta)\}\trans$, then the BLUP for $\xi_j(\bs_0)$ is
\begin{equation} \label{Equation: BLUP}
\wh{\xi}_{j}(\bs_{0}) = \BUpsilon_{\bs_0, j}\trans \BSigma_{\bs_0, \Delta}^{-1} (\BY_{\bs_0, \Delta}-\bdmu_{\bs_0, \Delta}),
\end{equation}
where $\bdmu_{\bs_0, \Delta}=\E(\BY_{\bs_0, \Delta})$ is the mean vector interpolated from the mean function $\mu(t)$. 
The BLUP in (\ref{Equation: BLUP}) depends on unknown functions such as $R(\cdot, \cdot, \cdot)$, $\Lambda(\cdot, \cdot)$, $\CC_j(\cdot)$, $\psi_j(\cdot)$ and $\mu(\cdot)$, which we replace with the estimators proposed in Sections \ref{Section: Method} and \ref{Section: Implementation}.

%
%
%
%
%
%
%
%
%
%
%
%
%
%
%

\section{Simulation studies} \label{Section: Simulation}

We now illustrate the proposed methodology using simulation studies. Data are generated from model (\ref{eq:model}) in the spatial domain $\CD = [0, 10]^2$ and time domain $T=[0,1]$, with $X(\bs,t)=\mu(t)+\sum_{j=1}^{3}\xi_{j}(\bs)\psi_{j}(t)$, $\mu(t)=2t\sin(2\pi t)$, $\psi_{1}(t) = \sqrt{2}\cos(2\pi t)$, $\psi_{2}(t) = \sqrt{2}\sin(2\pi t)$ and $\psi_{3}(t) = \sqrt{2}\cos(4\pi t)$. The principal component scores, $\xi_{j}(\bs)$, $j=1,2,3$, are Gaussian random fields generated using the \textit{RandomFields} package in R. 
The variances of $\xi_{j}$'s are $(\varpi_1, \varpi_2, \varpi_3) = (3, 2, 1)$. Their spatial covariance functions are members of the Mat\'{e}rn family, $\CC_j(u; \nu, \rho) = \varpi_j {2^{1-\nu} \over \Gamma(\nu)} (\sqrt{2\nu} u/ \rho)^{\nu} K_\nu( \sqrt{2\nu} u/\rho)$, where $K_{\nu}(\cdot)$ is the modified Bessel function of the second kind with degree $\nu$. We set the shape parameter $\nu$ to be $5.5$, $3.5$ and $1.5$ and range parameter $\rho$ to be $1$, $0.5$ and $0.5$ respectively for the three principal components. 
The spatial locations $\{\bs_i\}$ are sampled from a homogeneous spatial Poisson process over $\CD$, with the first-order intensity $\lambda_{\bs} \equiv 10$; time of repeated measures on each function are sampled from a Poisson process over $T$ with $\lambda_{t} = 10$. 
The measurement errors $\epsilon_{ij} $ are generated as iid $ \mathrm{Normal}(0, \sigma_{\epsilon}^2)$, where $\sigma_{\epsilon}^2 = 0.25$. We consider two scenarios for the functional nugget effect.
\begin{itemize}
	\item Scenario A: functional nugget effect $U_{i}(t) = \sum_{j=1}^2 \xi_{\mathrm{nug}, j}(\bs_i) \psi_{\mathrm{nug}, j}(t)$, where $\psi_{ \mathrm{nug},1}(t)$ and $\psi_{\mathrm{nug}, 2}(t)$ are the first two basis functions in the normalized Fourier-Bessel Series, $\xi_{\mathrm{nug}, j} \sim \mathrm{Normal}(0,\omega_{\mathrm{nug},j})$, $j=1,2$, and $(\omega_{\mathrm{nug},1}, \omega_{\mathrm{nug},2} )= (2,1)$.
	\item Scenario B: no functional nugget effect, i.e., $Y(\bs_{i},t_{ij})=X(\bs_{i},t_{ij}) + \epsilon_{ij}$.
\end{itemize}

\begin{figure}
	\subfloat[$\wh{\psi}_1(t)$]{\includegraphics[width = 0.33\linewidth]{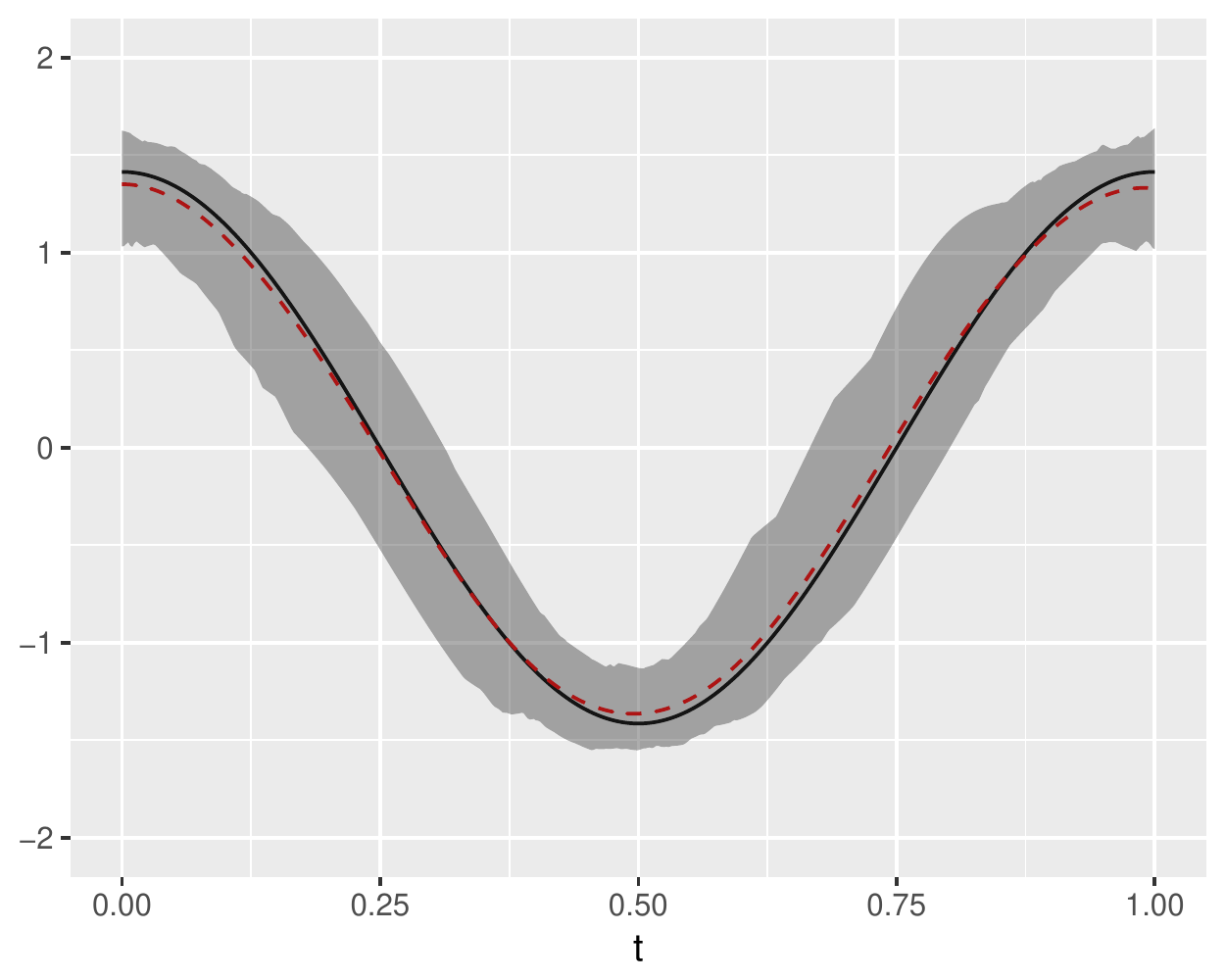}} 
	\subfloat[$\wh{\psi}_2(t)$]{\includegraphics[width = 0.33\linewidth]{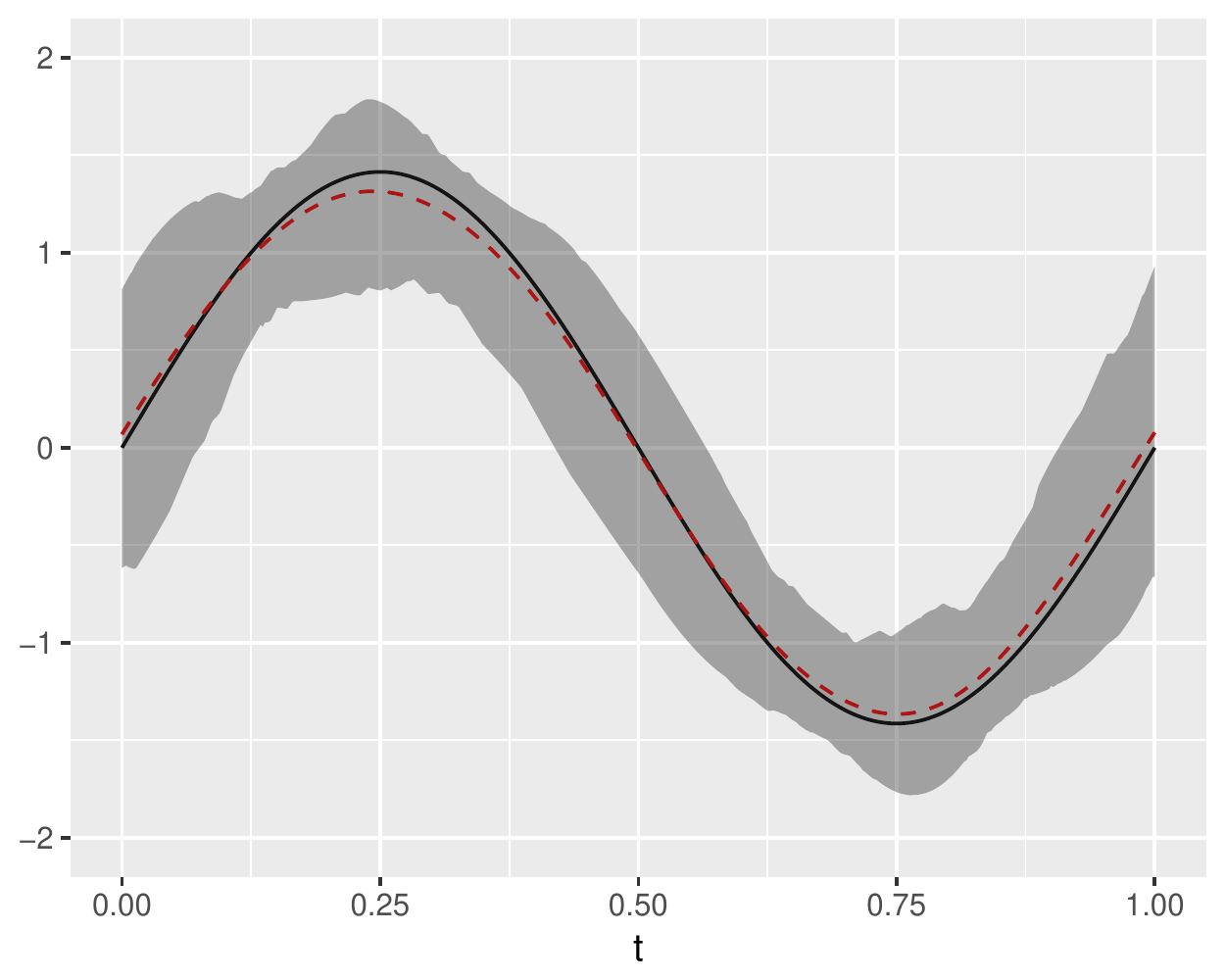}} 
	\subfloat[$\wh{\psi}_3(t)$]{\includegraphics[width = 0.33\linewidth]{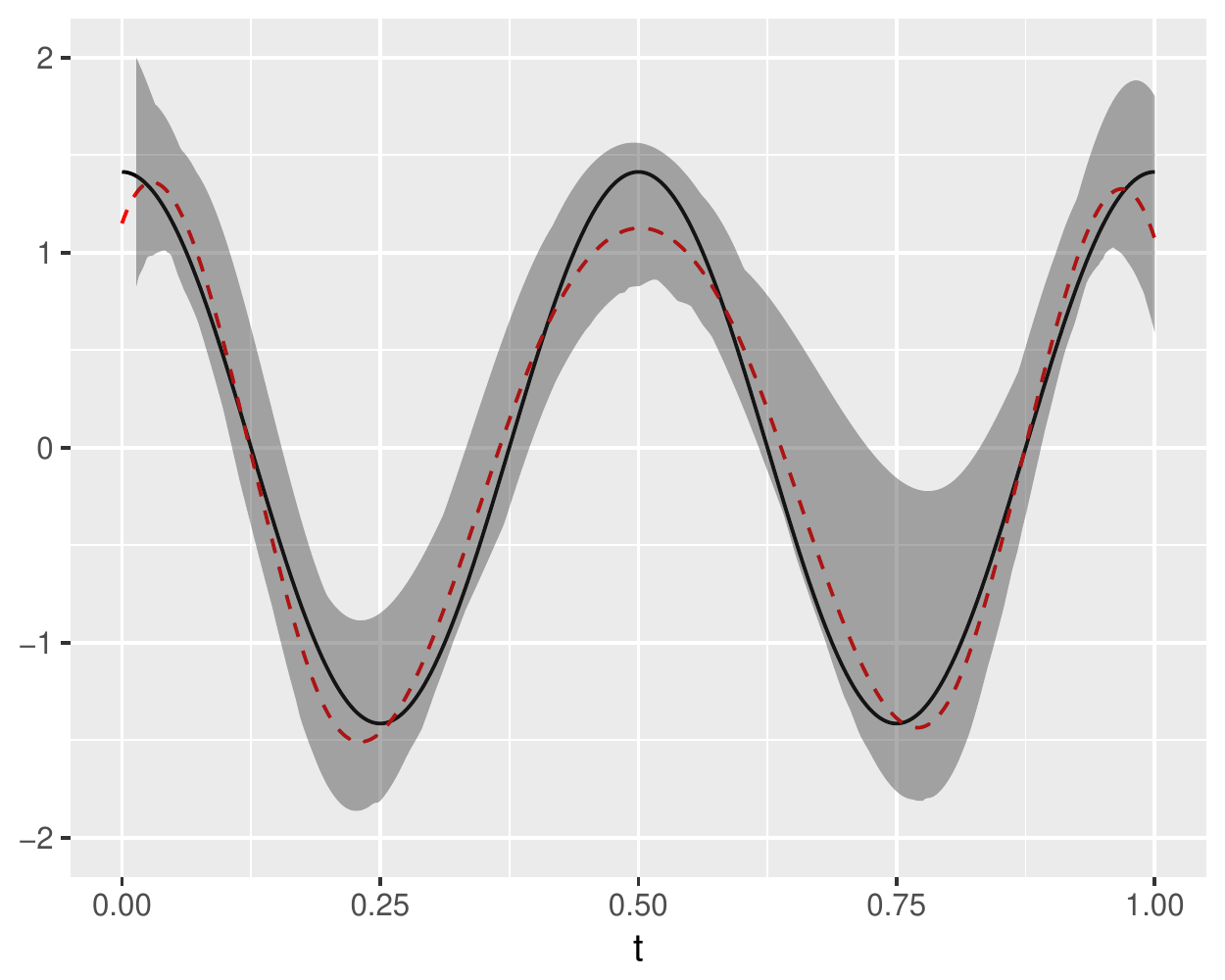}}\\
	\subfloat[$\wt{\CC}_1(u)$]{\includegraphics[width = 0.33\linewidth]{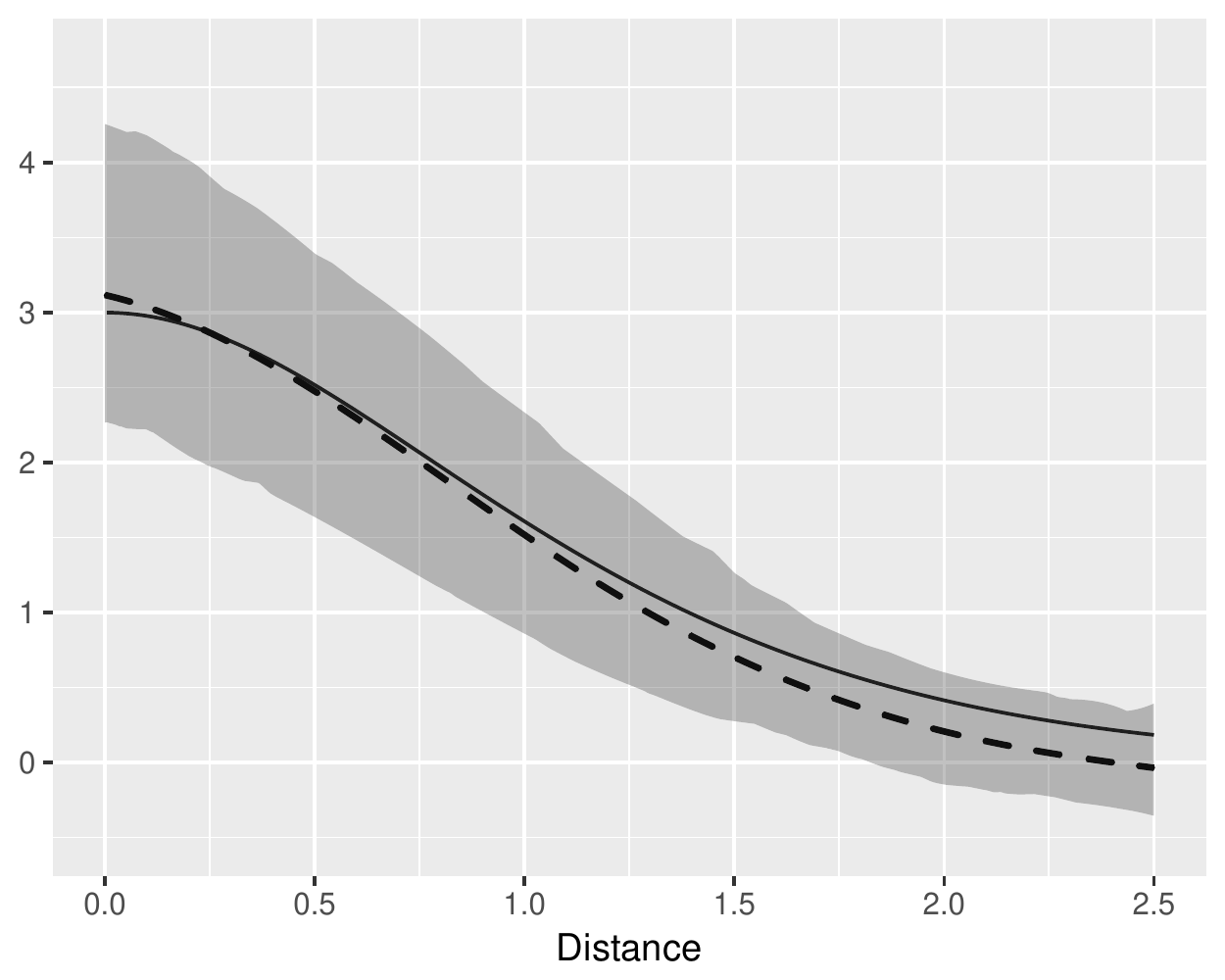}} 
	\subfloat[$\wt{\CC}_2(u)$]{\includegraphics[width = 0.33\linewidth]{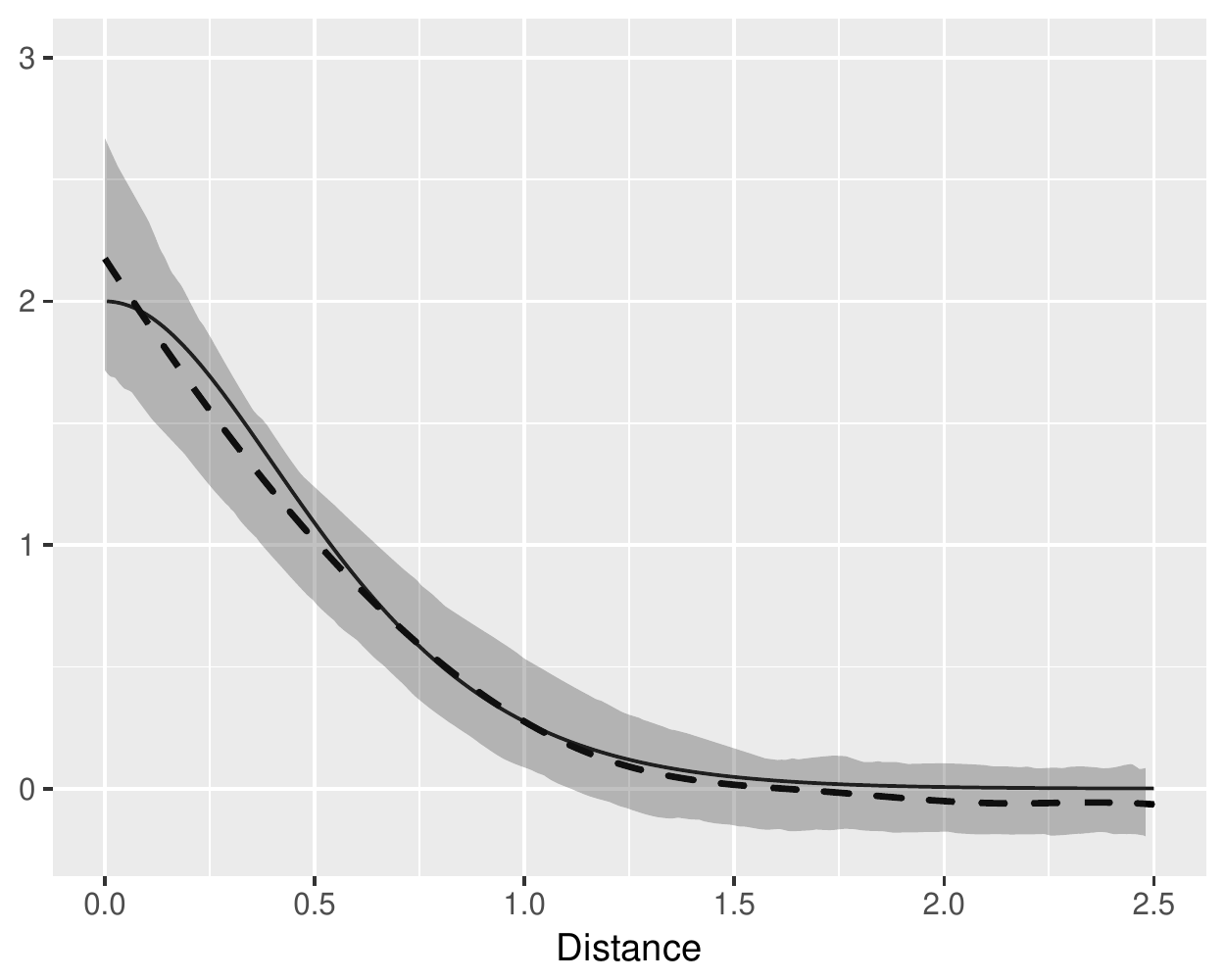}}
	\subfloat[$\wt{\CC}_3(u)$]{\includegraphics[width = 0.33\linewidth]{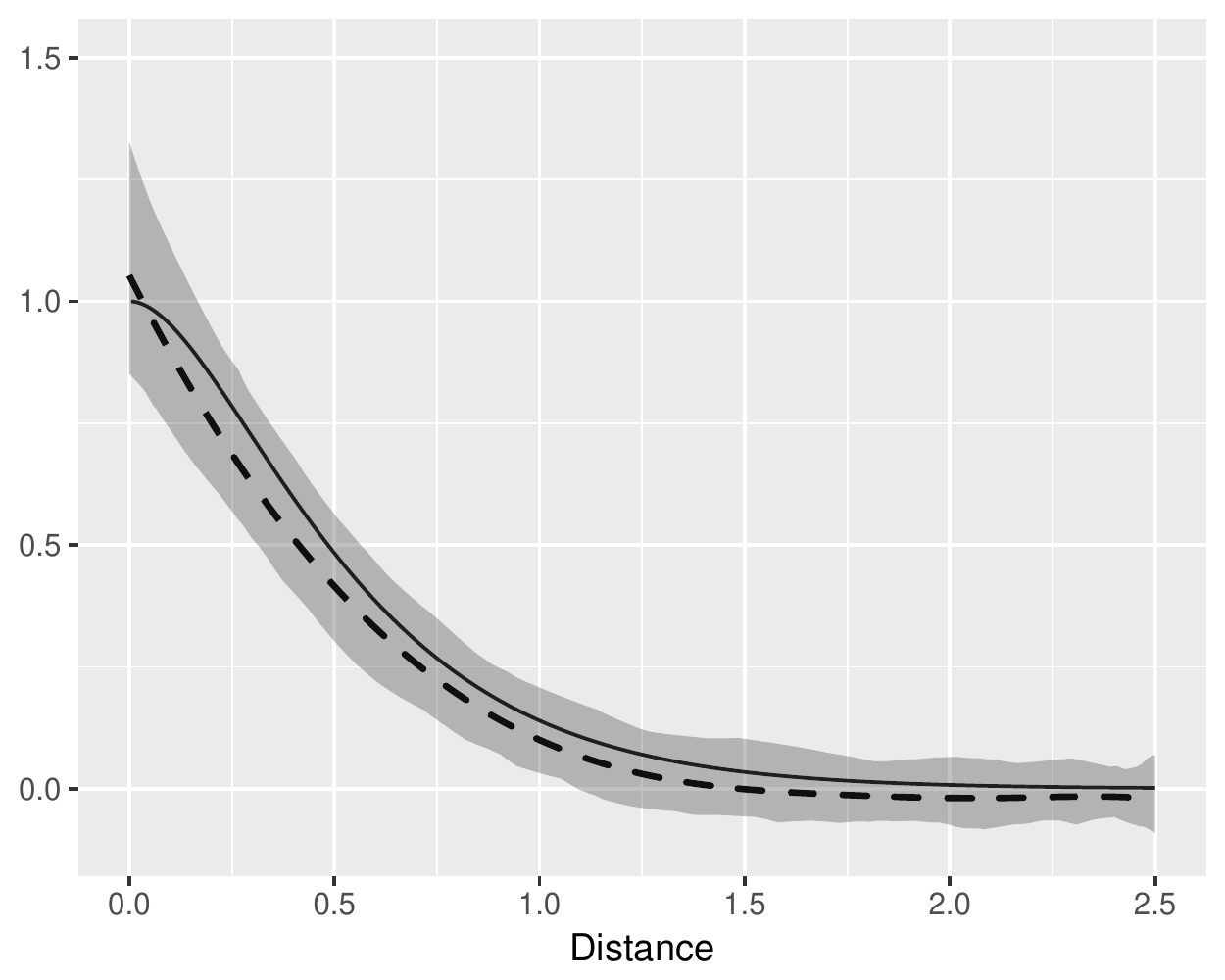}}\\
	\subfloat[$\wh{\psi}_{\mathrm{nug}, 1}(t)$]{\includegraphics[width =0.33\linewidth]{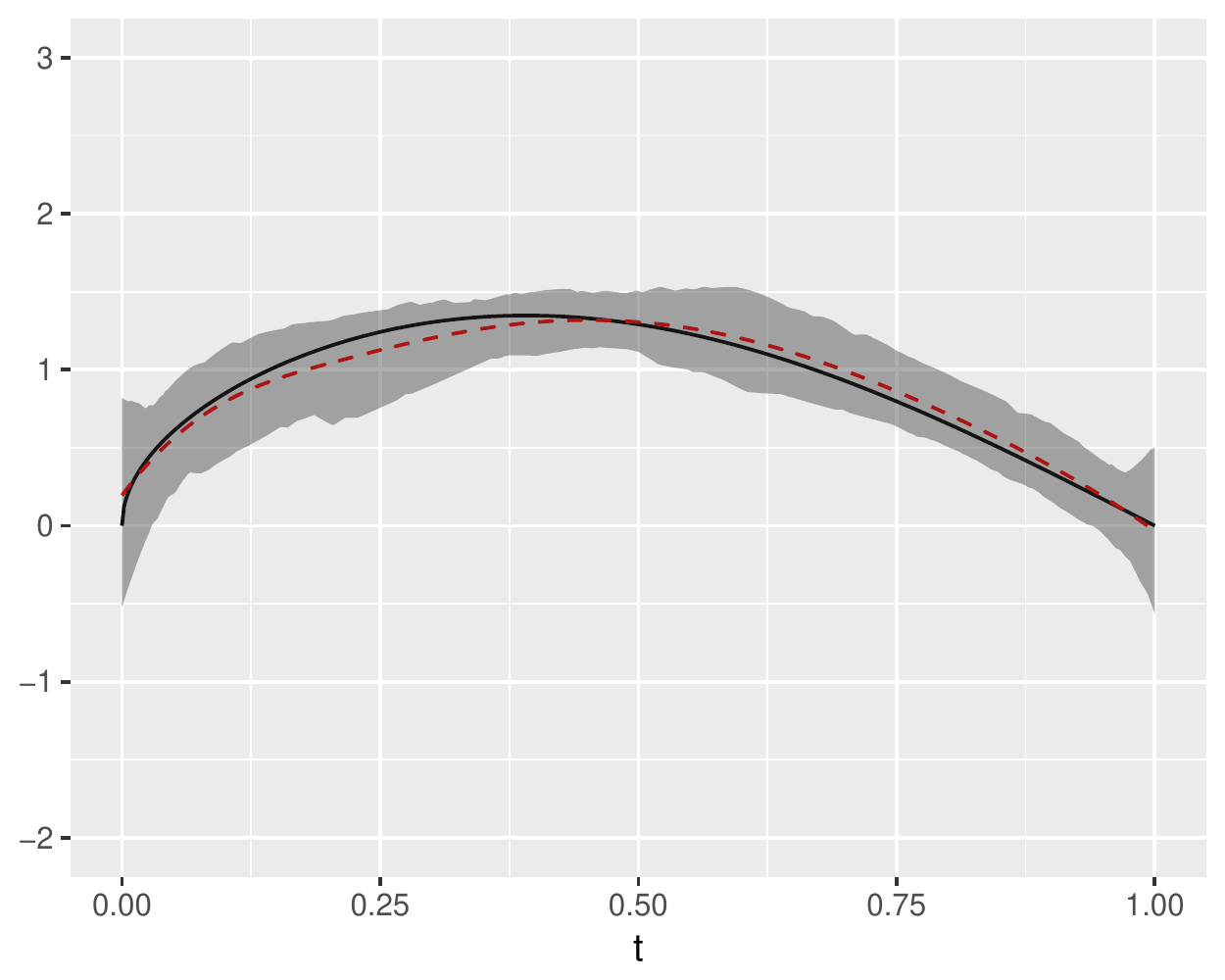}} 
	\subfloat[$\wh{\psi}_{\mathrm{nug}, 2}(t)$]{\includegraphics[width = 0.33\linewidth]{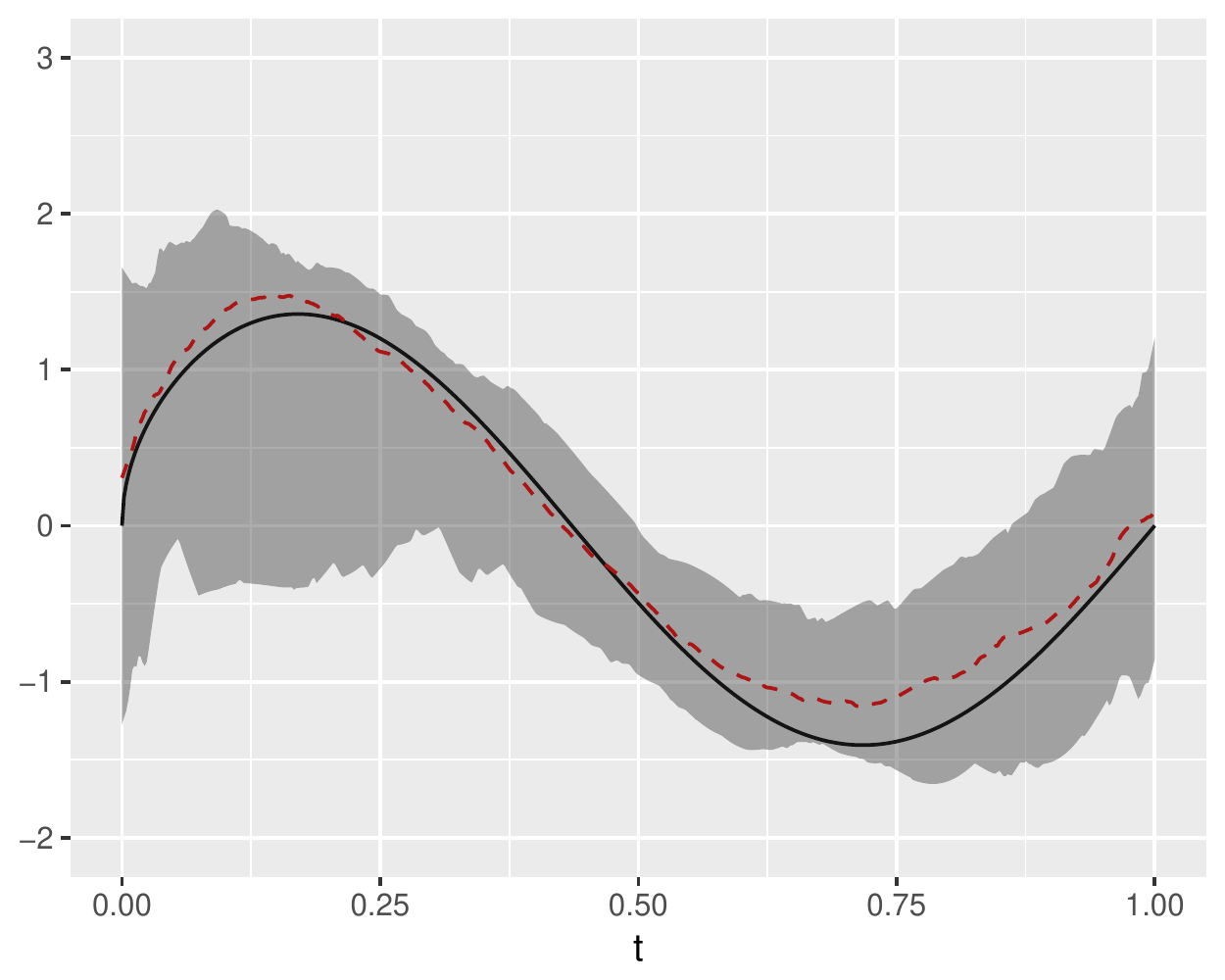}}
	\subfloat[$\wh{\varpi}_1, \wh{\varpi}_2, \wh{\varpi}_3, \wh{\omega}_{\mathrm{nug},1}, \wh{\omega}_{\mathrm{nug},2}$]{\includegraphics[width = 0.33\linewidth]{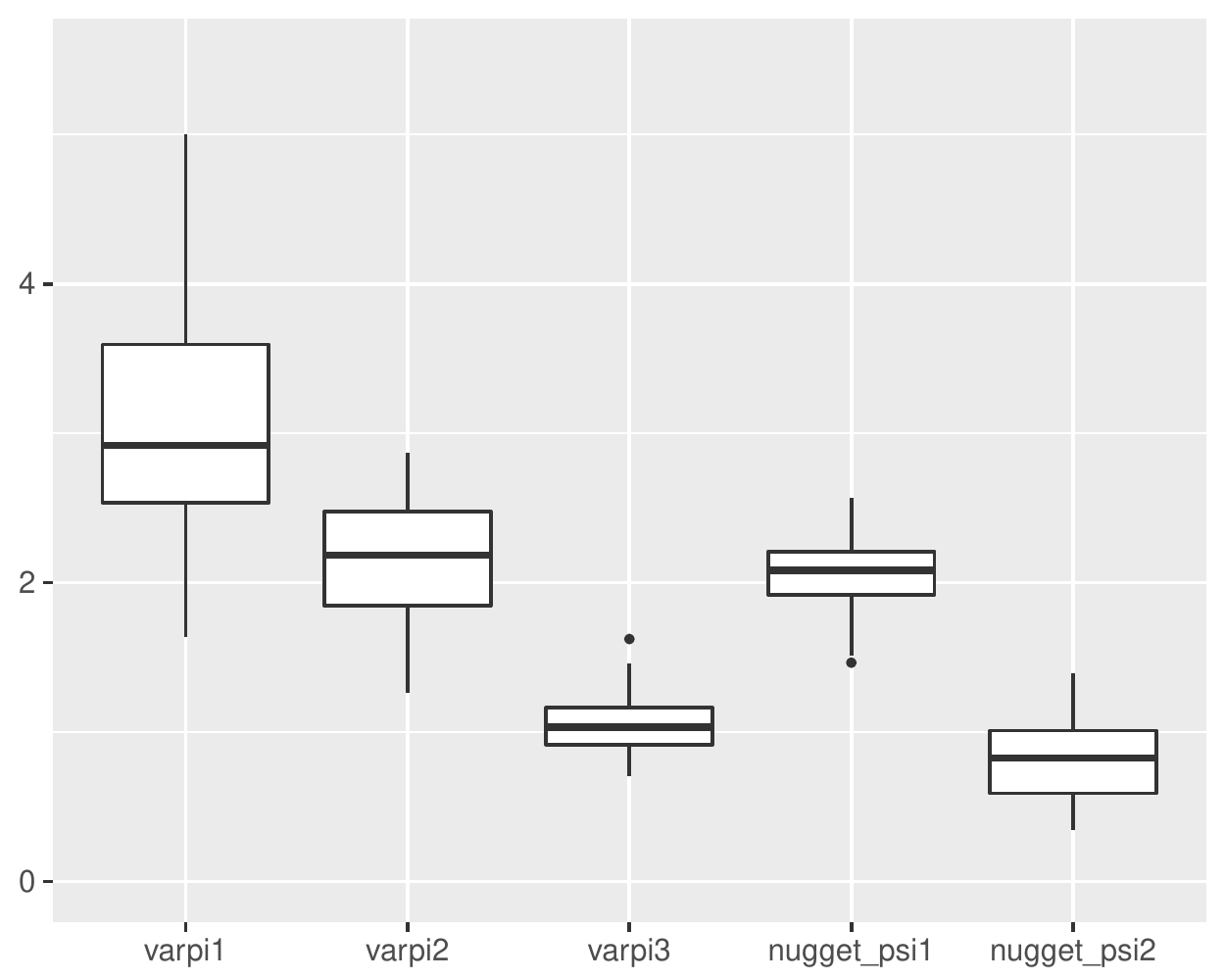}}
	\caption{Estimation results of \textit{sFPCA} under Scenario A. Panels (a) - (h) contain summaries of the functional estimators, as described in the labels. In each panel, the solid line is the true function; the dashed line is the mean of the functional estimator; and the shaded area illustrates the bands of pointwise 5\% and 95\% percentiles. 
		Panel (i) contains the boxplots of $\wh{\varpi}_1, \wh{\varpi}_2, \wh{\varpi}_3, \wh{\omega}_{\mathrm{nug},1}$, and $\wh{\omega}_{\mathrm{nug},2}$.}
	\label{figure: ScenarioA_spatialFDA}
\end{figure}

We simulate $200$ datasets for each scenario and apply the proposed estimation procedure (denoted as \textit{sFPCA}) to each simulated dataset. We use tensor product of cubic B-splines to estimate the spatial-temporal covariance function. The tuning parameters are selected using the BIC described in Section \ref{Section: Implementation} on some pilot datasets, then held fixed for massive simulations.
For comparison, we also apply the classic FPCA method \citep{yao2005functional} to the simulated datasets. 
To the best of our knowledge,  \cite{liu2017functional} is the only exiting work on FPCA for discretely-observed, spatially-dependent functional data, and their method is identical to the classic FPCA method when it comes to estimating the eigenvalues and eigenfunctions.
The classic FPCA, denoted as \textit{iFPCA}, is implemented using the R package \textit{fdapace}, which has built-in tuning parameter selection. Compared with our methods, \textit{iFPCA} only estimates a bivariate temporal covariance function using observations at the same location $\bs$, does not distinguish the functional nugget effect and does not borrow spatial information like what we do through integration in (\ref{eq:Omega}). Since our focus is on covariance estimation, estimation results for $\mu(t)$ are relegated to Figure S.1
in the Supplementary Material.


\begin{figure}
	\subfloat[$\wh{\psi}_1(t)$]{\includegraphics[width = 0.33\linewidth]{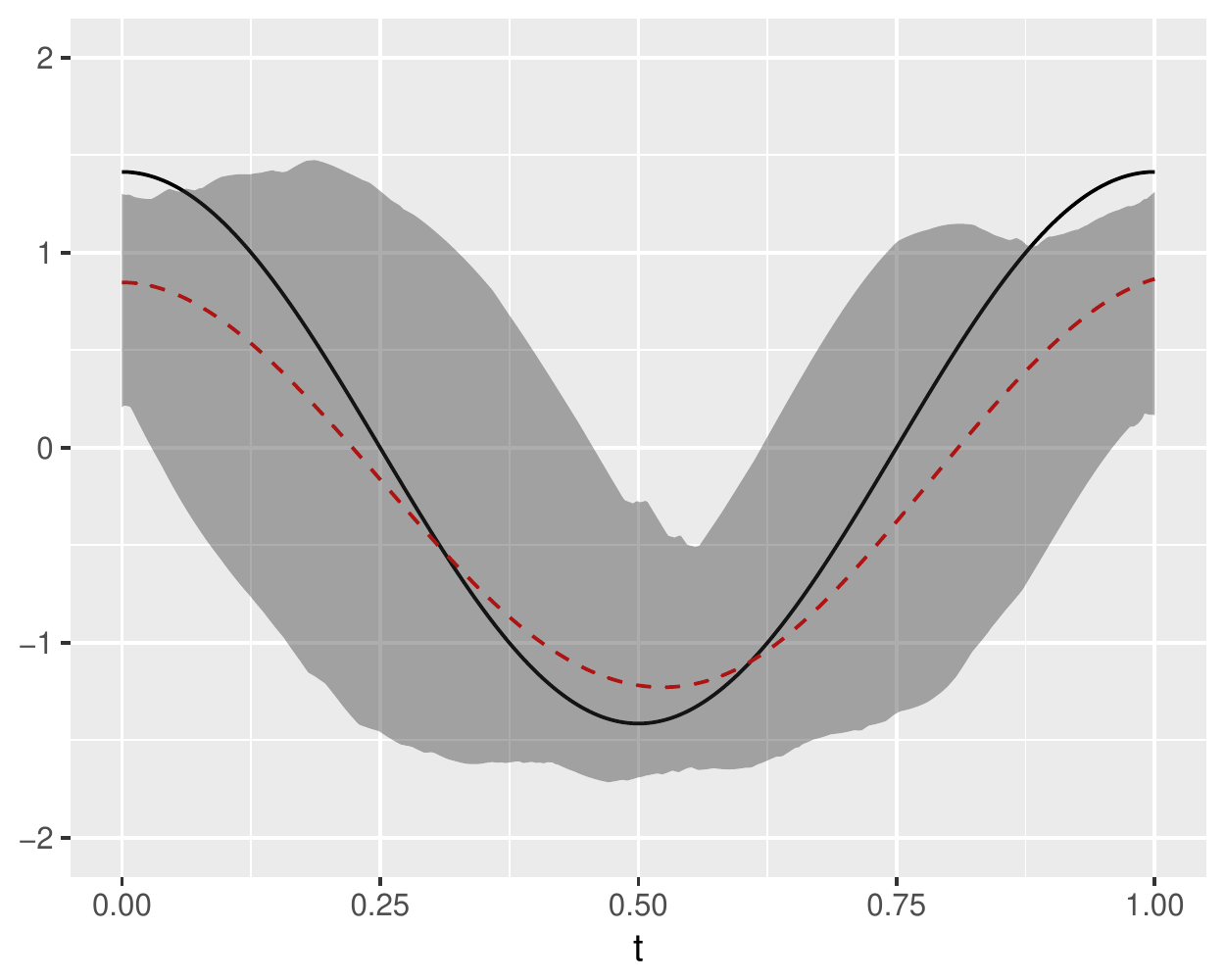}}
	\subfloat[$\wh{\psi}_2(t)$]{\includegraphics[width = 0.33\linewidth]{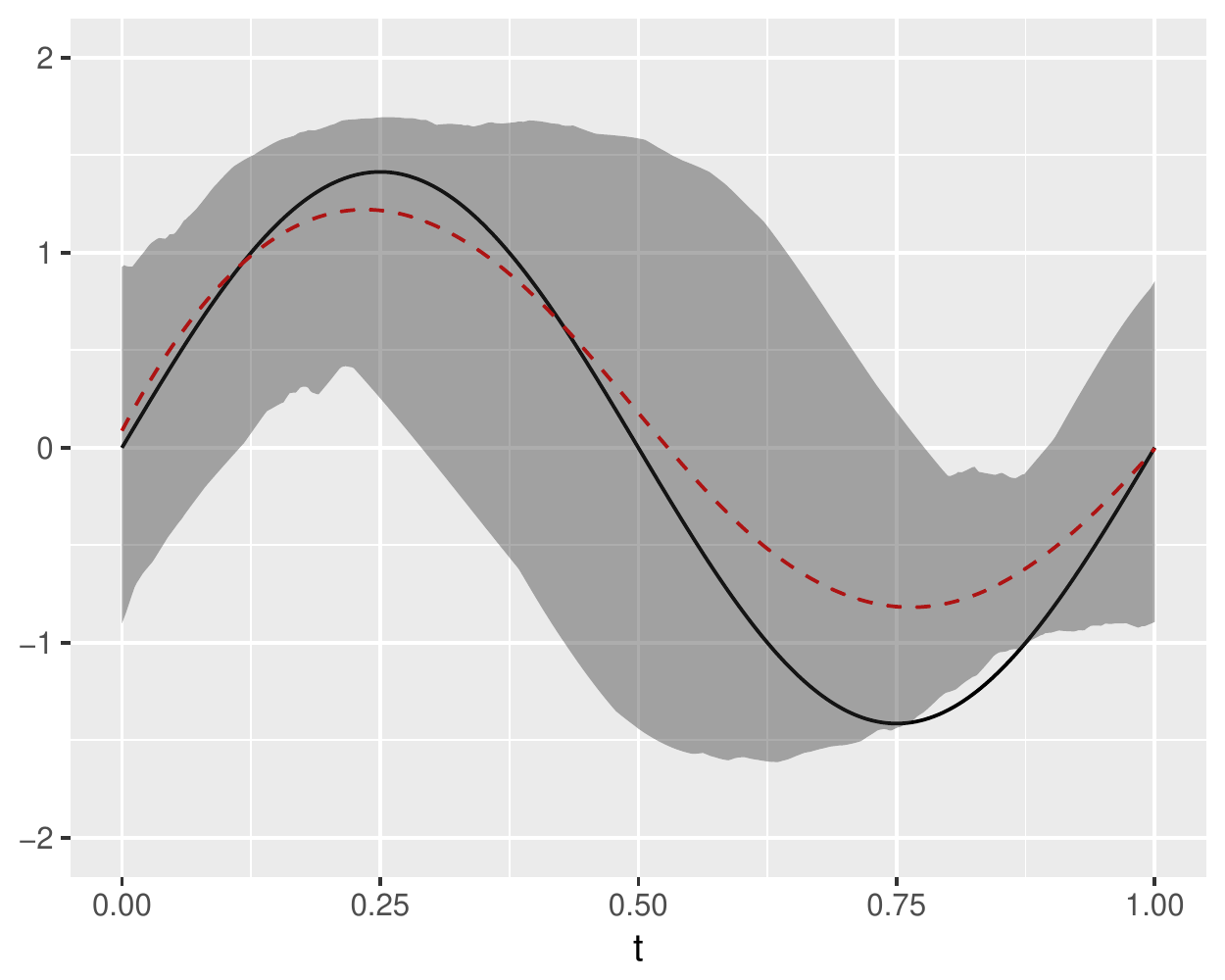}}
	\subfloat[$\wh{\psi}_3(t)$]{\includegraphics[width = 0.33\linewidth]{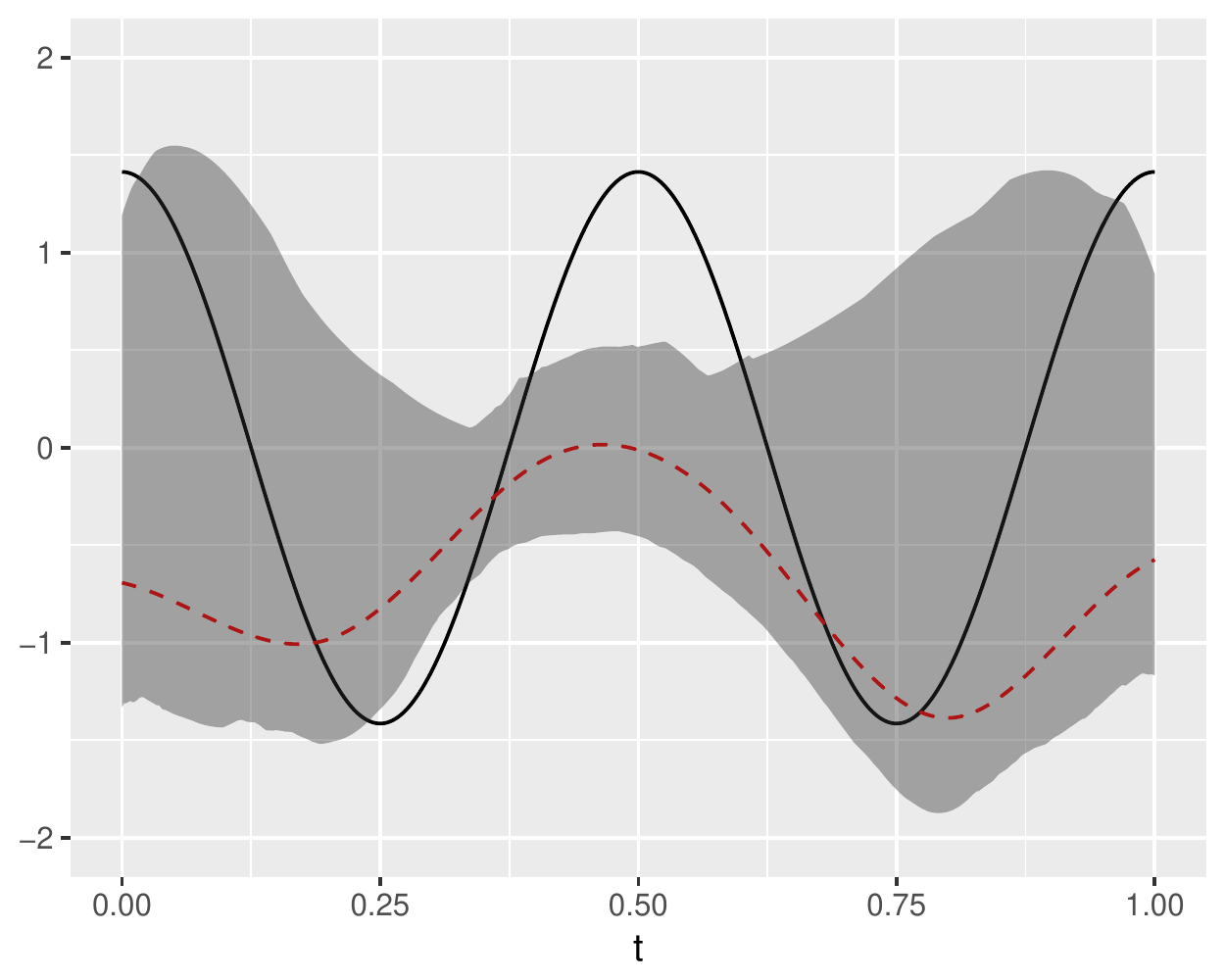}}
	\caption{Estimation results of \textit{iFPCA} under Scenario A. In each panel, the solid line is the true function; the dashed line is the mean of the functional estimator; and the shaded area illustrates the bands of pointwise 5\% and 95\% percentiles. 	}
	\label{figure: ScenarioA_independentFDA}
\end{figure}
In Panels (a) - (f) of Figure~\ref{figure: ScenarioA_spatialFDA}, we summarize the estimation results of \textit{sFPCA} under Scenario A for $\psi_j(\cdot)$ and $\CC_j (\cdot)$, $j=1,2,3$. In each plot, we compare the mean of our estimator with the true function and provide confidence bands formed by pointwise 5\% and 95\% percentiles of the estimator. By taking a spectral decomposition of $\wh \Lambda$ in (\ref{eq:Lambda_hat}), we also get estimators of $\psi_{\mathrm{nug},j}(\cdot)$ and $\omega_{\mathrm{nug},j}$. Graphical summaries of $\wh \psi_{\mathrm{nug},j}(t)$, $j=1,2$, are provided in Panels (g) and (h) of Figure~\ref{figure: ScenarioA_spatialFDA}; boxplots of scalar estimators $\wh \varpi_j$ and $\wh \omega_{\mathrm{nug},j}$ are provided in Panel (i). As we can see, the \textit{sFPCA} estimators behave reasonably well: all functional estimators exhibit very little bias and the confidence bands are tight around the true functions. The only functional estimator shows considerable variation is $\wh \psi_{\mathrm{nug}, 2}$, which is partially due to the fact that the convergence rate of $\wh \Gamma$ in Theorem \ref{thm: nugget effect} is much slower compared with that of $\wh \Omega$ in Theorem \ref{thm:2dim covariance}.

The \textit{iFPCA} method does not produce estimates for the spatial covariance functions nor the eigenfunctions of the functional nugget effect, we therefore only provide graphical summaries of $\wh \psi_j(t)$ for \textit{iFPCA} under Scenario A  in Figure~\ref{figure: ScenarioA_independentFDA}. As we can see, these functional estimators suffer from significant biases and large variation. The large biases can be explained by fact that \textit{iFPCA} does not distinguish the functional nugget effect from the spatially dependent functional effect; the large variations, on the other hand, are due to strong spatial dependence and the fact that \textit{iFPCA} does not borrow spatial information like we do through integration in (\ref{eq:Omega}).
Under Scenario B, which is a simpler setting by removing the functional nugget effect $U_i(t)$ from Scenario A,  both the classic \textit{iFPCA} and our \textit{sFPCA} methods provide consistent estimators for the eigenfunctions, and the differences between these methods are not as striking as in Scenario A. We therefore relegate graphical summaries under Scenario B to Figures S.2 and S.3
in the Supplementary Material. In theory, spectral decomposition of $\wh R(0, \cdot, \cdot)$ also provides consistent estimators of the eigenfunctions, however such a method suffers from the slow convergence rate of 3-dim spline smoothing and is not recommended. In simulation results not shown here, directly decomposing $\wh R(0, \cdot, \cdot)$ performs poorer than the proposed method under Scenario A and poorer than both methods under Scenario B. 



\begin{table}
	\caption{\label{tb:simulation_estimation_error}Simulation results on the mean and standard deviation of integrated square errors for functional principal components estimated by \textit{sFPCA} and \textit{iFPCA}. }
	\centering
	\begin{tabular}{cccc}
		\hline\hline
		Simulation Scenario & FPC &\textit{sFPCA} & \textit{iFPCA} \\
		\hline
		\multirow{ 5}{*}{Scenario A}& $\psi_1$ & $0.076(0.104)$ &$0.411(0.376)$\\
		&$\psi_2$ &$0.104(0.119)$ & $0.367(0.369)$\\
		&$\psi_3$ &$0.077(0.071)$ & $1.494(0.311)$\\
		&$\psi_{\mathrm{nug}, 1}$ &$0.035(0.031)$  & -- \\
		&$\psi_{\mathrm{nug}, 2}$ & $0.368(0.515)$ & -- \\
		\hline
		\multirow{ 3}{*}{Scenario B} &$\psi_1$ &  $0.073 (0.114) $& $0.134(0.232)$ \\
		&$\psi_2$ & $0.092(0.113)$  &   $0.123(0.232)$       \\
		&$\psi_3$ & $0.061(0.043)$  & $0.059(0.025)$ \\ 
		\hline\hline
	\end{tabular}
\end{table}

We also summarize, in Table~\ref{tb:simulation_estimation_error}, the mean and standard deviation of integrated square error (ISE) for the functional estimators of \textit{sFPCA} and  \textit{iFPCA}. These numerical summaries confirm our observations from the graphs that the \textit{sFPCA} estimators behave overwhelmingly better than those of \textit{iFPCA} under Scenario A, due to the existence of functional nugget effect. All estimators behave better under Scenario B due to smaller noises. However, even under Scenario B without functional nugget effects, \textit{sFPCA} estimators of the eigenfunctions are still better than \textit{iFPCA} because we borrow spatial information by including pairs of data in neighboring locations.


To illustrate the proposed \textit{sFPCA} kriging method in Section \ref{Section: Kriging}, we randomly sample new functions from 100 new locations in each simulated dataset, and use the training data and the estimated covariance structure to predict $X(\bs,t)$ at the new locations. The integrated square error (ISE), $\int \{\wh X(\bs, t) -X(\bs, t) \}^2 dt$, is averaged over all new locations and then repeated for each dataset.
For comparison, we apply the \textit{iFPCA+CoKriging} two-step procedure \citep{nerini2010cokriging} and the trace kriging method \citep{giraldo2011ordinary} to the simulated data. Both methods are implemented in R package \textit{fdagstat}. For the \textit{iFPCA+CoKriging} method, the number of principal components for \textit{iFPCA} is selected to explain $99\%$ of the variation and the spatial covariance functions are estimated using the Mat\'ern models based on the estimated \textit{iFPCA} scores. The trace kriging method requires fully observed functional data, we therefore treat the observed data as step functions with jumps at observed time points.
The kriging results are summarized in Table \ref{tb:simulation_kriging}, where we provide the mean and standard deviation of ISE for all competing methods. As we can see, our kriging method yields much smaller prediction errors than the two competing methods under both scenarios.

\begin{table}
	\caption{\label{tb:simulation_kriging} Kriging results in the simulation study: mean and standard deviation of the integrated 
		squared errors for \textit{sFPCA}, \textit{iFPCA+CoKriging} and \textit{Trace Kriging}. }
	\centering
	\begin{tabular}{cccc}
		\hline\hline
		Simulation Scenario & \textit{sFPCA} & \textit{iFPCA+CoKriging} & \textit{Trace Kriging}\\
		\hline
		Scenario A& 2.123(0.589)   & 5.147(0.989) & 5.224(4.941)\\
		\hline
		Scenario B & 1.563(0.704)  & 4.602(1.335) & 5.073(4.846)\\
		\hline\hline
	\end{tabular}
\end{table}


%
%
%
%
%
%
%
%
%
%
%
%
%
%
%
%
%
%
%
%

\section{Data analysis} \label{Section: Data Analysis}
We now analyze the two motivating datasets described in Section \ref{sec:introduction}. 
\vspace{-0.15in}
\subsection{Analysis of the London house price data}

This dataset consists of $10,980$ transaction records of $2013$ houses in the Greater London Area from Jan 1, 1995 to Dec 31, 2018. 
Figure S.4 in the Supplemental Material shows the empirical distributions for the number of transactions per house and the transaction dates. The estimated mean function, shown in Figure \ref{fig: london_data}, demonstrates an overall increasing trend.  Remarkably, the two dips on the mean curve reflect the impacts of the 2008 financial crisis and the 2016 Brexit.

\begin{figure}[ht]
	\centering
	\subfloat[]{\includegraphics[width = 0.33\linewidth]{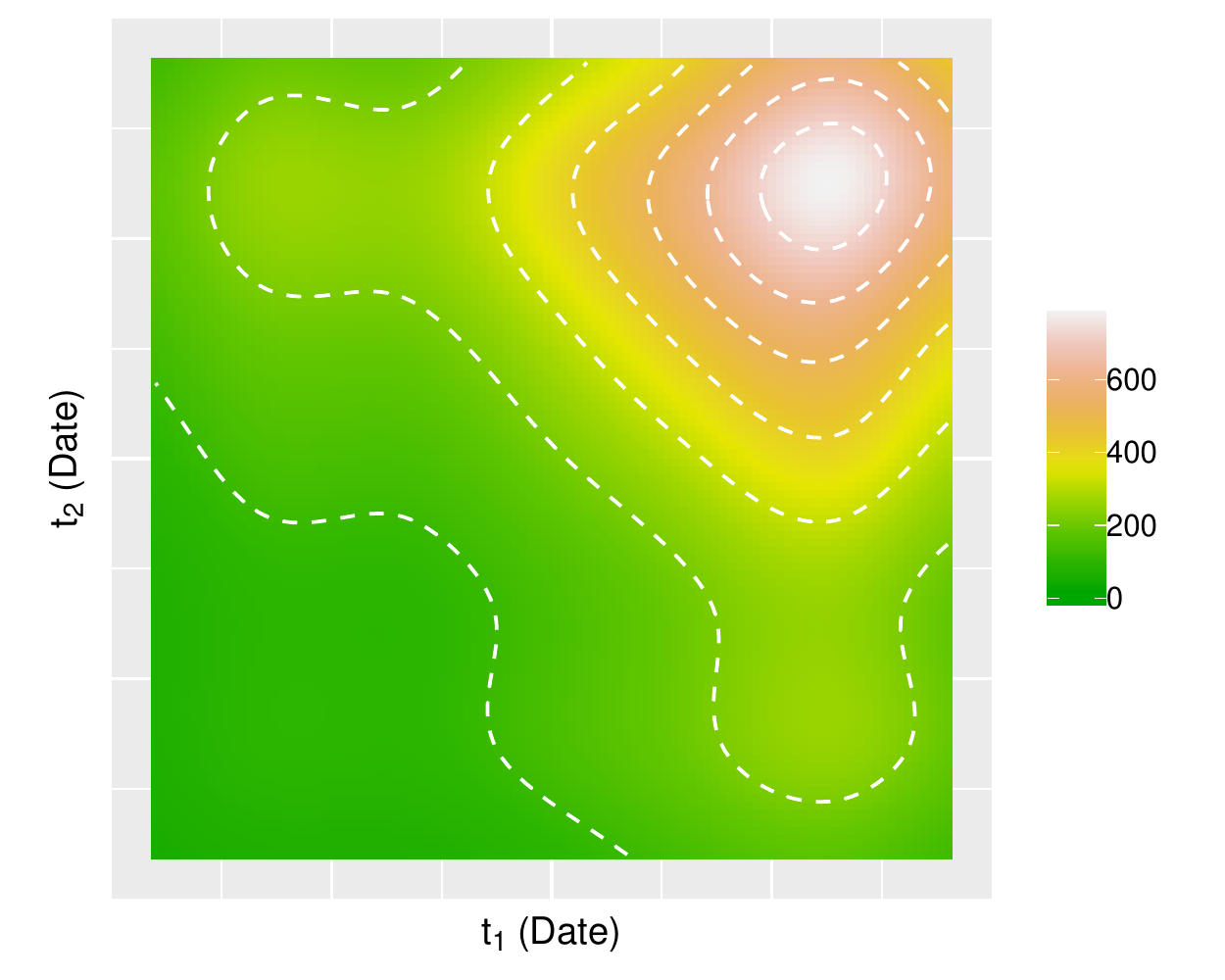}} 
	\subfloat[]{\includegraphics[width = 0.33\linewidth]{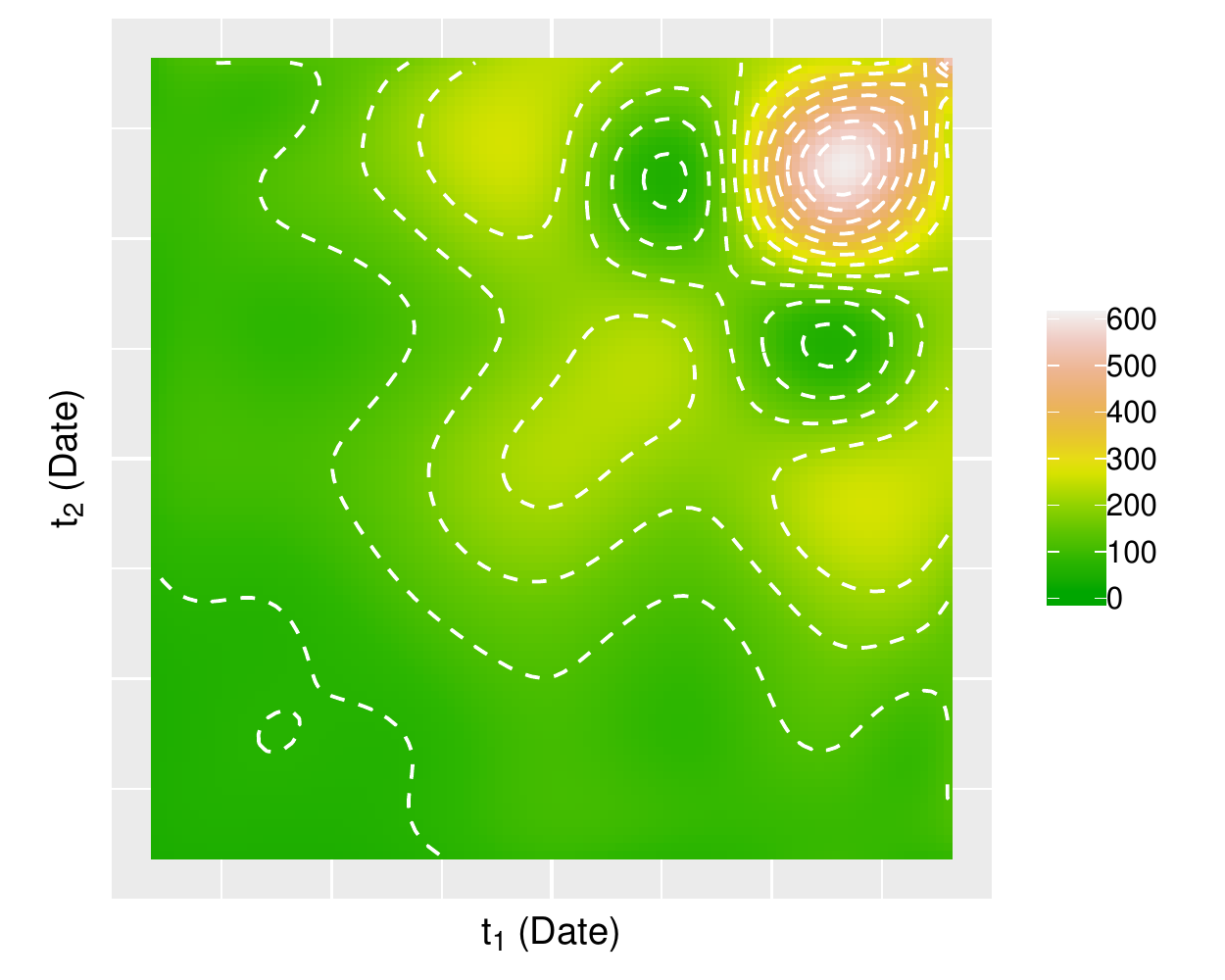}} 
	\subfloat[]{\includegraphics[width = 0.33\linewidth]{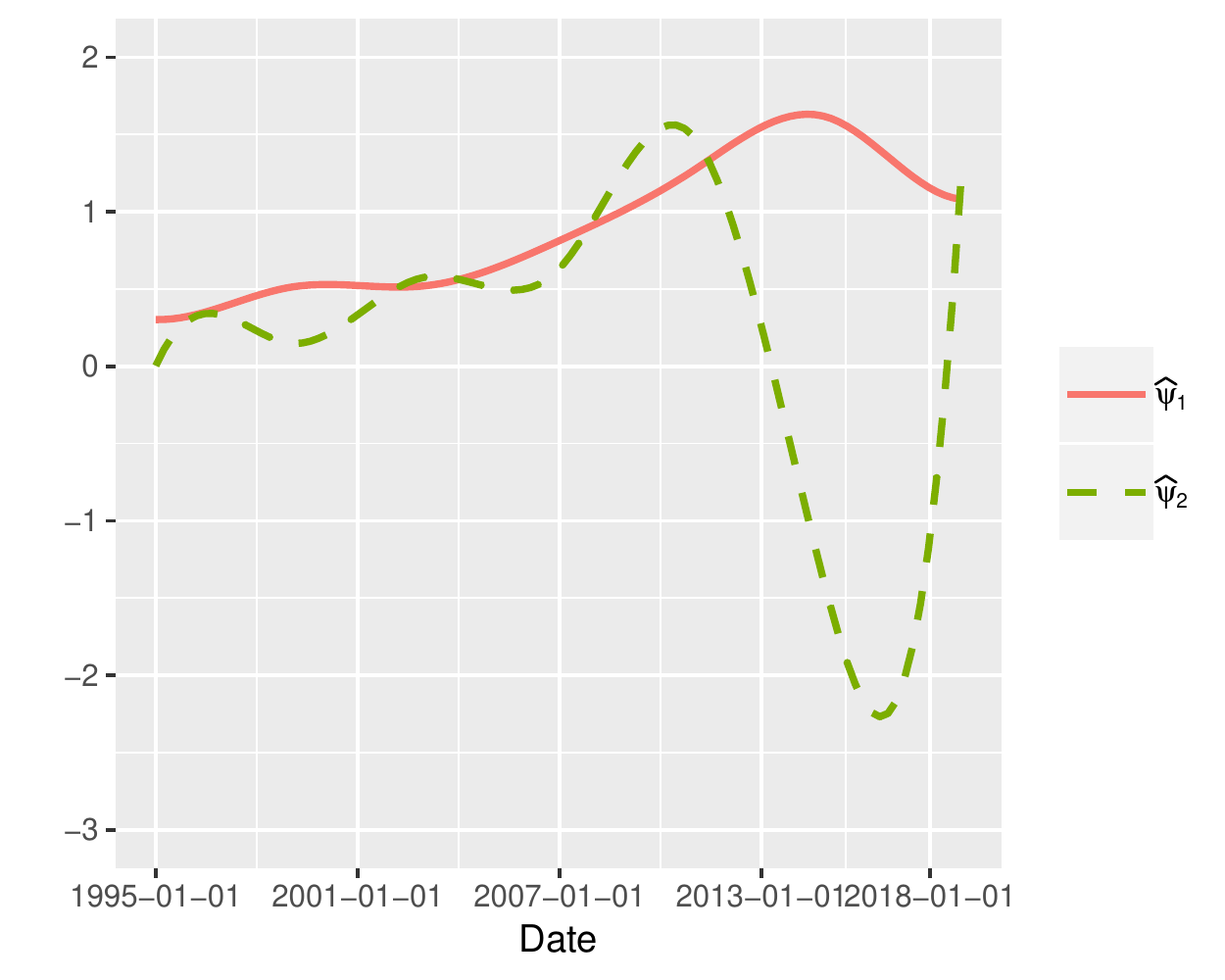}} \\
	\subfloat[]{\includegraphics[width = 0.33\linewidth]{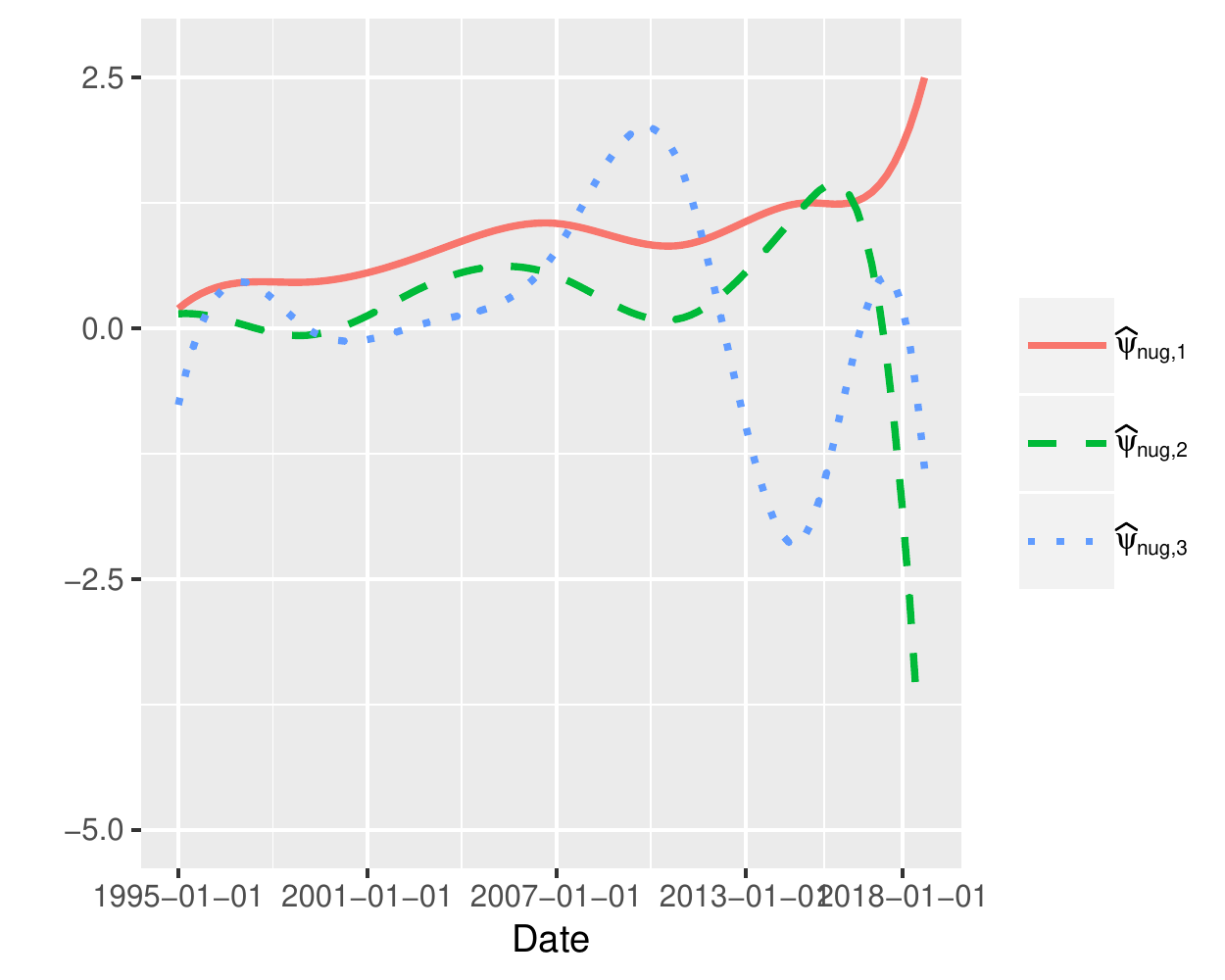}} 
	\subfloat[]{\includegraphics[width = 0.33\linewidth]{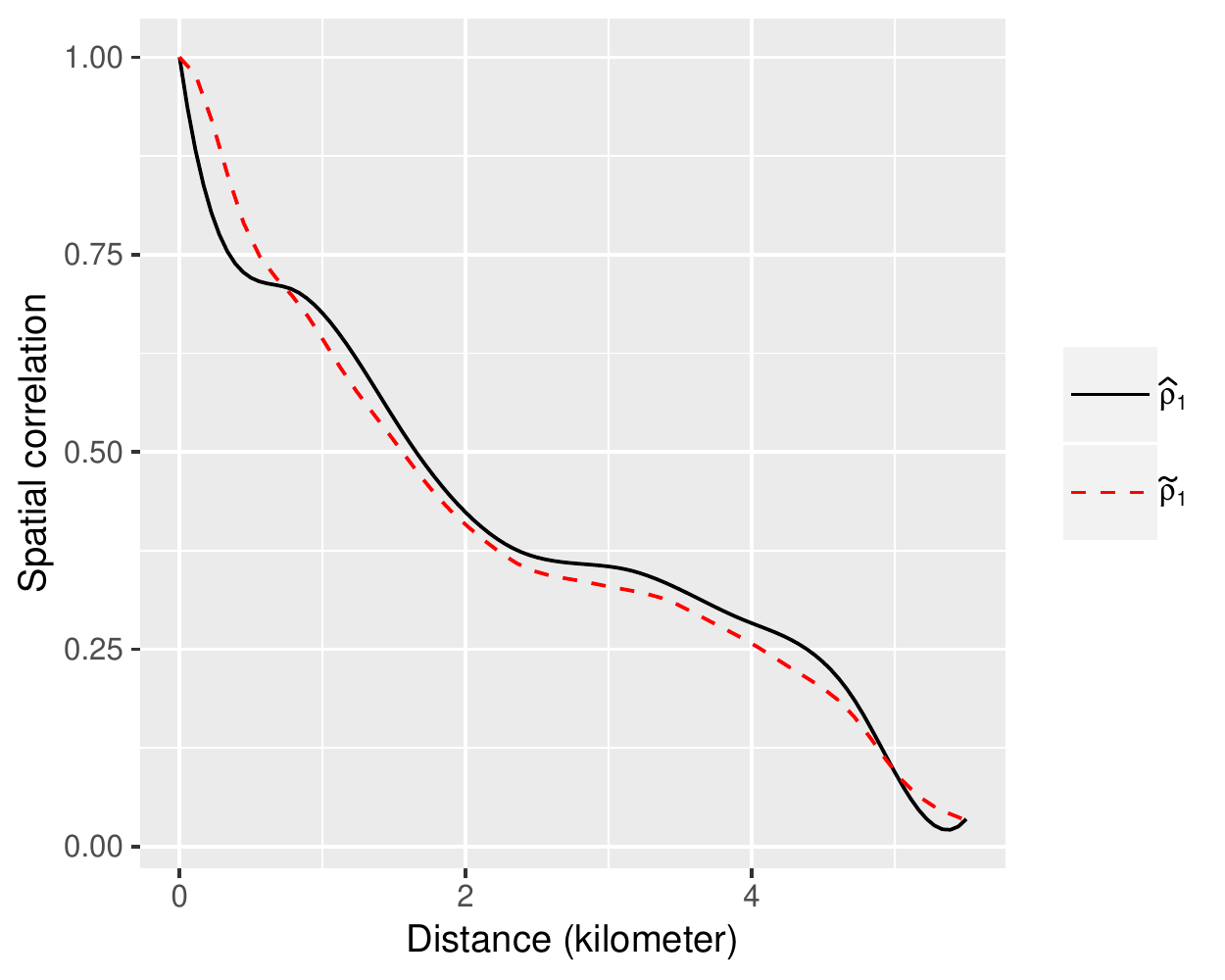}}
	\subfloat[]{\includegraphics[width = 0.33\linewidth]{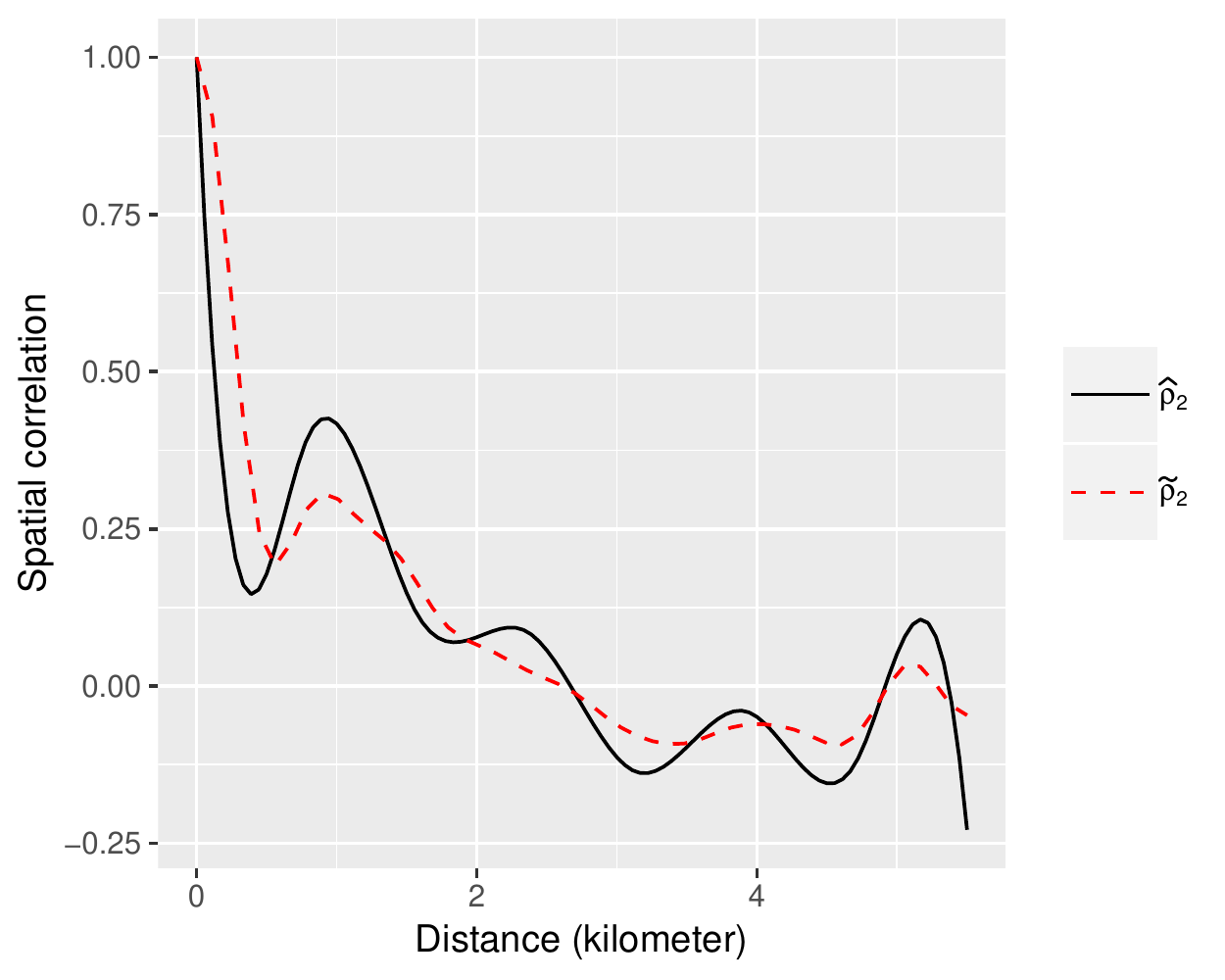}}
	\caption{Results on the London housing price data: (a) contour plot of $\wh{\Omega}(t_{1},t_{2})$; 
		(b) contour plot of $\wh{\Lambda}(t_{1},t_{2})$, covariance of the functional nugget effect;
		(c) the first two eigenfunctions of $\wh{\Omega}(\cdot,\cdot)$;
		(d) the first three eigenfunctions of $\wh{\Lambda}(\cdot,\cdot)$;
		(e) spatial correlation function $\wh{\rho}_1(\cdot)$ and its positive semi-definite adjustment $\wt{\rho}_1(\cdot)$;
		(f) $\wh{\rho}_2(\cdot)$ and $\wt{\rho}_2(\cdot)$.}
	\label{fig: london_estimation_result}
\end{figure}

A pilot study indicates that the range of spatial dependency is about $5.5$ kilometers, which is also confirmed by the final estimators of the spatial correlations in Figure \ref{fig: london_estimation_result}. We therefore estimate the spatio-temporal covariance function $R(\cdot,\cdot, \cdot)$ up to a spatial lag of $\Delta = 5.5$ km, using tensor product of cubic B-splines. The numbers of knots chosen by BIC are $K_s = 6$ and $K_t = 6$  in spatial and temporal directions, respectively. 

Next, we perform FPCA to the data by a spectral decomposition of $\wh{\Omega}$. 
The first two eigenvalues, $\wh{\omega}_1 = 285.80$ and $\wh{\omega}_2 = 21.52$, in total explain $99.42\%$ of variation in $\wh \Omega$. A contour plot of $\wh{\Omega}(\cdot, \cdot)$ and the first two estimated eigenfunctions are shown in Figure~\ref{fig: london_estimation_result} (a) and (c). The estimated spatial correlation functions and their positive semi-definite adjustments are shown in Figure~\ref{fig: london_estimation_result} (e) and (f). 
As we can see, $\wh \rho_2(u)$ decays to $0$ more sharply than $\wh\rho_1(u)$, indicating that the two principal components have different ranges of spatial dependence and the spatio-temporal covariance may not be separable.
We also estimate the covariance function $\Lambda(\cdot,\cdot)$ of the functional nugget effect and the nugget principal components, the results of which are shown in Figure \ref{fig: london_estimation_result} (b) and (d).  The noise-to-signal ratio of the functional nugget effect is $\|\wh{\Lambda}(\cdot,\cdot)\|_{L^2}\big /\|\wh{R}(0, \cdot,\cdot)\|_{L^2}=1.11$. 
The first three eigenvalues,  $\wh{\omega}_{\mathrm{nug},1} = 144.89$, $\wh{\omega}_{\mathrm{nug},2} = 80.50$, and $\wh{\omega}_{\mathrm{nug},3} = 29.23$, explain $98.77\%$ of the total variation in the functional nugget effect. These results show that, for the London housing market, the house-specific effect is more important than the spatial dependent effect. These house-specific effects might be explained by factors such as size, year built, number of bedrooms, number of bathrooms, etc. These variables are not available in public records, hence not included in our analysis. It would be interesting to include these covariates in our future analysis, should an external data source becomes available.


\subsection{Analysis of the Zillow real estate data}

The spatial locations in this dataset are sampled from six regions in the Bay Area: \textit{Fremont, Oakland, Palo Alto, San Francisco, San Jose}, and \textit{San Mateo}. The estimated region-specific mean functions are presented  in 
Figure S.5 of the Supplementary Material. To get rid of the regional effects, we center the trajectories in Figure \ref{fig: zillow_map_data} by subtracting their region-specific mean functions, and the residual trajectories are presented in Figure S.6. 
Our methodology is based on the spatially stationary assumption, but can be easily extended to piecewise-stationary settings, we therefore apply the proposed methodology to the residual trajectories.

\begin{figure}[htb]
	\centering
	\subfloat[]{\includegraphics[width = 0.33\linewidth]{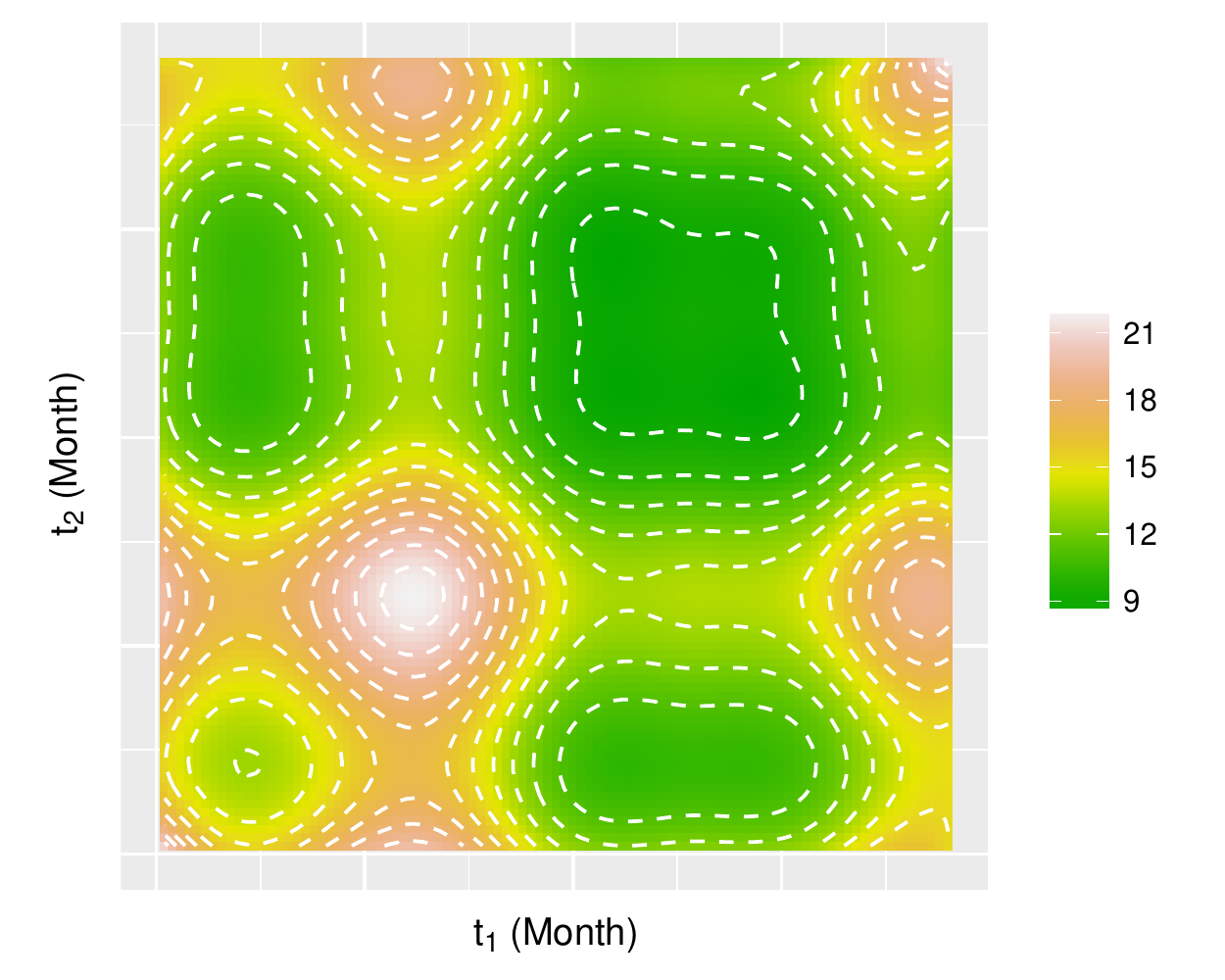}} 
	\subfloat[]{\includegraphics[width = 0.33\linewidth]{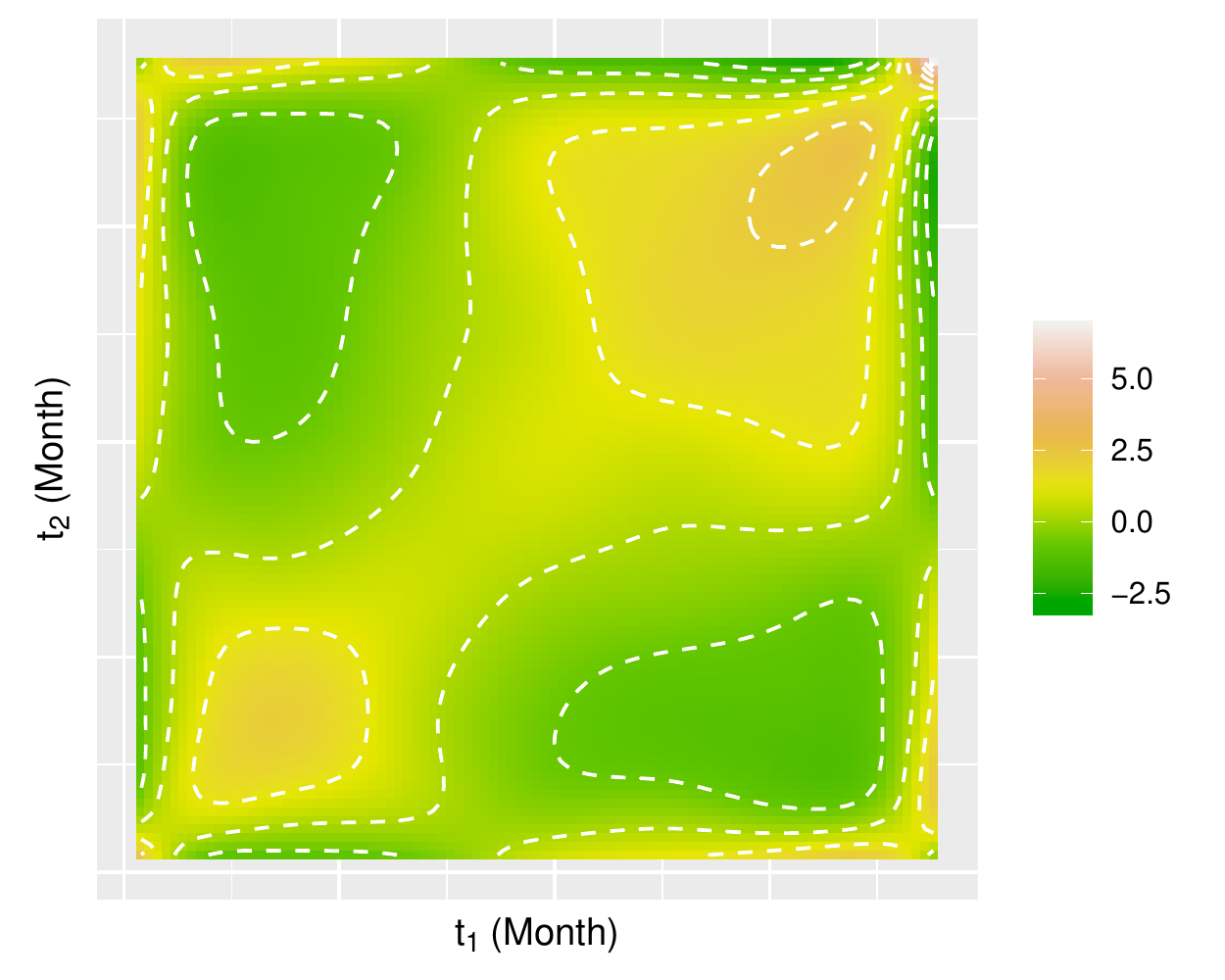}}
	\subfloat[]{\includegraphics[width = 0.33\linewidth]{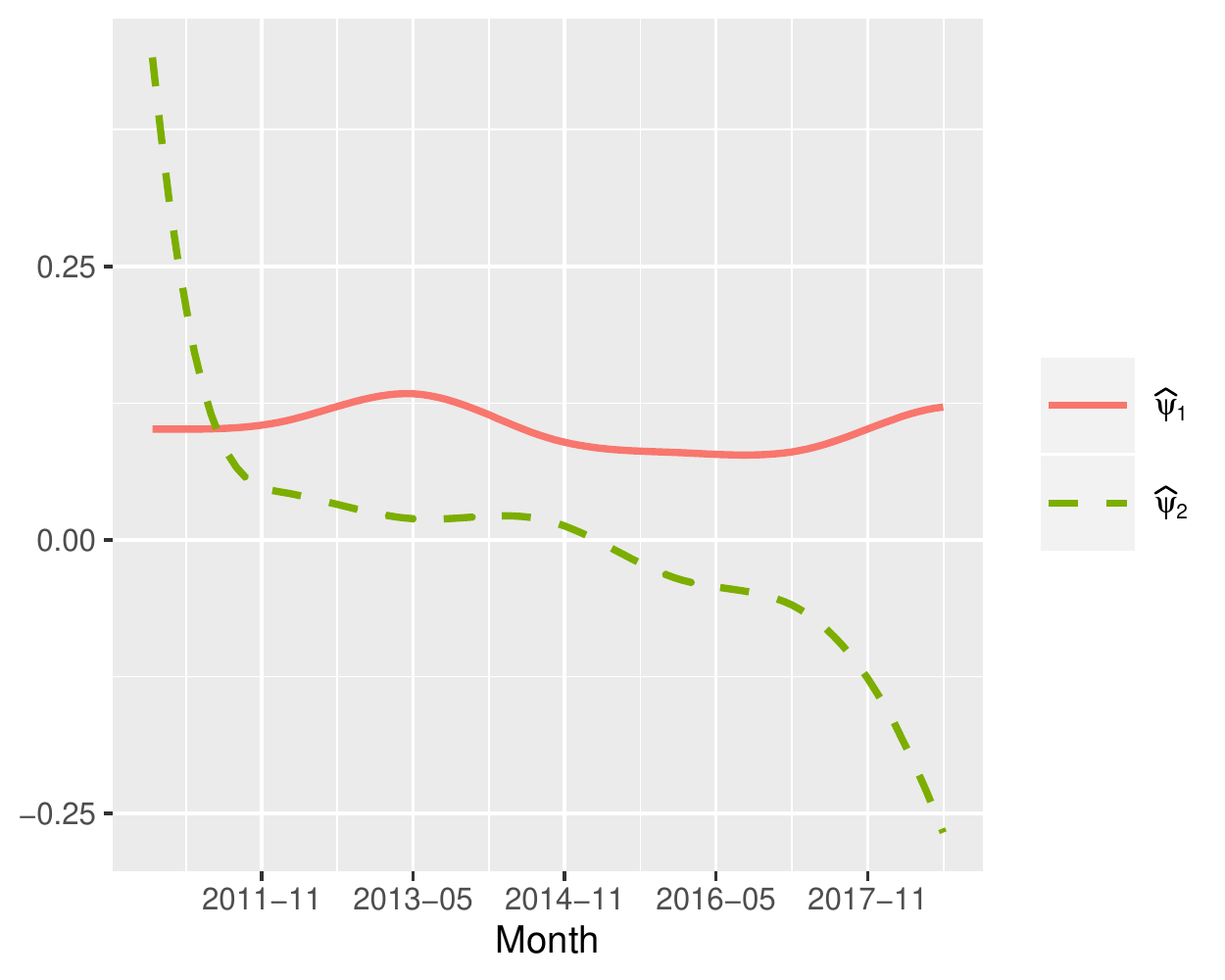}} \\
	\subfloat[]{\includegraphics[width = 0.33\linewidth]{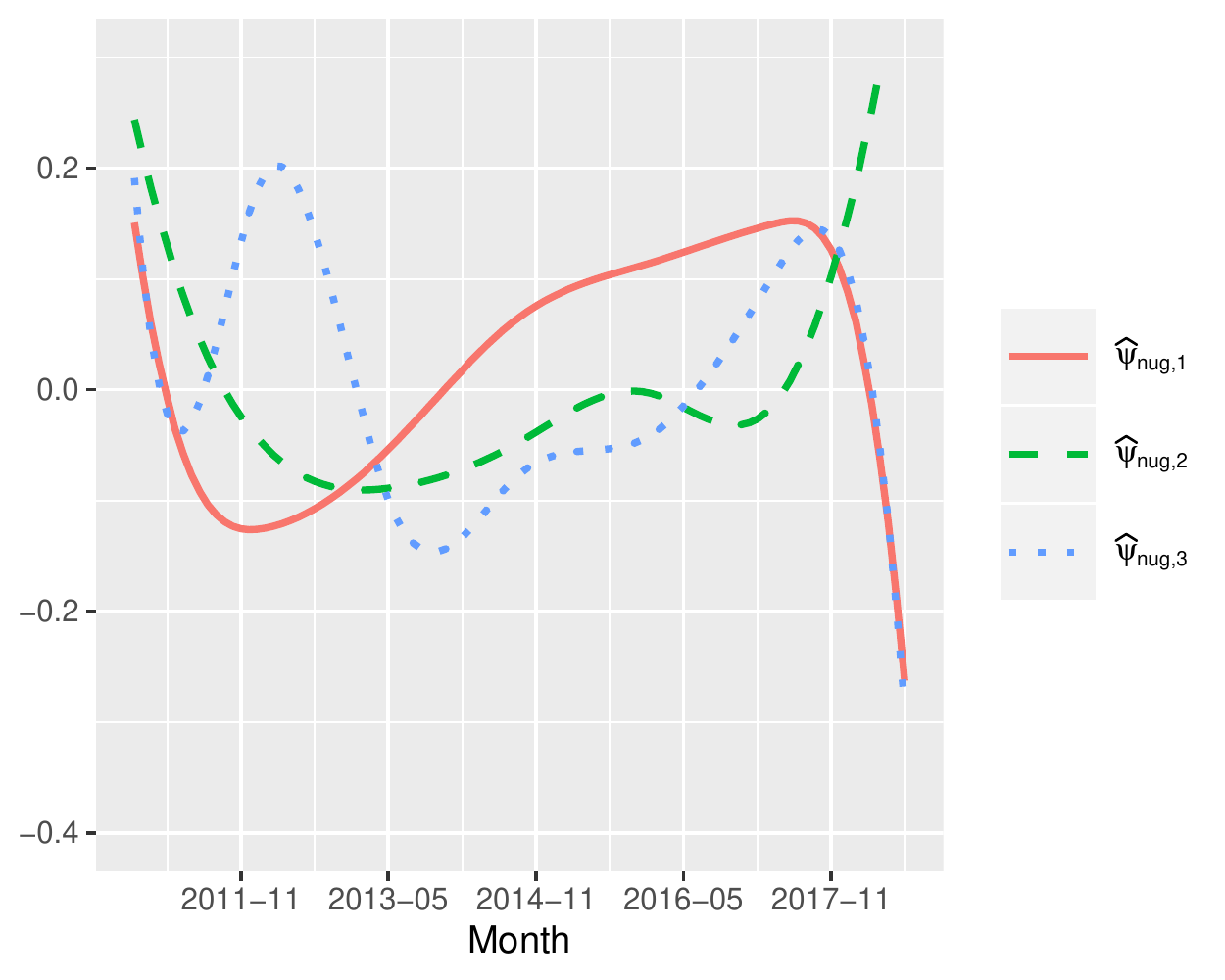}}
	\subfloat[]{\includegraphics[width = 0.33\linewidth]{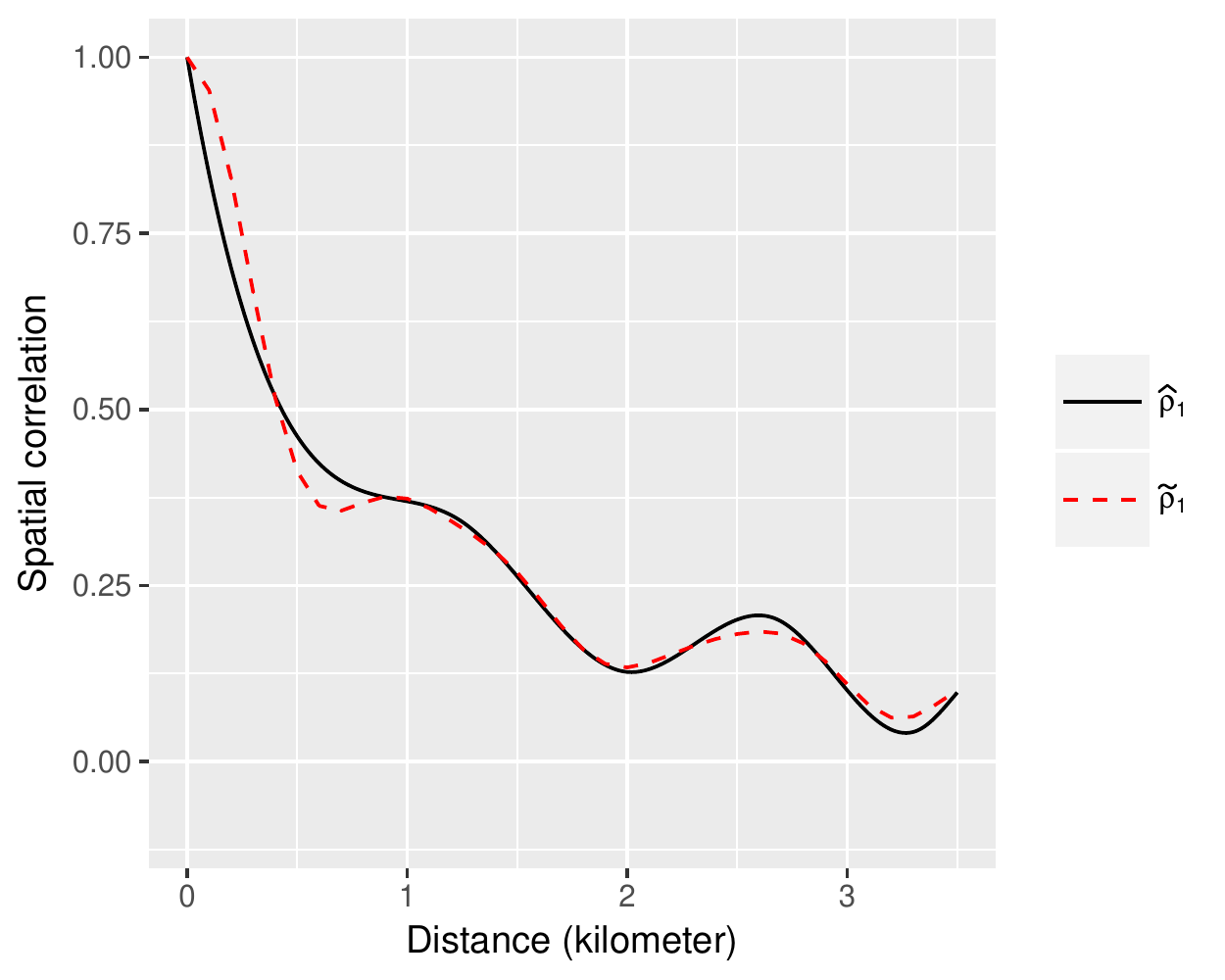}} 
	\subfloat[]{\includegraphics[width = 0.33\linewidth]{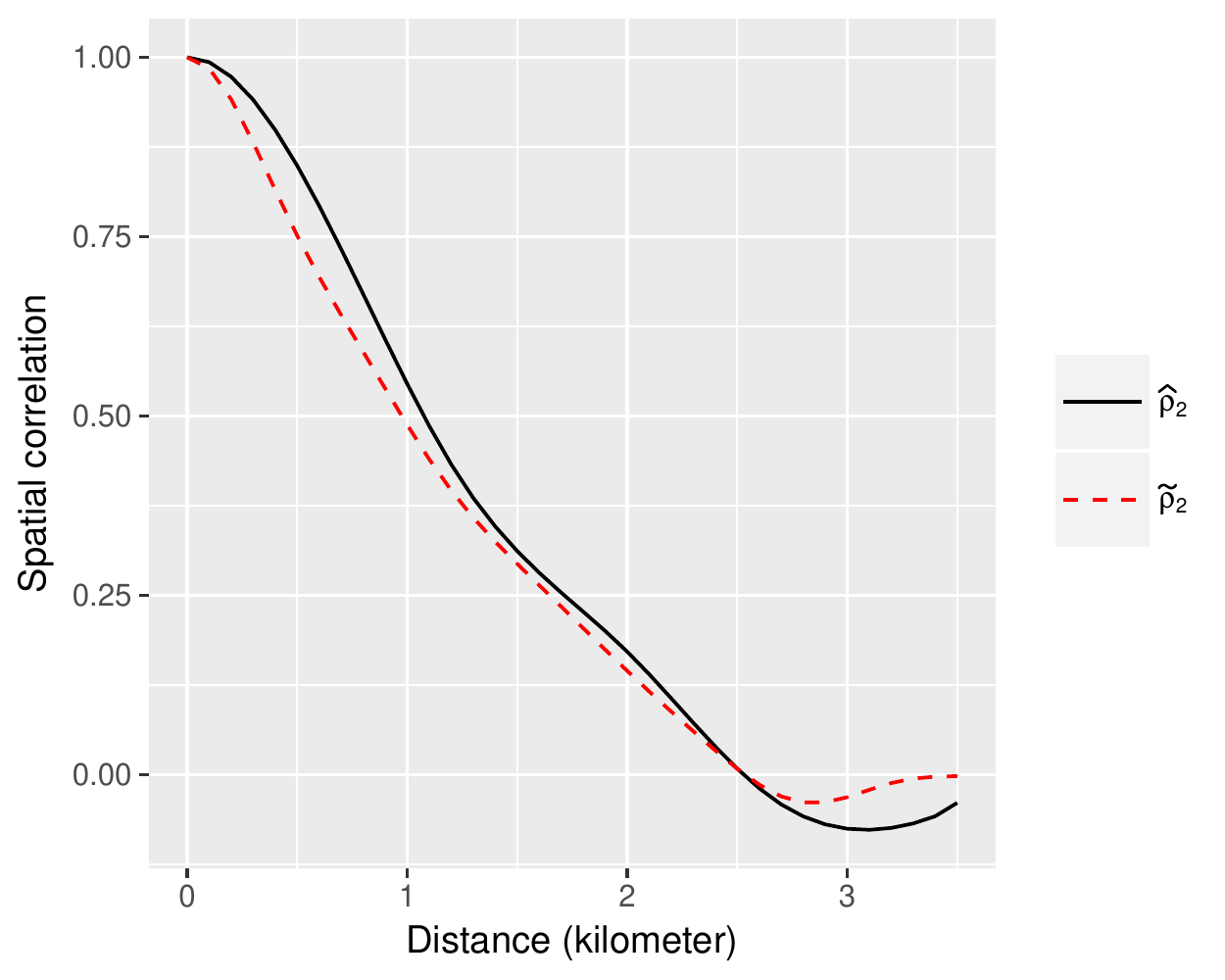}}
	\caption{Results on the Zillow real estate data: (a) contour plot of $\wh{\Omega}(t_{1},t_{2})$; (b) contour plot of $\wh{\Lambda}(t_{1},t_{2})$, covariance function of the functional nugget effect; 
		(c) the first two eigenfunctions 
		(d) the first three eigenfunctions of $\wh{\Lambda}(\cdot,\cdot)$
		(e) the estimated spatial correlation function $\wh{\rho}_1(\cdot)$ and its positive semi-definite adjustment $\wt{\rho}_1(\cdot)$; (f) $\wh{\rho}_2(\cdot)$ and $\wt{\rho}_2(\cdot)$.}
	\label{fig: zillow_estimation_result}
\end{figure}

Our pilot analysis on the Zillow data indicates that the spatial correlation diminishes at a distance of about $3$ km. 
We therefore estimate the spatio-temporal covariance function $R(\cdot,\cdot, \cdot)$ up to a spatial lag of $\Delta = 3.5$ km, using tensor-product cubic B-splines. The number of knots chosen by BIC are $K_s = 5$ and $K_t = 6$. 
Spectral analysis of $\wh \Omega$ yields that the first two eigenvalues, $\wh{\omega}_1 = 974.22$ and $\wh{\omega}_2 = 18.59$, explain $97.97\%$ of variation in $\wh \Omega$. A contour plot of $\wh{\Omega}(\cdot, \cdot)$ and the first two eigenfunctions are shown in Figure~\ref{fig: zillow_estimation_result}. Notice that $\wh{\psi}_{1}(t)$, given by the solid curve in Figure~\ref{fig: zillow_estimation_result} (c), is almost constant over time, which implies that the first FPC is a spatial random intercept -- locations with high scores $\xi_{1}(\bs)$ on the first FPC has higher than average price-to-rent ratio. On the other hand, $\wh{\psi}_{2}(t)$ represents a decreasing trend in time. Since the overall trend of price-to-rent ratio is increasing in Figure \ref{fig: zillow_map_data} (b), locations with high values of $\xi_{2}(\bs)$ has slower than average increase of price-to-rent ratio. The estimated spatial correlation functions and their positive semi-definite adjustments are shown in the lower panels of Figure ~\ref{fig: zillow_estimation_result}. 
We also estimate the covariance function $\Lambda(\cdot,\cdot)$ of the functional nugget effect and the nugget principal components, the results of which are shown in Figure~\ref{fig: zillow_estimation_result}.  
The first three eigenvalues,  $\wh{\omega}_{\mathrm{nug},1} =92.72$, $\wh{\omega}_{\mathrm{nug},2} = 20.75$, and $\wh{\omega}_{\mathrm{nug},3} = 10.43$, explain $91.12\%$ of the  total variation in the functional nugget effect. The estimated variance of measurement errors is $\wh{\sigma}_{\epsilon}^2 = 0.246$. 

We illustrate the performance of the proposed \textit{sFPCA} kriging method by a leave-one-curve-out kriging experiment: leave one curve out as test data, use the rest of the data and the fitted model to predict the curve on the left out location, calculate the integrated squared error (ISE) for the prediction, and repeat this experiment for all locations.
For comparison, we also perform the same kriging experiment for \textit{iFPCA+Co-kriging} and \textit{Trace Kriging}, described in Sections \ref{Section: Kriging} and \ref{Section: Simulation}.
After scaling the time domain to $[0,1]$, the median prediction ISE is $1.85$ for \textit{sFPCA} kriging, $2.91$ for \textit{Trace Kriging}, and $3.61$ for \textit{iFPCA+Co-kriging}, which confirms that our proposed kriging method has much smaller prediction error than existing functional kriging methods.

\subsection{Sensitivity Analysis}\label{sec:sensitivity}
In Figures S.7 and S.8, we show contour plots of $\wh{R}(u,\cdot,\cdot)$ at different values of $u$ for the two data examples, respectively.  To make different slices of this 3-dim function comparable, we standardize the contour plots by $\| \wh R(u, \cdot, \cdot) \|_1= \int |\wh R(u, t_1, t_2) | dt_1 dt_2/ |T|^2$. For both datasets, the differences in the standardized contour plots show some evidence that the covariance structures are non-separable.

In Section S.6 in the Supplementary Material, we perform sensitivity analyses on both datasets to verify the assumption of spatial stationarity. We compare the FPCA estimates obtained from the whole spatial domain with those obtained from sub-domains.  For the London data, we consider two sub-domains -- regions to the north and south of River Thames; for the Zillow data, we divide the domain into two sub-domains: areas on the peninsula (\textit{San Francisco, San Mateo} and \textit{Palo Alto}) and those outside (\textit{Fremont, Oakland} and \textit{San Jose}). The fact that the FPCA estimates from the whole domain agree well with those from subdomains suggests that there is no serious violation of the stationarity assumption.


\section{Discussion} \label{Section: Discussion}
As discussed in Section \ref{Section: Model}, spatial functional data analysis is deeply connected with spatio-temporal models, yet substantially different. In the two real data examples presented in this paper, our focus is to perform dimension reduction for temporal processes defined on real entities, which happen to be spatially correlated. We demonstrate how our model can be used for spatial prediction, but more importantly it extracts latent factors in the data, which can be used in further analysis, including a second stage regression.

We propose a three dimensional tensor product spline approach to estimate the spatio-temporal covariance function. Based on a coregionalization structural assumption, which is more flexible than the commonly used separable structure assumed in the literature, our three dimensional spline covariance estimator yields important byproducts, including nonparametric estimators of the principal components and the spatial covariance functions for the FPC scores. We also stress the importance of modeling the functional nugget effects, which model the local characteristics that are not dependent to the neighbors. We show in our simulation studies, ignoring the functional nugget effects can potentially cause large biases in the FPCA estimators. 
Our asymptotic study for the proposed methodology is quite comprehensive, where we combine both infill and increasing domain paradigms and accommodate both sparse and dense functional data. We found that, compared with the domain size, the effect of infilling locations in a unit spatial domain only has a secondary effect on the asymptotic convergence rate of the proposed estimators. We also establish phase transition in the convergence rates from sparse to dense functional data, which was not previously available for spatially dependent functional data.

Our method is based on three dimensional spline smoothing on the product of all data pairs within a prescribed distance, and hence computationally more intense than some of the existing method such as  the \textit{iFPCA} method implemented in the `fdaPACE' package. In the Scenario A of our simulation study reported in Section \ref{Section: Simulation}, the average running time of \textit{iFPCA} on a computer of 2.60GHz processor and 128 GB memory is 45.2 seconds, while the average running time for our method is 384.0 seconds. The extra computational cost is justifiable by the additional information we offer on the spatio-temporal covariance structure and being able to distinguish the functional nugget effect from the spatial functional effect. In our supplementary material, we also provide additional simulation results on the sensitivity of our method to the choice of $\Delta$. We recommend to use a $\Delta$ approximately equal to the range of spatial dependency, where the spatial correlation decays to $0$. In reality such a range is unknown and our results in Table S.1 suggest that our estimation results for the functional principal components are not sensitive to the choice of $\Delta$. On the other hand, Table S.1 also summarizes the running time of \textit{sFPCA} under different choices of $\Delta$, and a larger $\Delta$ results in a longer running time. This is understandable because more data pairs are included into the three dimensional smoothing when a larger $\Delta$ is used.

Our approach is based on moderate model assumptions, such as spatial stationarity. As we demonstrate in our real data analysis, the stationarity assumption can be easily relaxed to piecewise stationarity. The second order stationarity assumption on the principal component scores can also be relaxed: suppose $\CC_j(\bs_1, \bs_2)=\cov\{\xi_j(\bs_1), \xi_j(\bs_2)\}$ is non-stationary, but the averages of these covariance functions at distance $u$, $\CC_j^\ast(u)= \lim_{n\to \infty} {1\over  2\pi u |D_n|} \int_{\CD_n} \int_{\|\bdv\|=1} \CC_j (\bs, \bs + u \bdv) d\bdv d \bs$, exist and are uniformly bounded, then under some weak dependence assumptions the proposed tensor spline covariance estimator consistently estimates $\CR^\ast(u, t_1, t_2) =\sum_{j} \CC_j^\ast (u) \psi_j(t_1) \psi_j(t_2)$. We still get legitimate principal component estimates, but spatial covariance function estimates become less interpretable. Our work based on the stationary assumption also paves the way for extensions to more sophisticated models, such as the locally stationary models \citep{Kuusela-Stein2018}, which can be applied to data collected from a large spatial region.
Our methods also open up many new research questions, related to model selection and statistical inference for the proposed model. For instance, one important research question is how to select the number of principal components in the model. Aikaike information criterion such as that studied in \cite{li2013selecting} depends on evaluating the likelihood, which is difficult for spatially dependent functional data. 
It might also be possible to relax the isotropic assumption in our approach to a more flexible geometric anisotropy setting. All these questions and possible extensions call for future research.

%
%
%
%
%
%
%
%
%
%
%
%
%
%
%

\section*{Acknowledgement} 
Li's research was partially supported by National Institute on Aging, grant 5R21AG058198. We thank the two anonymous referees for their constructive comments and helpful suggestions, which lead to significant improvement of our paper.

\section*{Supplemental Materials}
The online Supplementary Material contains detailed proofs of the theoretical results, additional figures and tables for the simulation studies and real data analysis, and the codes implementing the proposed methods.


\nocite{ZhangH2016Identifying, liang2020modeling, zhangphdthesis}
\bibliographystyle{apalike}
\setlength{\bibsep}{5.5pt}
\bibliography{spatialFDA}

\begin{thebibliography}{}

\bibitem[Bosq, 2012]{bosq2012nonparametric}
Bosq, D. (2012).
\newblock {\em Nonparametric Statistics for Stochastic Processes: Estimation
  and Prediction}.
\newblock Springer Science \& Business Media, New York.

\bibitem[Demko, 1977]{demko1977inverses}
Demko, S. (1977).
\newblock Inverses of band matrices and local convergence of spline
  projections.
\newblock {\em SIAM Journal on Numerical Analysis}, 14(4):616--619.

\bibitem[Demko et~al., 1984]{demko1984decay}
Demko, S., Moss, W.~F., and Smith, P.~W. (1984).
\newblock Decay rates for inverses of band matrices.
\newblock {\em Mathematics of Computation}, 43(168):491--499.

\bibitem[Guan et~al., 2004]{guan2004nonparametric}
Guan, Y., Sherman, M., and Calvin, J.~A. (2004).
\newblock A nonparametric test for spatial isotropy using subsampling.
\newblock {\em Journal of the American Statistical Association},
  99(467):810--821.

\bibitem[Hall and Hosseini-Nasab, 2006]{hall2006aproperties}
Hall, P. and Hosseini-Nasab, M. (2006).
\newblock On properties of functional principal components analysis.
\newblock {\em Journal of the Royal Statistical Society: Series B},
  68(1):109--126.

\bibitem[Karr, 1986]{karr1986inference}
Karr, A.~F. (1986).
\newblock Inference for stationary random fields given poisson samples.
\newblock {\em Advances in Applied Probability}, 18(2):406--422.

\bibitem[Liang et~al., 2021]{liang2020modeling}
Liang, D., Zhang, H., Chang, X., and Huang, H. (2021).
\newblock Modeling and regionalization of {C}hina's {PM}$_{2.5}$ using
  spatial-functional mixture models.
\newblock {\em Journal of the American Statistical Association},
  116(533):116--132.

\bibitem[Mastronardi et~al., 2010]{mastronardi2010decay}
Mastronardi, N., Ng, M., and Tyrtyshnikov, E.~E. (2010).
\newblock Decay in functions of multiband matrices.
\newblock {\em SIAM Journal on Matrix Analysis and Applications},
  31(5):2721--2737.

\bibitem[Rudin, 1991]{rudin1991functional}
Rudin, W. (1991).
\newblock {\em Functional Analysis (International Series in Pure and Applied
  Mathematics)}.
\newblock McGraw-Hill, Inc., New York.

\bibitem[Schumaker, 1981]{schumaker1981spline}
Schumaker, L. (1981).
\newblock {\em Spline Functions: Basic Theory}.
\newblock Cambridge University Press, Cambridge.

\bibitem[Wang and Yang, 2009]{wang2009spline}
Wang, L. and Yang, L. (2009).
\newblock Spline estimation of single-index models.
\newblock {\em Statistica Sinica}, 19(2):765--783.

\bibitem[Zhang, 2019]{zhangphdthesis}
Zhang, H. (2019).
\newblock {\em Topics in functional data analysis and machine learning
  predictive inference}.
\newblock PhD thesis, Iowa State University.

\bibitem[Zhang et~al., 2016]{ZhangH2016Identifying}
Zhang, H., Zhu, Z., and Yin, S. (2016).
\newblock Identifying precipitation regimes in {C}hina using model-based
  clustering of spatial functional data.
\newblock In {\em Proceedings of the Sixth International Workshop on Climate
  Informatics}, pages 117--120.

\bibitem[Zhou et~al., 1998]{zhou1998local}
Zhou, S., Shen, X., and Wolfe, D. (1998).
\newblock Local asymptotics for regression splines and confidence regions.
\newblock {\em The Annals of Statistics}, 26(5):1760--1782.

\end{thebibliography}


\begin{thebibliography}{}

\bibitem[Aue et~al., 2015]{Aue2015jasa}
Aue, A., Norinho, D.~D., and H{\"o}rmann, S. (2015).
\newblock On the prediction of stationary functional time series.
\newblock {\em Journal of the American Statistical Association},
  110(509):378--392.

\bibitem[Banerjee et~al., 2004]{Banerjee2003}
Banerjee, S., Carlin, B.~P., and Gelfand, A.~E. (2004).
\newblock {\em Hierarchical Modeling and Analysis for Spatial Data}.
\newblock Chapman and Hall/CRC, New York.

\bibitem[Cai and Hall, 2006]{CaiHall06}
Cai, T.~T. and Hall, P. (2006).
\newblock Prediction in functional linear regression.
\newblock {\em The Annals of Statistics}, 34(5):2159--2179.

\bibitem[Campbell et~al., 2009]{campbell2009moves}
Campbell, S.~D., Davis, M.~A., Gallin, J., and Martin, R.~F. (2009).
\newblock What moves housing markets: A variance decomposition of the
  rent--price ratio.
\newblock {\em Journal of Urban Economics}, 66(2):90--102.

\bibitem[Crainiceanu et~al., 2009]{crainiceanu2009generalized}
Crainiceanu, C.~M., Staicu, A.-M., and Di, C.-Z. (2009).
\newblock Generalized multilevel functional regression.
\newblock {\em Journal of the American Statistical Association},
  104(488):1550--1561.

\bibitem[Cressie, 1993]{Cressie1993}
Cressie, N. A.~C. (1993).
\newblock {\em Statistics for Spatial Data}.
\newblock Wiley, New York.

\bibitem[de~Boor, 2001]{de2001practical}
de~Boor, C. (2001).
\newblock {\em A Practical Guide to Splines}.
\newblock Springer-Verlag, New York.

\bibitem[Fan et~al., 2018]{Fan2018}
Fan, J., Liu, H., and Wang, W. (2018).
\newblock Large covariance estimation through elliptical factor models.
\newblock {\em The Annals of Statistics}, 46:1383--1414.

\bibitem[Gelfand et~al., 2004]{gelfand2004nonstationary}
Gelfand, A.~E., Schmidt, A.~M., Banerjee, S., and Sirmans, C. (2004).
\newblock Nonstationary multivariate process modeling through spatially varying
  coregionalization.
\newblock {\em Test}, 13(2):263--312.

\bibitem[Giraldo et~al., 2011]{giraldo2011ordinary}
Giraldo, R., Delicado, P., and Mateu, J. (2011).
\newblock Ordinary kriging for function-valued spatial data.
\newblock {\em Environmental and Ecological Statistics}, 18(3):411--426.

\bibitem[Gromenko et~al., 2012]{gromenko2012estimation}
Gromenko, O., Kokoszka, P., Zhu, L., and Sojka, J. (2012).
\newblock Estimation and testing for spatially indexed curves with application
  to ionospheric and magnetic field trends.
\newblock {\em The Annals of Applied Statistics}, 6(2):669--696.

\bibitem[Guan et~al., 2004]{guan2004nonparametric}
Guan, Y., Sherman, M., and Calvin, J.~A. (2004).
\newblock A nonparametric test for spatial isotropy using subsampling.
\newblock {\em Journal of the American Statistical Association},
  99(467):810--821.

\bibitem[Guyon, 1995]{guyon1995random}
Guyon, X. (1995).
\newblock {\em Random Fields on a Network: Modeling, Statistics, and
  Applications}.
\newblock Springer-Verlag, New York.

\bibitem[Hall et~al., 1994]{hall1994nonparametric}
Hall, P., Fisher, N.~I., and Hoffmann, B. (1994).
\newblock On the nonparametric estimation of covariance functions.
\newblock {\em The Annals of Statistics}, 22(4):2115--2134.

\bibitem[Hall and Hosseini-Nasab, 2006]{hall2006aproperties}
Hall, P. and Hosseini-Nasab, M. (2006).
\newblock On properties of functional principal components analysis.
\newblock {\em Journal of the Royal Statistical Society: Series B},
  68(1):109--126.

\bibitem[Hall et~al., 2006]{hall2006bproperties}
Hall, P., M{\"u}ller, H.-G., and Wang, J.-L. (2006).
\newblock Properties of principal component methods for functional and
  longitudinal data analysis.
\newblock {\em The Annals of Statistics}, 34(3):1493--1517.

\bibitem[H{\"o}rmann and Kokoszka, 2010]{hormann2010weakly}
H{\"o}rmann, S. and Kokoszka, P. (2010).
\newblock Weakly dependent functional data.
\newblock {\em The Annals of Statistics}, 38(3):1845--1884.

\bibitem[H\"ormann and Kokoszka, 2013]{hormann2013consistency}
H\"ormann, S. and Kokoszka, P. (2013).
\newblock Consistency of the mean and the principal components of spatially
  distributed functional data.
\newblock {\em Bernoulli}, 19(5A):1535--1558.

\bibitem[Horv\'{a}th and Kokoszka, 2012]{HorvathKokoszka2012}
Horv\'{a}th, L. and Kokoszka, P. (2012).
\newblock {\em Inference for Functional Data with Applications}.
\newblock Springer, New York.

\bibitem[Hsing and Eubank, 2015]{Hsing-Eubank15}
Hsing, T. and Eubank, R. (2015).
\newblock {\em Theoretical Foundations of Functional Data Analysis, with an
  Introduction to Linear Operators}.
\newblock Wiley.

\bibitem[Huang and Yang, 2004]{huang2004identification}
Huang, J.~Z. and Yang, L. (2004).
\newblock Identification of non-linear additive autoregressive models.
\newblock {\em Journal of the Royal Statistical Society: Series B},
  66(2):463--477.

\bibitem[Kishor and Morley, 2015]{kishor2015factors}
Kishor, N.~K. and Morley, J. (2015).
\newblock What factors drive the price--rent ratio for the housing market? {A}
  modified present-value analysis.
\newblock {\em Journal of Economic Dynamics and Control}, 58:235--249.

\bibitem[Kokoszka and Reimherr, 2017]{KokoszkaReimherr2017}
Kokoszka, P. and Reimherr, M. (2017).
\newblock {\em Introduction to Functional Data Analysis}.
\newblock CRC Press, New York.

\bibitem[Kuenzer et~al., 2020]{Kokoszka2020}
Kuenzer, T., H{\"o}rmann, S., and Kokoszka, P. (2020).
\newblock Principal component analysis of spatially indexed functions.
\newblock {\em Journal of the American Statistical Association}, to appear.

\bibitem[Kuusela and Stein, 2018]{Kuusela-Stein2018}
Kuusela, M. and Stein, M.~L. (2018).
\newblock Locally stationary spatio-temporal interpolation of argo profiling
  float data.
\newblock {\em Proceedings of the Royal Society A: Mathematical, Physical and
  Engineering Sciences}, 474:20180400.

\bibitem[Li and Guan, 2014]{li2014functional}
Li, Y. and Guan, Y. (2014).
\newblock Functional principal component analysis of spatiotemporal point
  processes with applications in disease surveillance.
\newblock {\em Journal of the American Statistical Association},
  109(507):1205--1215.

\bibitem[Li and Hsing, 2010]{li2010uniform}
Li, Y. and Hsing, T. (2010).
\newblock Uniform convergence rates for nonparametric regression and principal
  component analysis in functional/longitudinal data.
\newblock {\em The Annals of Statistics}, 38(6):3321--3351.

\bibitem[Li et~al., 2013]{li2013selecting}
Li, Y., Wang, N., and Carroll, R.~J. (2013).
\newblock Selecting the number of principal components in functional data.
\newblock {\em Journal of the American Statistical Association},
  108(504):1284--1294.

\bibitem[Li et~al., 2007]{li2007nonparametric}
Li, Y., Wang, N., Hong, M., Turner, N.~D., Lupton, J.~R., and Carroll, R.~J.
  (2007).
\newblock Nonparametric estimation of correlation functions in longitudinal and
  spatial data, with application to colon carcinogenesis experiments.
\newblock {\em The Annals of Statistics}, 35(4):1608--1643.

\bibitem[Liang et~al., 2021]{liang2020modeling}
Liang, D., Zhang, H., Chang, X., and Huang, H. (2021).
\newblock Modeling and regionalization of {C}hina's {PM}$_{2.5}$ using
  spatial-functional mixture models.
\newblock {\em Journal of the American Statistical Association},
  116(533):116--132.

\bibitem[Liu et~al., 2017]{liu2017functional}
Liu, C., Ray, S., and Hooker, G. (2017).
\newblock Functional principal component analysis of spatially correlated data.
\newblock {\em Statistics and Computing}, 27(6):1639--1654.

\bibitem[Lu and Tj{\o}stheim, 2014]{Lu2014}
Lu, Z. and Tj{\o}stheim, D. (2014).
\newblock Nonparametric estimation of probability density functions for
  irregularly observed spatial data.
\newblock {\em Journal of the American Statistical Association},
  109(508):1546--1564.

\bibitem[Menafoglio et~al., 2016]{menafoglio2016universal}
Menafoglio, A., Grujic, O., and Caers, J. (2016).
\newblock Universal kriging of functional data: Trace-variography vs
  cross-variography? {A}pplication to gas forecasting in unconventional shales.
\newblock {\em Spatial Statistics}, 15:39--55.

\bibitem[Menafoglio et~al., 2013]{menafoglio2013}
Menafoglio, A., Secchi, P., and Dalla~Rosa, M. (2013).
\newblock A universal kriging predictor for spatially dependent functional data
  of a {H}ilbert space.
\newblock {\em Electronic Journal of Statistics}, 7:2209--2240.

\bibitem[Nerini et~al., 2010]{nerini2010cokriging}
Nerini, D., Monestiez, P., and Mant{\'e}, C. (2010).
\newblock Cokriging for spatial functional data.
\newblock {\em Journal of Multivariate Analysis}, 101(2):409--418.

\bibitem[Ramsay and Silverman, 2005]{Ramsay-Silverman05}
Ramsay, J.~O. and Silverman, B.~W. (2005).
\newblock {\em Functional Data Analysis}.
\newblock Springer, New York.

\bibitem[Rosenblatt, 1956]{rosenblatt1956remarks}
Rosenblatt, M. (1956).
\newblock Remarks on some nonparametric estimates of a density function.
\newblock {\em The Annals of Mathematical Statistics}, 27(3):832--837.

\bibitem[Schabenberger and Gotway, 2017]{schabenberger2017statistical}
Schabenberger, O. and Gotway, C.~A. (2017).
\newblock {\em Statistical Methods for Spatial Data Analysis}.
\newblock Chapman and Hall/CRC, Boca Raton.

\bibitem[Staicu et~al., 2010]{staicu2010fast}
Staicu, A.-M., Crainiceanu, C.~M., and Carroll, R.~J. (2010).
\newblock Fast methods for spatially correlated multilevel functional data.
\newblock {\em Biostatistics}, 11(2):177--194.

\bibitem[Stein, 2012]{stein2012interpolation}
Stein, M.~L. (2012).
\newblock {\em Interpolation of Spatial Data: Some Theory for Kriging}.
\newblock Springer-Verlag, New York.

\bibitem[Stone, 1994]{stone1994use}
Stone, C.~J. (1994).
\newblock The use of polynomial splines and their tensor products in
  multivariate function estimation.
\newblock {\em The Annals of Statistics}, 22(1):118--171.

\bibitem[Wang et~al., 2018]{Wang2018jrssb}
Wang, H., Zhong, P.-S., Cui, Y., and Li, Y. (2018).
\newblock Unified empirical likelihood ratio tests for functional concurrent
  linear models and the phase transition from sparse to dense functional data.
\newblock {\em Journal of the Royal Statistical Society: Series B},
  80(2):343--364.

\bibitem[Wong et~al., 2019]{Wong2019}
Wong, R., Li, Y., and Zhu, Z. (2019).
\newblock Partially linear functional additive models for multivariate
  functional data.
\newblock {\em Journal of the American Statistical Association}, 114:406--418.

\bibitem[Xiao et~al., 2013]{Xiao2013jrssb}
Xiao, L., Li, Y., and Ruppert, D. (2013).
\newblock Fast bivariate p‐splines: the sandwich smoother.
\newblock {\em Journal of the Royal Statistical Society: Series B},
  75(3):577--599.

\bibitem[Xu et~al., 2018]{XuLiNettleton2017}
Xu, Y., Li, Y., and Nettleton, D. (2018).
\newblock Nested hierarchical functional data modeling and inference for the
  analysis of functional plant phenotypes.
\newblock {\em Journal of the American Statistical Association},
  113(522):593--606.

\bibitem[Yao et~al., 2005]{yao2005functional}
Yao, F., M{\"u}ller, H.-G., and Wang, J.-L. (2005).
\newblock Functional data analysis for sparse longitudinal data.
\newblock {\em Journal of the American Statistical Association},
  100(470):577--590.

\bibitem[Zhang, 2019]{zhangphdthesis}
Zhang, H. (2019).
\newblock {\em Topics in functional data analysis and machine learning
  predictive inference}.
\newblock PhD thesis, Iowa State University.

\bibitem[Zhang et~al., 2016a]{ZhangH2016Identifying}
Zhang, H., Zhu, Z., and Yin, S. (2016a).
\newblock Identifying precipitation regimes in {C}hina using model-based
  clustering of spatial functional data.
\newblock In {\em Proceedings of the Sixth International Workshop on Climate
  Informatics}, pages 117--120.

\bibitem[Zhang and Zimmerman, 2005]{zhang2005towards}
Zhang, H. and Zimmerman, D.~L. (2005).
\newblock Towards reconciling two asymptotic frameworks in spatial statistics.
\newblock {\em Biometrika}, 92(4):921--936.

\bibitem[Zhang et~al., 2016b]{Zhang2016jasa}
Zhang, L., Baladandayuthapani, V., Zhu, H., Baggerly, K.~A., Majewski, T.,
  Czerniak, B.~A., and Morris, J.~S. (2016b).
\newblock Functional car models for large spatially correlated functional
  datasets.
\newblock {\em Journal of the American Statistical Association},
  111(514):772--786.

\bibitem[Zhang and Wang, 2016]{ZhangWang2016AOS}
Zhang, X. and Wang, J.~L. (2016).
\newblock From sparse to dense functional data and beyond.
\newblock {\em The Annals of Statistics}, 44(5):2281--2321.

\bibitem[Zhou et~al., 2010]{zhou2010reduced}
Zhou, L., Huang, J.~Z., Martinez, J.~G., Maity, A., Baladandayuthapani, V., and
  Carroll, R.~J. (2010).
\newblock Reduced rank mixed effects models for spatially correlated
  hierarchical functional data.
\newblock {\em Journal of the American Statistical Association},
  105(489):390--400.

\end{thebibliography}



\section*{Supplementary material}

# Supplemental Material to "Unified Principal Component Analysis for Sparse and Dense Functional Data under Spatial Dependency"

Haozhe Zhang

Microsoft Corporation, Redmond, WA 98052

Email: haozhe.zhang@microsoft.com

Yehua Li

Department of Statistics, University of California at Riverside, Riverside CA 92521

Email: yehua.li@ucr.edu

This supplementary article consists of the technical proofs to the theoretical results in the main paper and additional graphs for the simulation study and the real data analysis. It is organized as follows. We introduce some notation in Section S.1, present technical lemmas and their proofs in Section S.2, prove the main theorems in Section S.3, and provide additional figures for simulation studies and real data analysis in Sections S.4 and S.5, respectively. We also provide sensitivity analysis to the two real data examples in Section S.6.

## S.1 Notations

- We use $C$ (or any $C$ with a subscript) to denote a generic positive constant.

- Cumbersome notation on B-spline functions, such as $B_{j,K_t}^{p_t}(t)$ and $B_{j,K_s}^{p_s}(u)$ used in Section 3.1 of the main paper, are simplified as $B_j(t)$ and $B_j(u)$ for ease of exposition in our proofs, as long as no confusion is raised.

- For any vector $\boldsymbol{a} = (a_1, \ldots, a_p) \in \mathbb{R}^p$, denote vector norms $\|\boldsymbol{a}\|_r = (|a_1|^r + \cdots + |a_n|^r)^{1/r}$, $1 \leq r < +\infty$, and $\|\boldsymbol{a}\|_{\max} = \max(|a_1|, \ldots, |a_n|)$.

- For any $q \times p$ matrix $\boldsymbol{A} = (a_{ij})_{q \times p}$, denote $\|\boldsymbol{A}\|_r = \max_{\boldsymbol{a} \in \mathbb{R}^p, \boldsymbol{a} \neq \boldsymbol{0}} \|\boldsymbol{A}\boldsymbol{a}\|_r \|\boldsymbol{a}\|_r^{-1}$, for $1 \leq r < +\infty$, $\|\boldsymbol{A}\|_\infty = \max_{1 \leq i \leq q} \sum_{j=1}^p |a_{ij}|$, and $\|\boldsymbol{A}\|_{\max} = \max_{1 \leq i \leq q, 1 \leq j \leq p} |a_{ij}|$.

- Denote the distance between two locations as $u_{ij} = \|\boldsymbol{s}_i - \boldsymbol{s}_j\|$, the perimeter of the spatial domain $\mathcal{D}_n$ as $d_n := \max_{\boldsymbol{s}_1, \boldsymbol{s}_2 \in \mathcal{D}_n} \|\boldsymbol{s}_1 - \boldsymbol{s}_2\|_2$, and a disc of radius $h$ centered at the origin as $D_h := \{\boldsymbol{x} \in \mathbb{R}^2 : \|\boldsymbol{x}\|_2 \leq h\}$.

- Let $\mathcal{G} := \{(\boldsymbol{s}_i, t_{ij}) | i = 1, \ldots, N, j = 1, \ldots, M_i\}$ be the collection of all locations and times, which is a realization of the point process $\mathcal{N}(\cdot, \cdot)$, and denote $\boldsymbol{Y} := \{Y(\boldsymbol{s}, t) | (\boldsymbol{s}, t) \in \mathcal{G}\}$.

  Additionally, define $\mathcal{D}_n^{\otimes k} = \underbrace{\mathcal{D}_n \times \cdots \times \mathcal{D}_n}_{k}$ and $T^{\otimes k} = \underbrace{T \times \cdots \times T}_{k}$, for $k = 1, 2, 3, 4$.

- The tensor product spline coefficient in (8) is the vectorization of a three dimensional array, with dimensions $(K_s + p_s) \times (K_t + p_t) \times (K_t + p_t)$. For convenience, we use an index vector to denote the location of an entry in the 3-dim array and the corresponding location in the vectorization. For two index vectors $\boldsymbol{\ell} = (\ell_1, \ell_2, \ell_3)$ and $\boldsymbol{\ell}' = (\ell_1', \ell_2', \ell_3')$, where $\ell_1, \ell_1' \in \{1, \ldots, K_s + p_s\}$ and $\ell_2, \ell_2', \ell_3, \ell_3' \in \{1, \ldots, K_t + p_t\}$, let $a_{\boldsymbol{\ell}}$ represent the $\{(K_t + p_t)^2(\ell_1 - 1) + (K_t + p_t)(\ell_2 - 1) + \ell_3\}$th element of the vector $\boldsymbol{a} \in \mathbb{R}^{(K_s + p_s)(K_t + p_t)^2}$, and let $a_{\boldsymbol{\ell}, \boldsymbol{\ell}'}$ represent the element on the $\{(K_t + p_t)^2(\ell_1 - 1) + (K_t + p_t)(\ell_2 - 1) + \ell_3\}$th row and the $\{(K_t + p_t)^2(\ell_1' - 1) + (K_t + p_t)(\ell_2' - 1) + \ell_3'\}$th column of the matrix $\boldsymbol{A} \in \mathbb{R}^{\{(K_s + p_s)(K_t + p_t)^2\} \times \{(K_s + p_s)(K_t + p_t)^2\}}$.

## S.2 Technical Lemmas

<u>LEMMA</u> 1 *Under Assumption 7, there exists an $R^* \in \mathcal{S}_{[3]}$, the three-dimensional tensor product spline space defined in the Section 3.1, such that $\|R - R^*\|_\infty = O(K_s^{-p_s} + K_t^{-p_t})$ as $K_s, K_t \to \infty$, where $\|f\|_\infty = \sup_{(u, t_1, t_2) \in H} |f(u, t_1, t_2)|$.*

**Proof of Lemma 1:** Lemma 1 follows directly from Theorem 12.7 of (Schumaker, 1981, p.491). □

<u>LEMMA</u> 2 *Let $\{b_{j_1 j_2 j_3} : j_1 = 1, \ldots, K_s + p_s; j_2, j_3 = 1, \ldots, K_t + p_t\}$ be a 3-dim array of spline coefficients, and $\mathbf{b}$ be its vectorization so that $\boldsymbol{B}_{[3]}^T(u, t_1, t_2) \cdot \mathbf{b} = \sum_{j_1=1}^{K_s+p_s} \sum_{j_2=1}^{K_t+p_t} \sum_{j_3=1}^{K_t+p_t} b_{j_1 j_2 j_3} B_{j_1 j_2 j_3}(u, t_1, t_2)$. There exist constants $C_1, C_2, C_3, C_4$, and $C_5$, such that,*

$$\frac{C_1 \|\mathbf{b}\|_2^2}{K_s K_t^2} \leq \int_{[0,\Delta] \times T \times T} \{\boldsymbol{B}_{[3]}^T(u, t_1, t_2) \cdot \mathbf{b}\}^2 du dt_1 dt_2 \leq \frac{C_2 \|\mathbf{b}\|_2^2}{K_s K_t^2}, \tag{S.1}$$

$$\int_{T^{\otimes 2}} \left\{ \int_0^\Delta \boldsymbol{B}_{[3]}^T(u, t_1, t_2) du \cdot \mathbf{b} \right\}^2 dt_1 dt_2 \leq \frac{C_3}{K_t^2} \sum_{j_2, j_3} \left( \sum_{j_1} b_{j_1 j_2 j_3} \cdot \int_0^\Delta B_{j_1}(u) du \right)^2, \tag{S.2}$$

$$\int_T \left\{ \int_0^\Delta \int_T \boldsymbol{B}_{[3]}^T(u, t_1, t_2) dt_1 du \cdot \mathbf{b} \right\}^2 dt_2 \leq \frac{C_4}{K_t} \sum_{j_3} \left( \sum_{j_1, j_2} b_{j_1 j_2 j_3} \cdot \int_0^\Delta \int_T B_{j_1}(u) B_{j_2}(t_1) dt_1 du \right)^2 \tag{S.3}$$

$$\int_{T^{\otimes 2}} \left\{ \boldsymbol{B}_{[3]}^T(0, t_1, t_2) dt_1 dt_2 \cdot \mathbf{b} \right\}^2 dt_1 dt_2 \leq \frac{C_5}{K_t^2} \sum_{j_2, j_3} \left( \sum_{1 \leq j_1 \leq p_s} b_{j_1 j_2 j_3} \cdot B_{j_1}(0) \right)^2. \tag{S.4}$$

**Proof of Lemma 2:** By applying inequality (13) in Zhou et al. (1998, p.1770) repeatedly,

$$\int_{[0,\Delta]\times T\times T} \{\boldsymbol{B}_{[3]}^T(u,t_1,t_2)\cdot \mathbf{b}\}^2 \, dudt_1dt_2$$

$$= \int_{[0,\Delta]\times T\times T} \left[\sum_{j_1}\left\{\sum_{j_2,j_3} b_{j_1j_2j_3} B_{j_2}(t_1)B_{j_3}(t_2)\right\} B_{j_1}(u)\right]^2 dudt_1dt_2$$

$$\leq \frac{C}{K_s}\sum_{j_1}\int_{T^{\otimes 2}}\left\{\sum_{j_2,j_3} b_{j_1j_2j_3} B_{j_2}(t_1)B_{j_3}(t_2)\right\}^2 dt_1dt_2$$

$$\leq \frac{C^2}{K_sK_t}\sum_{j_1,j_2}\int_T\left\{\sum_{j_3} b_{j_1j_2j_3} B_{j_3}(t_2)\right\}^2 dt_2 \leq \frac{C^3}{K_sK_t^2}\|\boldsymbol{b}\|_2^2,$$

and hence the right hand side of (S.1) follows. Since inequality (13) of Zhou et al. (1998) provides both the upper bound and lower bound of the squared $L^2$ norm of a spline function, the left hand side of (S.1) is obtained following a similar argument by repeatedly applying the lower bound inequality of Zhou et al. (1998).

Similarly, inequality (S.2) is derived as follows

$$\int_{T^{\otimes 2}}\left\{\int_{[0,\Delta]} \boldsymbol{B}_{[3]}^T(u,t_1,t_2)du \cdot \mathbf{b}\right\}^2 dt_1dt_2$$

$$= \int_{T^{\otimes 2}}\left\{\sum_{j_1,j_2,j_3} b_{j_1j_2j_3} \cdot \int_{[0,\Delta]} B_{j_1}(u)du \cdot B_{j_2}(t_1)B_{j_3}(t_2)\right\}^2 dt_1dt_2$$

$$\leq \frac{C_3}{K_t^2}\sum_{j_2,j_3}\left(\sum_{j_1} b_{j_1j_2j_3} \cdot \int_{[0,\Delta]} B_{j_1}(u)du\right)^2.$$

Inequality (S.3) follows similar arguments as (S.2), and (S.4) follows from the fact that $B_{j_1j_2j_3}(0,t_1,t_2) \equiv 0$ if $j_1 > p_s$. □

LEMMA 3 *Define*

$$\boldsymbol{\xi}_n := \frac{1}{|\mathcal{D}_n|M_n^2L_n^2}\int_{\mathcal{D}_n^{\otimes 2}}\int_{T^{\otimes 2}} \boldsymbol{B}_{[3]}(\|\boldsymbol{s}_1-\boldsymbol{s}_2\|,t_1,t_2)\cdot \{Y(\boldsymbol{s}_1,t_1)Y(\boldsymbol{s}_2,t_2)-R(\|\boldsymbol{s}_1-\boldsymbol{s}_2\|,t_1,t_2)\}$$
$$\times I\left(\|\boldsymbol{s}_1-\boldsymbol{s}_2\|\leq \Delta\right)\mathcal{N}_{s,2}(d\boldsymbol{s}_1,d\boldsymbol{s}_2)\mathcal{N}_t(dt_1|\boldsymbol{s}_1)\mathcal{N}_t(dt_2|\boldsymbol{s}_2).$$

*Following the same index convention in the previous lemma, the* $(j_1,j_2,j_3)$*th entry in* $\boldsymbol{\xi}_n$ *is*

$$\xi_{j_1j_2j_3} = \frac{1}{|\mathcal{D}_n|M_n^2L_n^2}\int_{\mathcal{D}_n^{\otimes 2}}\int_{T^{\otimes 2}} B_{j_1}\left(\|\boldsymbol{s}_1-\boldsymbol{s}_2\|\right) B_{j_2}(t_1)B_{j_3}(t_2) I\left(\|\boldsymbol{s}_1-\boldsymbol{s}_2\|\leq \Delta\right)$$
$$\times\{Y(\boldsymbol{s}_1,t_1)Y(\boldsymbol{s}_2,t_2)-R(\|\boldsymbol{s}_1-\boldsymbol{s}_2\|,t_1,t_2)\}\mathcal{N}_{s,2}(d\boldsymbol{s}_1,d\boldsymbol{s}_2)\mathcal{N}_t(dt_1|\boldsymbol{s}_1)\mathcal{N}_t(dt_2|\boldsymbol{s}_2),$$

for $j_1 \in \{1, \ldots, K_s + p_s\}$ and $j_2, j_3 \in \{1, \ldots, K_t + p_t\}$. Denote $\mathcal{G} := \{(\boldsymbol{s}_i, t_{ij}) : i = 1, \ldots, N, j = 1, \ldots, M_i\}$ as the collection of the realizations of the spatial point process $\mathcal{N}_s(d\boldsymbol{s})$ and the temporal point process $\mathcal{N}_t(dt|\boldsymbol{s})$. Under Assumptions 1–7, there exists some constant $C > 0$ not depending on n or any substripts $(j_1 j_2 j_3, j_1' j_2' j_3')$, such that

- for $|j_1 - j_1'| > p_s$ and $\min\left(|j_2 - j_2'|, |j_3 - j_3'|, |j_2 - j_3'|, |j_3 - j_2'|\right) \leq p_t$,

$$\mathrm{E}\left\{\left|\mathrm{E}\left(\xi_{j_1 j_2 j_3} \xi_{j_1' j_2' j_3'} | \mathcal{G}\right)\right|\right\} \leq C \left(\frac{1}{|\mathcal{D}_n| K_s^2 K_t^4} + \frac{1}{|\mathcal{D}_n| M_n L_n K_s^2 K_t^3}\right); \quad (\text{S}.5)$$

- for $|j_1 - j_1'| > p_s$ and $\min\left(|j_2 - j_2'|, |j_3 - j_3'|, |j_2 - j_3'|, |j_3 - j_2'|\right) > p_t$,

$$\mathrm{E}\left\{\left|\mathrm{E}\left(\xi_{j_1 j_2 j_3} \xi_{j_1' j_2' j_3'} | \mathcal{G}\right)\right|\right\} \leq \frac{C}{|\mathcal{D}_n| K_s^2 K_t^4}; \quad (\text{S}.6)$$

- for $|j_1 - j_1'| \leq p_s$ and $\min\left(|j_2 - j_2'|, |j_3 - j_3'|, |j_2 - j_3'|, |j_3 - j_2'|\right) \leq p_t$,

$$\mathrm{E}\left\{\left|\mathrm{E}\left(\xi_{j_1 j_2 j_3} \xi_{j_1' j_2' j_3'} | \mathcal{G}\right)\right|\right\} \leq C \left(\frac{1}{|\mathcal{D}_n| K_s K_t^4} + \frac{1}{|\mathcal{D}_n| M_n L_n K_s K_t^3} + \frac{1}{|\mathcal{D}_n| M_n^2 L_n^2 K_s K_t^2}\right); \quad (\text{S}.7)$$

- for $|j_1 - j_1'| \leq p_s$, $\min\left(|j_2 - j_2'|, |j_3 - j_3'|\right) > p_t$ and $\min\left(|j_2 - j_3'|, |j_3 - j_2'|\right) > p_t$,

$$\mathrm{E}\left\{\left|\mathrm{E}\left(\xi_{j_1 j_2 j_3} \xi_{j_1' j_2' j_3'} | \mathcal{G}\right)\right|\right\} \leq C \left(\frac{1}{|\mathcal{D}_n| K_s K_t^4} + \frac{1}{|\mathcal{D}_n| M_n L_n K_s K_t^3}\right); \quad (\text{S}.8)$$

- for $|j_1 - j_1'| \leq p_s$ and $\min\left(|j_2 - j_2'|, |j_3 - j_3'|, |j_2 - j_3'|, |j_3 - j_2'|\right) > p_t$

$$\mathrm{E}\left\{\left|\mathrm{E}\left(\xi_{j_1 j_2 j_3} \xi_{j_1' j_2' j_3'} | \mathcal{G}\right)\right|\right\} \leq \frac{C}{|\mathcal{D}_n| K_s K_t^4}. \quad (\text{S}.9)$$

It also holds that,

$$\|\boldsymbol{\xi}_n\|_2^2 = O_p\left(\frac{1}{|\mathcal{D}_n| K_t^2} + \frac{1}{|\mathcal{D}_n| M_n L_n K_t} + \frac{1}{|\mathcal{D}_n| M_n^2 L_n^2}\right). \quad (\text{S}.10)$$

**Proof of Lemma 3:** We first show (S.5). By the definition,

$$\mathrm{E}\left\{\left|\mathrm{E}\left(\xi_{j_1 j_2 j_3} \xi_{j_1' j_2' j_3'} | \mathcal{G}\right)\right|\right\}$$

$$\leq \frac{1}{|\mathcal{D}_n|^2 M_n^4 L_n^4} \mathrm{E}\Bigg\{ \int_{\mathcal{D}_n^{\otimes 4}} \int_{T^{\otimes 4}} B_{j_1}(\|\boldsymbol{s}_1 - \boldsymbol{s}_2\|) B_{j_1'}(\|\boldsymbol{s}_3 - \boldsymbol{s}_4\|) B_{j_2}(t_1) B_{j_2'}(t_3) B_{j_3}(t_2) B_{j_3'}(t_4)$$
$$\times |\mathrm{cov}\{Y(\boldsymbol{s}_1, t_1) Y(\boldsymbol{s}_2, t_2), Y(\boldsymbol{s}_3, t_3) Y(\boldsymbol{s}_4, t_4)\}| \times I(\|\boldsymbol{s}_1 - \boldsymbol{s}_2\| \leq \Delta) I(\|\boldsymbol{s}_3 - \boldsymbol{s}_4\| \leq \Delta)$$
$$\times \mathcal{N}_{s,2}(d\boldsymbol{s}_1, d\boldsymbol{s}_2) \mathcal{N}_{s,2}(d\boldsymbol{s}_3, d\boldsymbol{s}_4) \mathcal{N}_t(dt_1|\boldsymbol{s}_1) \mathcal{N}_t(dt_2|\boldsymbol{s}_2) \mathcal{N}_t(dt_3|\boldsymbol{s}_3) \mathcal{N}_t(dt_4|\boldsymbol{s}_4) \Bigg\}. \tag{S.11}$$

Following similar calculations as in Guan et al. (2004),

$$\mathrm{E}\{\mathcal{N}_{s,2}(d\boldsymbol{s}_1, d\boldsymbol{s}_2) \mathcal{N}_{s,2}(d\boldsymbol{s}_3, d\boldsymbol{s}_4) \mathcal{N}_t(dt_1|\boldsymbol{s}_1) \mathcal{N}_t(dt_2|\boldsymbol{s}_2) \mathcal{N}_t(dt_3|\boldsymbol{s}_3) \mathcal{N}_t(dt_4|\boldsymbol{s}_4)\}$$
$$= \lambda_{s,4}(\boldsymbol{s}_1, \boldsymbol{s}_2, \boldsymbol{s}_3, \boldsymbol{s}_4) \lambda_{t,1}(t_1) \lambda_{t,1}(t_2) \lambda_{t,1}(t_3) \lambda_{t,1}(t_4) d\boldsymbol{s}_1 d\boldsymbol{s}_2 d\boldsymbol{s}_3 d\boldsymbol{s}_4 dt_1 dt_2 dt_3 dt_4$$
$$+ \lambda_{s,3}(\boldsymbol{s}_1, \boldsymbol{s}_2, \boldsymbol{s}_4) \lambda_{t,2}(t_1, t_3) \lambda_{t,1}(t_2) \lambda_{t,1}(t_4) \epsilon_{\boldsymbol{s}_1}(d\boldsymbol{s}_3) d\boldsymbol{s}_1 d\boldsymbol{s}_2 d\boldsymbol{s}_4 dt_1 dt_2 dt_3 dt_4$$
$$+ \lambda_{s,3}(\boldsymbol{s}_1, \boldsymbol{s}_2, \boldsymbol{s}_3) \lambda_{t,2}(t_1, t_4) \lambda_{t,1}(t_2) \lambda_{t,1}(t_3) \epsilon_{\boldsymbol{s}_1}(d\boldsymbol{s}_4) d\boldsymbol{s}_1 d\boldsymbol{s}_2 d\boldsymbol{s}_3 dt_1 dt_2 dt_3 dt_4$$
$$+ \lambda_{s,3}(\boldsymbol{s}_1, \boldsymbol{s}_2, \boldsymbol{s}_4) \lambda_{t,2}(t_2, t_3) \lambda_{t,1}(t_1) \lambda_{t,1}(t_4) \epsilon_{\boldsymbol{s}_2}(d\boldsymbol{s}_3) d\boldsymbol{s}_1 d\boldsymbol{s}_2 d\boldsymbol{s}_4 dt_1 dt_2 dt_3 dt_4$$
$$+ \lambda_{s,3}(\boldsymbol{s}_1, \boldsymbol{s}_2, \boldsymbol{s}_3) \lambda_{t,2}(t_2, t_4) \lambda_{t,1}(t_1) \lambda_{t,1}(t_3) \epsilon_{\boldsymbol{s}_2}(d\boldsymbol{s}_4) d\boldsymbol{s}_1 d\boldsymbol{s}_2 d\boldsymbol{s}_3 dt_1 dt_2 dt_3 dt_4$$
$$+ \lambda_{s,3}(\boldsymbol{s}_1, \boldsymbol{s}_2, \boldsymbol{s}_4) \lambda_{t,1}(t_1) \lambda_{t,1}(t_2) \lambda_{t,1}(t_4) \epsilon_{\boldsymbol{s}_1}(d\boldsymbol{s}_3) \epsilon_{t_1}(t_3) d\boldsymbol{s}_1 d\boldsymbol{s}_2 d\boldsymbol{s}_4 dt_1 dt_2 dt_4$$
$$+ \lambda_{s,3}(\boldsymbol{s}_1, \boldsymbol{s}_2, \boldsymbol{s}_3) \lambda_{t,1}(t_1) \lambda_{t,1}(t_2) \lambda_{t,1}(t_3) \epsilon_{\boldsymbol{s}_1}(d\boldsymbol{s}_4) \epsilon_{t_1}(t_4) d\boldsymbol{s}_1 d\boldsymbol{s}_2 d\boldsymbol{s}_3 dt_1 dt_2 dt_3$$
$$+ \lambda_{s,3}(\boldsymbol{s}_1, \boldsymbol{s}_2, \boldsymbol{s}_4) \lambda_{t,1}(t_1) \lambda_{t,1}(t_2) \lambda_{t,1}(t_4) \epsilon_{\boldsymbol{s}_2}(d\boldsymbol{s}_3) \epsilon_{t_2}(t_3) d\boldsymbol{s}_1 d\boldsymbol{s}_2 d\boldsymbol{s}_4 dt_1 dt_2 dt_4$$
$$+ \lambda_{s,3}(\boldsymbol{s}_1, \boldsymbol{s}_2, \boldsymbol{s}_3) \lambda_{t,1}(t_1) \lambda_{t,2}(t_2) \lambda_{t,1}(t_3) \epsilon_{\boldsymbol{s}_2}(d\boldsymbol{s}_4) \epsilon_{t_2}(t_4) d\boldsymbol{s}_1 d\boldsymbol{s}_2 d\boldsymbol{s}_3 dt_1 dt_2 dt_3$$
$$+ \lambda_{s,2}(\boldsymbol{s}_1, \boldsymbol{s}_2) \lambda_{t,1}(t_1) \lambda_{t,1}(t_2) \epsilon_{\boldsymbol{s}_1}(d\boldsymbol{s}_3) \epsilon_{\boldsymbol{s}_2}(d\boldsymbol{s}_4) \epsilon_{t_1}(t_3) \epsilon_{t_2}(t_4) d\boldsymbol{s}_1 d\boldsymbol{s}_2 dt_1 dt_2$$
$$+ \lambda_{s,2}(\boldsymbol{s}_1, \boldsymbol{s}_2) \lambda_{t,1}(t_1) \lambda_{t,1}(t_2) \epsilon_{\boldsymbol{s}_2}(d\boldsymbol{s}_3) \epsilon_{\boldsymbol{s}_1}(d\boldsymbol{s}_4) \epsilon_{t_2}(t_3) \epsilon_{t_1}(t_4) d\boldsymbol{s}_1 d\boldsymbol{s}_2 dt_1 dt_2$$
$$+ \lambda_{s,2}(\boldsymbol{s}_1, \boldsymbol{s}_2) \lambda_{t,1}(t_1) \lambda_{t,1}(t_2) \lambda_{t,1}(t_3) \epsilon_{\boldsymbol{s}_1}(d\boldsymbol{s}_3) \epsilon_{\boldsymbol{s}_2}(d\boldsymbol{s}_4) \epsilon_{t_2}(t_4) d\boldsymbol{s}_1 d\boldsymbol{s}_2 dt_1 dt_2 dt_3$$
$$+ \lambda_{s,2}(\boldsymbol{s}_1, \boldsymbol{s}_2) \lambda_{t,1}(t_1) \lambda_{t,1}(t_2) \lambda_{t,1}(t_3) \epsilon_{\boldsymbol{s}_2}(d\boldsymbol{s}_3) \epsilon_{\boldsymbol{s}_1}(d\boldsymbol{s}_4) \epsilon_{t_1}(t_4) d\boldsymbol{s}_1 d\boldsymbol{s}_2 dt_1 dt_2 dt_3$$
$$+ \lambda_{s,2}(\boldsymbol{s}_1, \boldsymbol{s}_2) \lambda_{t,1}(t_1) \lambda_{t,1}(t_2) \lambda_{t,1}(t_4) \epsilon_{\boldsymbol{s}_1}(d\boldsymbol{s}_3) \epsilon_{\boldsymbol{s}_2}(d\boldsymbol{s}_4) \epsilon_{t_1}(t_3) d\boldsymbol{s}_1 d\boldsymbol{s}_2 dt_1 dt_2 dt_4$$
$$+ \lambda_{s,2}(\boldsymbol{s}_1, \boldsymbol{s}_2) \lambda_{t,1}(t_1) \lambda_{t,1}(t_2) \lambda_{t,1}(t_4) \epsilon_{\boldsymbol{s}_2}(d\boldsymbol{s}_3) \epsilon_{\boldsymbol{s}_1}(d\boldsymbol{s}_4) \epsilon_{t_2}(t_3) d\boldsymbol{s}_1 d\boldsymbol{s}_2 dt_1 dt_2 dt_4$$
$$+ \lambda_{s,2}(\boldsymbol{s}_1, \boldsymbol{s}_2) \lambda_{t,2}(t_1, t_3) \lambda_{t,2}(t_2, t_4) \epsilon_{\boldsymbol{s}_1}(d\boldsymbol{s}_3) \epsilon_{\boldsymbol{s}_2}(d\boldsymbol{s}_4) d\boldsymbol{s}_1 d\boldsymbol{s}_2 dt_1 dt_2 dt_3 dt_4$$
$$+ \lambda_{s,2}(\boldsymbol{s}_1, \boldsymbol{s}_2) \lambda_{t,2}(t_2, t_3) \lambda_{t,2}(t_1, t_4) \epsilon_{\boldsymbol{s}_2}(d\boldsymbol{s}_3) \epsilon_{\boldsymbol{s}_1}(d\boldsymbol{s}_4) d\boldsymbol{s}_1 d\boldsymbol{s}_2 dt_1 dt_2 dt_3 dt_4, \tag{S.12}$$

where $\epsilon_x(\cdot)$ is a point measure defined in Karr (1986), such that $\epsilon_x(dy) = 1$ if $x \in dy$, 0 otherwise. Here $dy$ is defined to be a small disc centered at $y$.

Thus, the right hand side of (S.11) can be decomposed into 17 integrals, denoted in order as $\mathcal{Q}_1 - \mathcal{Q}_{17}$ according to the 17 terms in (S.12). We first derive the upper bound of $\mathcal{Q}_1$. By Assumptions 4 and 5, $\lambda_{s,4}(\boldsymbol{s}_1, \boldsymbol{s}_2, \boldsymbol{s}_3, \boldsymbol{s}_4) \lambda_{t,1}(t_1) \lambda_{t,1}(t_2) \lambda_{t,1}(t_3) \lambda_{t,1}(t_4) / (M_n^4 L_n^4)$ is positive and bounded above by some constant uniformly for all $\boldsymbol{s}_1, \boldsymbol{s}_2, \boldsymbol{s}_3, \boldsymbol{s}_4 \in \mathcal{D}_n$ and $t_1, t_2, t_3, t_4 \in T$.

Thus, for some constant $C_1 > 0$,

$$\begin{aligned}
\mathcal{Q}_1 &= \int_{\mathcal{D}_n^{\otimes 4}} \int_{T^{\otimes 4}} \frac{1}{|\mathcal{D}_n|^2 M_n^4 L_n^4} B_{j_1}(\|\boldsymbol{s}_1 - \boldsymbol{s}_2\|) B_{j_1'}(\|\boldsymbol{s}_3 - \boldsymbol{s}_4\|) B_{j_2}(t_1) B_{j_3}(t_2) B_{j_2'}(t_3) B_{j_3'}(t_4) \\
&\quad \times |\text{cov}\{Y(\boldsymbol{s}_1, t_1) Y(\boldsymbol{s}_2, t_2), Y(\boldsymbol{s}_3, t_3) Y(\boldsymbol{s}_4, t_4)\}| \cdot I(\|\boldsymbol{s}_1 - \boldsymbol{s}_2\| \leq \Delta) I(\|\boldsymbol{s}_3 - \boldsymbol{s}_4\| \leq \Delta) \\
&\quad \times \lambda_{s,4}(\boldsymbol{s}_1, \boldsymbol{s}_2, \boldsymbol{s}_3, \boldsymbol{s}_4) \lambda_{t,1}(t_1) \lambda_{t,1}(t_2) \lambda_{t,1}(t_3) \lambda_{t,1}(t_4) d\boldsymbol{s}_1 d\boldsymbol{s}_2 d\boldsymbol{s}_3 d\boldsymbol{s}_4 dt_1 dt_2 dt_3 dt_4 \\
&\leq \frac{C_1}{|\mathcal{D}_n|^2} \int_{T^{\otimes 4}} B_{j_2}(t_1) B_{j_3}(t_2) B_{j_2'}(t_3) B_{j_3'}(t_4) \Bigg[ \int_{\mathcal{D}_n} \int_{D_\Delta} \int_{D_\Delta} B_{j_1}(\|\boldsymbol{u}\|) B_{j_1'}(\|\boldsymbol{v}\|) \times \\
&\quad \int_{\mathcal{D}_{d_N}} |\text{cov}\{Y(\boldsymbol{s}, t_1) Y(\boldsymbol{s} + \boldsymbol{u}, t_2), Y(\boldsymbol{s} + \boldsymbol{\omega}, t_3) Y(\boldsymbol{s} + \boldsymbol{\omega} + \boldsymbol{\nu}, t_4)\}| d\boldsymbol{\omega} d\boldsymbol{u} d\boldsymbol{\nu} d\boldsymbol{s} \Bigg] dt_1 dt_2 dt_3 dt_4 \\
&= \frac{C_1}{|\mathcal{D}_n|^2} \int_{T^{\otimes 4}} B_{j_2}(t_1) B_{j_3}(t_2) B_{j_2'}(t_3) B_{j_3'}(t_4) \int_{\mathcal{D}_n} \int_{D_\Delta} \int_{D_\Delta} B_{j_1}(\|\boldsymbol{u}\|) B_{j_1'}(\|\boldsymbol{v}\|) \\
&\quad \times \Bigg[ \int_{\|\boldsymbol{\omega}\| > 2\Delta} |\text{cov}\{Y(\boldsymbol{s}, t_1) Y(\boldsymbol{s} + \boldsymbol{u}, t_2), Y(\boldsymbol{s} + \boldsymbol{\omega}, t_3) Y(\boldsymbol{s} + \boldsymbol{\omega} + \boldsymbol{\nu}, t_4)\}| d\boldsymbol{\omega} \\
&\quad + \int_{\|\boldsymbol{\omega}\| < 2\Delta} |\text{cov}\{Y(\boldsymbol{s}, t_1) Y(\boldsymbol{s} + \boldsymbol{u}, t_2), Y(\boldsymbol{s} + \boldsymbol{\omega}, t_3) Y(\boldsymbol{s} + \boldsymbol{\omega} + \boldsymbol{\nu}, t_4)\}| d\boldsymbol{\omega} \Bigg] \\
&\quad \times d\boldsymbol{u} d\boldsymbol{\nu} d\boldsymbol{s} dt_1 dt_2 dt_3 dt_4.
\end{aligned}$$

On one hand, Assumption 2 implies $|\text{cov}\{Y(\boldsymbol{s}, t_1) Y(\boldsymbol{s} + \boldsymbol{u}, t_2), Y(\boldsymbol{s} + \boldsymbol{\omega}, t_3) Y(\boldsymbol{s} + \boldsymbol{\omega} + \boldsymbol{\nu}, t_4)\}| \leq C_2$ for some constant $C_2 > 0$, when $\|\boldsymbol{\omega}\| \leq 2\Delta$. On the other hand, let $\nu > 4$ be the constant defined in Assumption 2 and put $\kappa = \nu/(\nu - 4) > 1$, then by Davydov's Inequality (Bosq, 2012), when $\|\boldsymbol{\omega}\| > 2\Delta$, $\|\boldsymbol{u}\| \leq \Delta$ and $\|\boldsymbol{v}\| \leq \Delta$,

$$\begin{aligned}
&|\text{cov}\{Y(\boldsymbol{s}, t_1) Y(\boldsymbol{s} + \boldsymbol{u}, t_2), Y(\boldsymbol{s} + \boldsymbol{\omega}, t_3) Y(\boldsymbol{s} + \boldsymbol{\omega} + \boldsymbol{\nu}, t_4)\}| \\
&\leq 2\kappa \{2\alpha_X(\|\boldsymbol{\omega}\|)\}^{1/\kappa} \{E|Y(\boldsymbol{s}_1, t_1) Y(\boldsymbol{s} + \boldsymbol{u}, t_2)|^{\nu/2}\}^{2/\nu} \left[ E\{|Y(\boldsymbol{s} + \boldsymbol{\omega}, t_3) Y(\boldsymbol{s} + \boldsymbol{\omega} + \boldsymbol{\nu}, t_4)|\}^{\nu/2} \right]^{2/\nu} \\
&\leq C_3 \{\alpha_X(\|\boldsymbol{\omega}\|)\}^{1/\kappa},
\end{aligned}$$

for some $C_3 > 0$. Here $\alpha_X(\cdot)$ is the $\alpha$-mixing coefficient define in (17). It follows that

$$\begin{aligned}
\mathcal{Q}_1 &\leq \frac{C_4}{|\mathcal{D}_n|^2} \int_{T^{\otimes 4}} B_{j_2}(t_1) B_{j_3}(t_2) B_{j_2'}(t_3) B_{j_3'}(t_4) \int_{\mathcal{D}_n} \int_{D_\Delta} \int_{D_\Delta} B_{j_1}(\|\boldsymbol{u}\|) B_{j_1'}(\|\boldsymbol{v}\|) \\
&\quad \times \left[ \int_{\|\boldsymbol{\omega}\| > 2\Delta} C_3 \cdot \{\alpha_X(\|\boldsymbol{\omega}\|)\}^{1/\kappa} d\boldsymbol{\omega} + \int_{\|\boldsymbol{\omega}\| < 2\Delta} C_2 d\boldsymbol{\omega} \right] \cdot d\boldsymbol{u} d\boldsymbol{\nu} d\boldsymbol{s} dt_1 dt_2 dt_3 dt_4,
\end{aligned}$$

for some constant $C_4 > 0$. By Assumption 3, $\alpha_X(\|\boldsymbol{\omega}\|) \leq C\|\boldsymbol{\omega}\|^{-\delta_1}$ for all $\boldsymbol{\omega}$ and $\delta_1/\kappa > 2$, $\int_{\|\boldsymbol{\omega}\| \geq 2\Delta} \{\alpha_X(\|\boldsymbol{\omega}\|)\}^{1/\kappa} d\boldsymbol{\omega} \leq C \int_{\|\boldsymbol{\omega}\| \geq 2\Delta} \|\boldsymbol{\omega}\|^{-\delta_1/\kappa} d\boldsymbol{\omega} < \infty$. Since $\int_{\|\boldsymbol{\omega}\| < 2\Delta} C_2 d\boldsymbol{\omega}$ is also bounded, there exists a constant $C_5 > 0$ such that

$$\mathcal{Q}_1 \leq \frac{C_5}{|\mathcal{D}_n| K_s^2 K_t^4}.$$

Following similar arguments, we can show that, for some constant $C_6 > 0$,

$$\mathcal{Q}_2 = \mathcal{Q}_3 = \mathcal{Q}_4 = \mathcal{Q}_5 \leq \frac{C_6}{|\mathcal{D}_n|L_n K_s^2 K_t^4},$$

$$\text{and} \quad \mathcal{Q}_6 = \mathcal{Q}_7 = \mathcal{Q}_8 = \mathcal{Q}_9 \leq \frac{C_6}{|\mathcal{D}_n|L_n M_n K_s^2 K_t^3}.$$

For $\mathcal{Q}_{10}$ - $\mathcal{Q}_{17}$, either $(s_1, s_2) = (s_3, s_4)$ or $(s_1, s_2) = (s_4, s_3)$, then $\|s_1 - s_2\| = \|s_3 - s_4\|$ and

$$B_{j_1}(\|s_1 - s_2\|) B_{j_1'}(\|s_3 - s_4\|) = 0 \quad \text{if } |j_1 - j_1'| > p_s.$$

As a result,

$$\mathcal{Q}_{10} = \mathcal{Q}_{11} = \mathcal{Q}_{12} = \mathcal{Q}_{13} = \mathcal{Q}_{14} = \mathcal{Q}_{15} = \mathcal{Q}_{16} = \mathcal{Q}_{17} = 0.$$

Thus, for $|j_1 - j_1'| > p_s$ and $\min\left(|j_2 - j_2'|, |j_3 - j_3'|, |j_2 - j_3'|, |j_3 - j_2'|\right) \leq p_t$,

$$\mathrm{E}\left\{\left|\mathrm{E}\left(\xi_{j_1 j_2 j_3} \xi_{j_1' j_2' j_3'} | \mathcal{G}\right)\right|\right\} \leq \frac{C_7}{|\mathcal{D}_n| K_s^2 K_t^4} + \frac{C_7}{|\mathcal{D}_n| M_n L_n K_s^2 K_t^3},$$

for some constant $C_7 > 0$. The proof for (S.6) – (S.9) is omitted because it follows similar arguments as the proof of (S.5). From (S.7), we have

$$\mathrm{E}\|\boldsymbol{\xi}_n\|_2^2 = \sum_{j_1 j_2 j_3} \mathrm{E}\left(\xi_{j_1 j_2 j_3}^2\right) \lesssim \sum_{j_1 j_2 j_3} \left(\frac{1}{|\mathcal{D}_n| K_s K_t^4} + \frac{1}{|\mathcal{D}_n| M_n L_n K_s K_t^3} + \frac{1}{|\mathcal{D}_n| M_n^2 L_n^2 K_s K_t^2}\right)$$

$$= O\left(\frac{1}{|\mathcal{D}_n| K_t^2} + \frac{1}{|\mathcal{D}_n| M_n L_n K_t} + \frac{1}{|\mathcal{D}_n| M_n^2 L_n^2}\right),$$

and (S.10) follows. □

LEMMA 4 *Define*

$$\boldsymbol{G}_n := \frac{1}{|\mathcal{D}_n| M_n^2 L_n^2} \int_{\mathcal{D}_n^{\otimes 2}} \int_{T^{\otimes 2}} \boldsymbol{B}_{[3]}(\|s_1 - s_2\|, t_1, t_2) \cdot \boldsymbol{B}_{[3]}^T(\|s_1 - s_2\|, t_1, t_2)$$
$$\times I\left(\|s_1 - s_2\| \leq \Delta\right) \mathcal{N}_{s,2}(ds_1, ds_2) \mathcal{N}_t(dt_1|s_1) \mathcal{N}_t(dt_2|s_2),$$

*and $\boldsymbol{G} = E(\boldsymbol{G}_n)$. Under Assumptions 1 – 6,*

$$\|\boldsymbol{G}_n - \boldsymbol{G}\|_{\max} = O\left\{\frac{\log(n)}{\sqrt{K_s K_t^2 |\mathcal{D}_n|}}\right\} \quad \text{with probability 1.}$$

**Proof of Lemma 4:** Our proof is an extension of Lemma A.2 in Wang and Yang (2009) from univariate spline to multivariate spline and from time series data to spatio-temporal data. We use index vector $\boldsymbol{\ell} = (\ell_1, \ell_2, \ell_3)^{\mathrm{T}}$ to denote the location of basis function $B_{\ell_1 \ell_2 \ell_3}(u, t_1, t_2)$

in the tensor product vector $\boldsymbol{B}_{[3]}(u, t_1, t_2)$. Then, for $\boldsymbol{\ell}' = (\ell_1', \ell_2', \ell_3')^{\mathrm{T}}$, the $(\boldsymbol{\ell}, \boldsymbol{\ell}')$th entry in $\boldsymbol{G}_n$ is

$$g_{\boldsymbol{\ell},\boldsymbol{\ell}'} := \frac{1}{|\mathcal{D}_n| M_n^2 L_n^2} \int_{\mathcal{D}_n} \int_{\mathcal{D}_n} \int_T \int_T B_{\ell_1}(\|\boldsymbol{s}_1 - \boldsymbol{s}_2\|) B_{\ell_1'}(\|\boldsymbol{s}_1 - \boldsymbol{s}_2\|) B_{\ell_2}(t_1) B_{\ell_2'}(t_1) \\ \times B_{\ell_3}(t_2) B_{\ell_3'}(t_2) I(\|\boldsymbol{s}_1 - \boldsymbol{s}_2\| \leq \Delta) \mathcal{N}_{s,2}(d\boldsymbol{s}_1, d\boldsymbol{s}_2) \mathcal{N}_t(dt_1|\boldsymbol{s}_1) \mathcal{N}_t(dt_2|\boldsymbol{s}_2).$$

Under the increasing domain framework described in Assumption 1, $\mathcal{D}_n$ can be split into $n_1 \times n_2$ subsets, i.e.,

$$\mathcal{D}_n = \bigcup_{i=1}^{n_1} \bigcup_{j=1}^{n_2} \mathcal{D}_n(i,j), \quad n_1 n_2 \asymp |\mathcal{D}_n|, \quad \sqrt{|\mathcal{D}_n|} \lesssim n_1,$$

such that $C \leq |\mathcal{D}_n(i,j)| \leq C'$, $C \leq |\partial \mathcal{D}_n(i,j)| \leq C'$ and $\mathrm{dist}\{\mathcal{D}_n(i,j), \mathcal{D}_n(i',j')\} \geq C \min(|i - i'| - 1, |j - j'| - 1)$, for some $C, C' > 0$. we define that

$$\mathcal{D}_{n,\Delta}(i,j) = \{\boldsymbol{x} \in \mathbb{R}^2 : \min_{\boldsymbol{y} \in \mathcal{D}_n(i,j)} |\boldsymbol{x} - \boldsymbol{y}| \leq \Delta\}.$$

Then we rewrite $g_{\boldsymbol{\ell},\boldsymbol{\ell}'}$ as a summation of $n_1 \times n_2$ components:

$$g_{\boldsymbol{\ell},\boldsymbol{\ell}'} = \frac{1}{|\mathcal{D}_n|} \sum_{i=1}^{n_1} \sum_{j=1}^{n_2} g_{\boldsymbol{\ell},\boldsymbol{\ell}'}(i,j),$$

where

$$g_{\boldsymbol{\ell},\boldsymbol{\ell}'}(i,j) := \frac{1}{M_n^2 L_n^2} \int_{\mathcal{D}_n(i,j)} \int_{\mathcal{D}_{n,\Delta}(i,j)} \int_T \int_T B_{\ell_1}(\|\boldsymbol{s}_1 - \boldsymbol{s}_2\|) B_{\ell_1'}(\|\boldsymbol{s}_1 - \boldsymbol{s}_2\|) B_{\ell_2}(t_1) B_{\ell_2'}(t_1) \\ \times B_{\ell_3}(t_2) B_{\ell_3'}(t_2) I(\|\boldsymbol{s}_1 - \boldsymbol{s}_2\| \leq \Delta) \mathcal{N}_{s,2}(d\boldsymbol{s}_1, d\boldsymbol{s}_2) \mathcal{N}_t(dt_1|\boldsymbol{s}_1) \mathcal{N}_t(dt_2|\boldsymbol{s}_2).$$

Since $|\mathcal{D}_n(i,j)| \asymp |\mathcal{D}_{n,\Delta}(i,j)| \asymp 1$, by moment calculations similar to Lemma 3,

$$\mathrm{E}\left\{g_{\boldsymbol{\ell},\boldsymbol{\ell}'}^2(i,j)\right\} \lesssim \frac{1}{M_n^2 K_t^2 K_s} + \frac{1}{K_t^4 K_s}, \text{ and } \left\{\mathrm{E}\, g_{\boldsymbol{\ell},\boldsymbol{\ell}'}(i,j)\right\}^2 \lesssim \frac{1}{K_t^4 K_s^2}.$$

Denote $\zeta_{\boldsymbol{\ell},\boldsymbol{\ell}'}(i,j) = g_{\boldsymbol{\ell},\boldsymbol{\ell}'}(i,j) - \mathrm{E}\{g_{\boldsymbol{\ell},\boldsymbol{\ell}'}(i,j)\}$, then

$$\mathrm{E}\left\{\zeta_{\boldsymbol{\ell},\boldsymbol{\ell}'}^2(i,j)\right\} = \mathrm{E}\left\{g_{\boldsymbol{\ell},\boldsymbol{\ell}'}^2(i,j)\right\} - \left\{\mathrm{E}\, g_{\boldsymbol{\ell},\boldsymbol{\ell}'}(i,j)\right\}^2 \lesssim \frac{1}{K_t^4 K_s} + \frac{1}{M_n^2 L_n^2 K_t^2 K_s}.$$

Similar calculations on higher moments shows that for $k \geq 3$,

$$\mathrm{E}\left|g_{\boldsymbol{\ell},\boldsymbol{\ell}'}(i,j)\right|^k \lesssim \frac{1}{K_t^{2k} K_s} + \frac{1}{(M_n L_n)^{2k-2} K_t^2 K_s}$$

and

$$\mathrm{E}\left|\zeta_{\boldsymbol{\ell},\boldsymbol{\ell}'}(i,j)\right|^k = \mathrm{E}\left|g_{\boldsymbol{\ell},\boldsymbol{\ell}'}(i,j) - \mathrm{E}\left\{g_{\boldsymbol{\ell},\boldsymbol{\ell}'}(i,j)\right\}\right|^k \leq 2^{k-1}\left[\mathrm{E}\left|g_{\boldsymbol{\ell},\boldsymbol{\ell}'}(i,j)\right|^k + \left|\mathrm{E}\left\{g_{\boldsymbol{\ell},\boldsymbol{\ell}'}(i,j)\right\}\right|^k\right].$$

Hence, there exists a constant $C^* > 0$ such that

$$\mathrm{E}\left|\zeta_{\boldsymbol{\ell},\boldsymbol{\ell}'}(i,j)\right|^k \leq C^* 2^{k-1} k! \mathrm{E}\left\{\zeta^2_{\boldsymbol{\ell},\boldsymbol{\ell}'}(i,j)\right\},$$

and therefore the Cramér's condition is satisfied. By the Bernstein's inequality under the case $k = 3$ (see Bosq (2012), Theorem 1.4, page 29), we have, for any $q > 0$,

$$P\left[\left|\frac{1}{n_1}\sum_{i=1}^{n_1}\zeta_{\boldsymbol{\ell},\boldsymbol{\ell}'}(i,j)\right| > \varepsilon_n\right] \leq a_1 \exp\left(-\frac{q\varepsilon_n^2}{25m_2^2 + 5C^*\varepsilon_n}\right) + a_2\alpha_{\mathcal{N}}\left(\left\lfloor\frac{n_1}{q+1}\right\rfloor\right)^{6/7},$$

where $C^*$ is the Cramér's constant, $\alpha_{\mathcal{N}}(\cdot)$ is the $\alpha$-mixing coefficient for the point process defined in Assumption 4,

$$\varepsilon_n = \frac{\log(n)}{\sqrt{K_s K_t^2 n}} \cdot \varepsilon, \ a_1 = \frac{2n_1}{q} + 2\left(1 + \frac{\varepsilon_n^2}{25m_2^2 + 5C^*\varepsilon_n}\right), \ m_2^2 = \max_{1 \leq i \leq n_1} \mathrm{E}\left\{\zeta^2_{\boldsymbol{\ell},\boldsymbol{\ell}'}(i,j)\right\},$$

$$a_2 = 11n_1\left(1 + \frac{5m_3^{6/7}}{\varepsilon_n}\right), \text{ and } m_3 = \max_{1 \leq i \leq n_1}\left\{\mathrm{E}\left|\zeta_{\boldsymbol{\ell},\boldsymbol{\ell}'}(i,j)\right|^3\right\}^{1/3}.$$

By the moment calculations above for $\zeta_{\boldsymbol{\ell},\boldsymbol{\ell}'}(i,j)$, $m_2^2 \lesssim \frac{1}{K_t^4 K_s} + \frac{1}{M_n^2 L_n^2 K_t^2 K_s}$, $m_3 \lesssim \left(\frac{1}{K_s K_t^2}\right)^{1/3}$. By taking $q$ such that

$$\left\lfloor\frac{n_1}{q+1}\right\rfloor \geq C_1\log(n_1) \quad \text{and} \quad q \geq \frac{C_2 n_1}{\log(n_1)}$$

for some constants $C_1, C_2 > 0$, one has $a_1 \lesssim \log(n)$, $a_2 \lesssim \frac{n^{3/14}n_1}{\log^{3/7}(n)}$ via Assumption 6. The Assumption 4 yields that for some constant $C_3$,

$$\alpha_{\mathcal{N}}\left(\left\lfloor\frac{n_1}{q+1}\right\rfloor\right)^{6/7} \leq C_3\left\{\exp(-\delta_2\sqrt{C_1\log(n_1)})\right\}^{6/7} \leq C_3 n_1^{-6\delta_2\sqrt{C_1}/7}.$$

Thus, when $C_1$, $\varepsilon$, and $n$ are large enough,

$$P\left[\left|\frac{1}{n_1}\sum_{i=1}^{n_1}\zeta_{\boldsymbol{\ell},\boldsymbol{\ell}'}(i,j)\right| > \varepsilon_n\right] \leq C_4\log^2(n_1)\exp(-C_5\varepsilon^2\log(n_1)) + C_3 n_1^{3/7 - 6\delta_2\sqrt{C_1}/7} \leq n^{-10}.$$

A known property of B-splines of order $p$ is that they are non-zero only between $p$ adjacent

knots, and hence $B_\ell(x)B_{\ell'}(x) = 0$ if $|\ell - \ell'| > p$. As a result, $g_{\ell_1\ell_2\ell_3,\ell'_1\ell'_2\ell'_3} = 0$ for $|\ell_1 - \ell'_1| > p_s$ or $|\ell_2 - \ell'_2| > p_t$ or $|\ell_3 - \ell'_3| > p_t$. For any $\varepsilon > 0$,

$$P\left\{\|\boldsymbol{G}_n - \boldsymbol{G}\|_{\max} > \frac{\log(n)}{\sqrt{K_s K_t^2 n}} \cdot \varepsilon\right\}$$

$$\leq \sum_{|\ell_1 - \ell'_1| \leq p_s} \sum_{|\ell_2 - \ell'_2| \leq p_t} \sum_{|\ell_3 - \ell'_3| \leq p_t} P\left\{|g_{\boldsymbol{\ell},\boldsymbol{\ell}'} - \mathrm{E}(g_{\boldsymbol{\ell},\boldsymbol{\ell}'})| > \frac{\log(n)}{\sqrt{K_s K_t^2 n}} \cdot \varepsilon\right\}$$

$$\leq \sum_{|\ell_1 - \ell'_1| \leq p_s} \sum_{|\ell_2 - \ell'_2| \leq p_t} \sum_{|\ell_3 - \ell'_3| \leq p_t} P\left[\left|\frac{1}{|\mathcal{D}_n|}\sum_{i=1}^{n_1}\sum_{j=1}^{n_2} \zeta_{\boldsymbol{\ell},\boldsymbol{\ell}'}(i,j)\right| > \varepsilon_n\right]$$

$$\leq \sum_{|\ell_1 - \ell'_1| \leq p_s} \sum_{|\ell_2 - \ell'_2| \leq p_t} \sum_{|\ell_3 - \ell'_3| \leq p_t} \sum_{j=1}^{n_2} P\left[\frac{1}{|\mathcal{D}_n|/n_2}\left|\sum_{i=1}^{n_1} \zeta_{\boldsymbol{\ell},\boldsymbol{\ell}'}(i,j)\right| > \varepsilon_n\right]$$

which implies that

$$\sum_{n=1}^{\infty} P\left\{\|\boldsymbol{G}_n - \boldsymbol{G}\|_{\max} > \frac{\log(n)}{\sqrt{K_s K_t^2 n}} \cdot \varepsilon\right\} \leq C_6 \sum_{n=1}^{\infty} K_s K_t^2 n_2 \times n^{-10} \leq C_7 \sum_{n=1}^{\infty} \log^2(n) \times n^{-8} < \infty.$$

Thus, Borel-Cantelli Lemma entails that, with probability 1,

$$\|\boldsymbol{G}_n - \boldsymbol{G}\|_{\max} = O\left\{\frac{\log(n)}{\sqrt{K_s K_t^2 |\mathcal{D}_n|}}\right\}.$$

□

<u>Lemma</u> 5 *Let $\lambda_{\min}(\cdot)$ and $\lambda_{\max}(\cdot)$ be the operators returning the minimal and maximal eigenvalues of a matrix. Under Assumptions 1–6, there exist two positive constants $C_1$ and $C_2$, such that, as $n \to \infty$,*

$$\frac{C_1}{K_s K_t^2} \leq \lambda_{\min}(\boldsymbol{G}_n) \leq \lambda_{\max}(\boldsymbol{G}_n) \leq \frac{C_2}{K_s K_t^2} \quad \text{with probability 1.} \tag{S.13}$$

**Proof of Lemma 5:** Following the index convention of Lemma 2, let $\boldsymbol{\theta} = (\theta_{j_1 j_2 j_3}) \in \mathbb{R}^{(K_s+p_s)(K_t+p_t)^2}$ be a vector of coefficients for the tensor product spline basis, then

$$\mathrm{E}(\boldsymbol{\theta}^T \boldsymbol{G}_n \boldsymbol{\theta}) = \frac{1}{|\mathcal{D}_n| M_n^2 L_n^2}\mathrm{E}\int_{\mathcal{D}_n^{\otimes 2}}\int_{T^{\otimes 2}}\left\{\sum_{j_1 j_2 j_3}\theta_{j_1 j_2 j_3}B_{j_1}(\|\boldsymbol{s}_1 - \boldsymbol{s}_2\|)B_{j_2}(t_1)B_{j_3}(t_2)\right\}^2$$
$$\times I(\|\boldsymbol{s}_1 - \boldsymbol{s}_2\| \leq \Delta)\mathcal{N}_{s,2}(d\boldsymbol{s}_1, d\boldsymbol{s}_2)\mathcal{N}_{t,2}(dt_1, dt_2)$$
$$= \frac{1}{|\mathcal{D}_n| M_n^2 L_n^2}\int_{\mathcal{D}_n^{\otimes 2}}\int_{T^{\otimes 2}}\left\{\sum_{j_1 j_2 j_3}\theta_{j_1 j_2 j_3}B_{j_1}(\|\boldsymbol{s}_1 - \boldsymbol{s}_2\|)B_{j_2}(t_1)B_{j_3}(t_2)\right\}^2$$

45

$$\times I(\|\boldsymbol{s}_1 - \boldsymbol{s}_2\| \leq \Delta)\lambda_{s,2}(\boldsymbol{s}_1, \boldsymbol{s}_2)\lambda_{t,1}(t_1)\lambda_{t,1}(t_2) d\boldsymbol{s}_1 d\boldsymbol{s}_2 dt_1 dt_2$$

$$\lesssim \frac{1}{|\mathcal{D}_n|K_t^2} \int_{\mathcal{D}_n^{\otimes 2}} \sum_{j_2 j_3} \left\{\sum_{j_1} \theta_{j_1 j_2 j_3} B_{j_1}(\|\boldsymbol{s}_1 - \boldsymbol{s}_2\|)\right\}^2 I(\|\boldsymbol{s}_1 - \boldsymbol{s}_2\| \leq \Delta) d\boldsymbol{s}_1 d\boldsymbol{s}_2$$

$$= \frac{1}{|\mathcal{D}_n|K_t^2} \int_{\mathcal{D}_n} \int_{[0,\Delta]} \sum_{j_2 j_3} \left\{\sum_{j_1} \theta_{j_1 j_2 j_3} B_{j_1}(\|\boldsymbol{u}\|)\right\}^2 d\boldsymbol{u} d\boldsymbol{s}_2$$

$$\lesssim \frac{1}{|\mathcal{D}_n| K_s K_t^2} \int_{\mathcal{D}_n} d\boldsymbol{s}_2 \sum_{j_1 j_2 j_3} \theta_{j_1 j_2 j_3}^2 \lesssim \frac{1}{K_s K_t^2} \|\boldsymbol{\theta}\|^2.$$

Following the same argument, the lower bound of $\mathrm{E}\left(\boldsymbol{\theta}^T \boldsymbol{G}_n \boldsymbol{\theta}\right)$ can also be proved. Thus, there exist two positive constants $\widetilde{C}_1$ and $\widetilde{C}_2$, such that,

$$\frac{\widetilde{C}_1}{K_s K_t^2} \leq \frac{\mathrm{E}\left(\boldsymbol{\theta}^T \boldsymbol{G}_n \boldsymbol{\theta}\right)}{\|\boldsymbol{\theta}\|^2} \leq \frac{\widetilde{C}_2}{K_s K_t^2}.$$

Lemma 4 and Assumption 6 entails that, with probability 1,

$$\|\boldsymbol{G}_n - \boldsymbol{G}\|_{\max} = O\left\{\frac{\log(n)}{\sqrt{K_s K_t^2 |\mathcal{D}_n|}}\right\} = o\left(\frac{1}{K_s K_t^2}\right).$$

Hence, by the Cauchy-Schwartz Inequality, with probability 1,

$$|\boldsymbol{\theta}^T \boldsymbol{G}_n \boldsymbol{\theta} - \mathrm{E}\left(\boldsymbol{\theta}^T \boldsymbol{G}_n \boldsymbol{\theta}\right)| = |\boldsymbol{\theta}^T (\boldsymbol{G}_n - \boldsymbol{G}) \boldsymbol{\theta}|$$

$$\leq \|\boldsymbol{G}_n - \boldsymbol{G}\|_{\max} \sum_{j_1 j_2 j_3} |\theta_{j_1 j_2 j_3}| \left(\sum_{|j_1 - j_1'| \leq p_s} \sum_{|j_2 - j_2'| \leq p_t} \sum_{|j_3 - j_3'| \leq p_t} |\theta_{j_1' j_2' j_3'}|\right)$$

$$\leq \sqrt{p_s p_t^2} \|\boldsymbol{G}_n - \boldsymbol{G}\|_{\max} \sum_{j_1 j_2 j_3} |\theta_{j_1 j_2 j_3}| \left(\sum_{|j_1 - j_1'| \leq p_s} \sum_{|j_2 - j_2'| \leq p_t} \sum_{|j_3 - j_3'| \leq p_t} |\theta_{j_1' j_2' j_3'}|\right)^{1/2}$$

$$\leq \sqrt{p_s p_t^2} \|\boldsymbol{G}_n - \boldsymbol{G}\|_{\max} \left(\sum_{j_1 j_2 j_3} \theta_{j_1 j_2 j_3}^2\right)^{1/2} \left(\sum_{j_1 j_2 j_3} \sum_{\substack{|j_1 - j_1'| \leq p_s \\ |j_2 - j_2'| \leq p_t \\ |j_3 - j_3'| \leq p_t}} \theta_{j_1' j_2' j_3'}^2\right)^{1/2}$$

$$= o\left(\frac{\|\boldsymbol{\theta}\|^2}{K_s K_t^2}\right),$$

which completes the proof of (S.13). $\square$

46

LEMMA 6 Let $\boldsymbol{j} = (j_1, j_2, j_3)$ and $\boldsymbol{j}' = (j_1', j_2', j_3')$ be two index vectors for tensor product spline functions, $j_1, j_1' \in \{1, \ldots, K_s + p_s\}$, $j_2, j_3, j_2', j_3' \in \{1, \ldots, K_t + p_t\}$. Following the convention in Lemma 4, let $g^*_{j_1 j_2 j_3, j_1' j_2' j_3'}$ be the $(\boldsymbol{j}, \boldsymbol{j}')$th entry of the matrix $\boldsymbol{G}_n^{-1}$. Under Assumptions 1 – 7, there exist constants $C > 0$ and $\tau \in (0, 1)$, such that, when $n$ is large enough,

$$\sup_{\substack{j_1, j_1' \in \{1, \ldots, K_s + p_s\}, \\ j_2, j_3, j_2', j_3' \in \{1, \ldots, K_t + p_t\}}} \left| \frac{g^*_{j_1 j_2 j_3, j_1' j_2' j_3'}}{\tau^{|j_1 - j_1'| + |j_2 - j_2'| + |j_3 - j_3'|}} \right| \leq C \cdot K_s K_t^2 \quad \text{with probability 1.} \tag{S.14}$$

**Proof of Lemma 6:** Here, let spect$(\boldsymbol{G}_n)$ denote the spectrum of $\boldsymbol{G}_n$, i.e., the set of eigenvalues of $\boldsymbol{G}_n$. Let $\mathcal{P}_k$ denote the set of all polynomial functions of degree less than or equal to $k$. Proposition 2.1 in Demko et al. (1984) shows that,

$$\inf_{f_k \in \mathcal{P}_k} \left\{ \sup_{x \in \text{spect}(\boldsymbol{G}_n)} \left| \frac{1}{x} - f_k(x) \right| \right\} \leq \frac{\left\{ \sqrt{\lambda_{\max}(\boldsymbol{G}_n)} + \sqrt{\lambda_{\min}(\boldsymbol{G}_n)} \right\}^2}{2 \lambda_{\max}(\boldsymbol{G}_n) \lambda_{\min}(\boldsymbol{G}_n)} \left\{ \frac{\sqrt{\lambda_{\max}(\boldsymbol{G}_n)} - \sqrt{\lambda_{\min}(\boldsymbol{G}_n)}}{\sqrt{\lambda_{\max}(\boldsymbol{G}_n)} + \sqrt{\lambda_{\min}(\boldsymbol{G}_n)}} \right\}^k. \tag{S.15}$$

By Lemma 5, there exist two constants $C_2 > C_1 > 0$ such that, when $n$ is sufficiently large,

$$\frac{C_1}{K_s K_t^2} \leq \lambda_{\min}(\boldsymbol{G}_n) \leq \lambda_{\max}(\boldsymbol{G}_n) \leq \frac{C_2}{K_s K_t^2} \quad \text{with probability 1.}$$

Let $\tau = \frac{\sqrt{C_2} - \sqrt{C_1}}{\sqrt{C_2} + \sqrt{C_1}} \in (0, 1)$, and $C_3 = \frac{(\sqrt{C_2} + \sqrt{C_1})^2}{2 C_2 C_1}$. It follows that, when $n$ is sufficiently large,

$$\inf_{f_k \in \mathcal{P}_k} \left\{ \sup_{x \in \text{spect}(\boldsymbol{G}_n)} \left| \frac{1}{x} - f_k(x) \right| \right\} \leq C_3 \tau^k K_s K_t^2 \quad \text{with probability 1.} \tag{S.16}$$

An application of the spectral theory (Rudin, 1991) yields that, for any $f_k \in \mathcal{P}_k$,

$$\|\boldsymbol{G}_n^{-1} - f_k(\boldsymbol{G}_n)\|_{\max} = \sup_{x \in \text{spect}(\boldsymbol{G}_n)} \left| \frac{1}{x} - f_k(x) \right|.$$

Note that $\boldsymbol{G}_n$ is a multiband matrix of multiwidth $(2(p_s + 1), 2(p_t + 1), 2(p_t + 1))$ (refer to Mastronardi et al. (2010) for the definitions of a multiband matrix and multiwidth). For $\boldsymbol{j} = (j_1, j_2, j_3)^{\mathrm{T}}$ and $\boldsymbol{j}' = (j_1', j_2', j_3')^{\mathrm{T}}$, let $k^\dagger = \max\left( \left\lfloor \frac{|j_1 - j_1'|}{p_s + 1} \right\rfloor, \left\lfloor \frac{|j_2 - j_2'|}{p_t + 1} \right\rfloor, \left\lfloor \frac{|j_3 - j_3'|}{p_t + 1} \right\rfloor \right)$, where $\lfloor x \rfloor$ is the floor of $x$. For any $f_{k^\dagger} \in \mathcal{P}_{k^\dagger}$, we write $f_{k^\dagger}(\boldsymbol{G}_n) = \left( g^\dagger_{j_1 j_2 j_3, j_1' j_2' j_3'} \right)$, where $g^\dagger_{j_1 j_2 j_3, j_1' j_2' j_3'}$ is the $(\boldsymbol{j}, \boldsymbol{j}')$th entry of $f_{k^\dagger}(\boldsymbol{G}_n)$.

47

- If $k^\dagger \geq 1$, then $g^\dagger_{j_1 j_2 j_3, j'_1 j'_2 j'_3} = 0$ for any $f_{k^\dagger} \in \mathcal{P}_{k^\dagger}$,

$$\begin{aligned}
\left| g^*_{j_1 j_2 j_3, j'_1 j'_2 j'_3} \right| &= \left| g^*_{j_1 j_2 j_3, j'_1 j'_2 j'_3} - g^\dagger_{j_1 j_2 j_3, j'_1 j'_2 j'_3} \right| \\
&= \inf_{f_{k^\dagger} \in \mathcal{P}_{k^\dagger}} \left| g^*_{j_1 j_2 j_3, j'_1 j'_2 j'_3} - g^\dagger_{j_1 j_2 j_3, j'_1 j'_2 j'_3} \right| \\
&\leq \inf_{f_{k^\dagger} \in \mathcal{P}_{k^\dagger}} \left\| \boldsymbol{G}_n^{-1} - f_{k^\dagger}(\boldsymbol{G}_n) \right\|_{\max} \\
&= \inf_{f_{k^\dagger} \in \mathcal{P}_{k^\dagger}} \left\{ \sup_{x \in \mathrm{spect}(\boldsymbol{G}_n)} \left| \frac{1}{x} - f_{k^\dagger}(x) \right| \right\}.
\end{aligned} \quad \text{(S.17)}$$

- If $k^\dagger = 0$, let $f_{k^\dagger} \equiv 0 \in \mathcal{P}_{k^\dagger}$, then

$$\left| g^*_{j_1 j_2 j_3, j'_1 j'_2 j'_3} \right| \leq \left\| \boldsymbol{G}_n^{-1} \right\|_{\max} = \left\| \boldsymbol{G}_n^{-1} - f_{k^\dagger}(\boldsymbol{G}_n) \right\|_{\max} = \sup_{x \in \mathrm{spect}(\boldsymbol{G}_n)} \left| \frac{1}{x} \right| = \frac{1}{\lambda_{\min}(\boldsymbol{G}_n)}. \quad \text{(S.18)}$$

Thus, by (S.16), (S.17), (S.18), and Lemma 5, with probability 1, when $n$ sufficiently large,

$$\sup_{j_1 j_2 j_3, j'_1 j'_2 j'_3} \left| \frac{g^*_{j_1 j_2 j_3, j'_1 j'_2 j'_3}}{\tau^{k^\dagger}} \right| \leq C_3 \cdot K_s K_t^2.$$

Since $k^\dagger = \max\left( \left\lfloor \frac{|j_1 - j'_1|}{p_s + 1} \right\rfloor, \left\lfloor \frac{|j_2 - j'_2|}{p_t + 1} \right\rfloor, \left\lfloor \frac{|j_3 - j'_3|}{p_t + 1} \right\rfloor \right) \geq \left\lfloor \frac{|j_1 - j'_1| + |j_2 - j'_2| + |j_3 - j'_3|}{3(p_s + p_t + 1)} \right\rfloor$, we conclude that, when $n$ is sufficiently large,

$$\sup_{j_1 j_2 j_3, j'_1 j'_2 j'_3} \left| \frac{g^*_{j_1 j_2 j_3, j'_1 j'_2 j'_3}}{\tau^{\left\lfloor \frac{|j_1 - j'_1| + |j_2 - j'_2| + |j_3 - j'_3|}{3(p_s + p_t + 1)} \right\rfloor}} \right| \leq C_3 \cdot K_s K_t^2 \quad \text{with probability 1,}$$

which completes the proof of (S.14). □

<u>REMARK</u> 1 *Lemma 6 implies that entries of $\boldsymbol{G}_n^{-1}$ decay exponentially. Similar results on the inverse of band matrices have been used to establish asymptotic properties of spline estimators in independent data (Demko et al., 1984). However, due to the random design under the geostatistics setting in this paper, $\boldsymbol{G}_n$ can not be exactly written as the Kronecker product of band matrices, hence Theorem 2.2 of Demko (1977) can not be directly applied. Our proof of Lemma 6 utilizes properties of multi-band matrices and advanced results from spectral theory and approximation theory (Mastronardi et al., 2010; Demko et al., 1984).*

## S.3 Proofs of the Main Theorems

### S.3.1 Proof of Theorem 4.1

By Lemma 1, there exists an $R^* \in \mathcal{S}_{[3]}$ such that $\|R - R^*\|_\infty = O(K_s^{-p_s} + K_t^{-p_t})$, as $K_s, K_t \to \infty$. Hence there exists a vector $\boldsymbol{\beta}^* \in \mathbb{R}^{(K_s+p_s)(K_t+p_t)^2}$ such that $R^*(u, t_1, t_2) = \boldsymbol{B}_{[3]}^{\mathrm{T}}(u, t_1, t_2)\boldsymbol{\beta}^*$.
Then, $\left\|\widehat{R} - R\right\|_{L^2} \leq \left\|\widehat{R} - R^*\right\|_{L^2} + \|R^* - R\|_{L^2} \leq \left\|\boldsymbol{B}_{[3]}^{\mathrm{T}}\left(\widehat{\boldsymbol{\beta}} - \boldsymbol{\beta}^*\right)\right\|_{L^2} + C_1\|R - R^*\|_\infty$, for some $C_1 > 0$. We write $\left(\widehat{\boldsymbol{\beta}} - \boldsymbol{\beta}^*\right)$ as a sum of two parts: $\left(\widehat{\boldsymbol{\beta}} - \boldsymbol{\beta}^*\right) = \boldsymbol{G}_n^{-1}\boldsymbol{\xi}_n + \boldsymbol{G}_n^{-1}\boldsymbol{\eta}_n$, where $\boldsymbol{G}_n$ and $\boldsymbol{\xi}_n$ are defined in Lemmas 4 and 3 respectively, and

$$\boldsymbol{\eta}_n := \frac{1}{|\mathcal{D}_n|M_n^2 L_n^2} \int_{\mathcal{D}_n^{\otimes 2}} \int_T \int_T \boldsymbol{B}_{[3]}(\|\boldsymbol{s}_1 - \boldsymbol{s}_2\|, t_1, t_2)\{R(\|\boldsymbol{s}_1 - \boldsymbol{s}_2\|, t_1, t_2) - R^*(\|\boldsymbol{s}_1 - \boldsymbol{s}_2\|, t_1, t_2)\}$$
$$\times I\left(\|\boldsymbol{s}_1 - \boldsymbol{s}_2\| \leq \Delta\right) \mathcal{N}_{s,2}(d\boldsymbol{s}_1, d\boldsymbol{s}_2) \mathcal{N}_t(dt_1|\boldsymbol{s}_1)\mathcal{N}_t(dt_2|\boldsymbol{s}_2).$$

By (S.1) in Lemma 2,

$$\left\|\boldsymbol{B}_{[3]}^{\mathrm{T}}\boldsymbol{G}_n^{-1}\boldsymbol{\xi}_n\right\|_{L^2}^2 = \int_T \int_T \int_{[0,\Delta]} \left\{\boldsymbol{B}_{[3]}^{\mathrm{T}}(u, t_1, t_2)\boldsymbol{G}_n^{-1}\boldsymbol{\xi}_n\right\}^2 du\, dt_1\, dt_2$$
$$\leq \frac{C_2}{K_s K_t^2}\|\boldsymbol{G}_n^{-1}\boldsymbol{\xi}_n\|_2^2 \leq \frac{C_2}{K_s K_t^2}\left\{\lambda_{\min}(\boldsymbol{G}_n)\right\}^{-2}\|\boldsymbol{\xi}_n\|_2^2.$$

Lemma 5 shows that $\lambda_{\min}(\boldsymbol{G}_n) \asymp \left(\frac{1}{K_s K_t^2}\right)$ and Lemma 3 suggests that $\|\boldsymbol{\xi}_n\|_2^2 = O_p\left(\frac{1}{|\mathcal{D}_n|K_t^2} + \frac{1}{|\mathcal{D}_n|M_n L_n K_t} + \frac{1}{|\mathcal{D}_n|M_n^2 L_n^2}\right)$, we therefore conclude that,

$$\left\|\boldsymbol{B}_{[3]}^{\mathrm{T}}\boldsymbol{G}_n^{-1}\boldsymbol{\xi}_n\right\|_{L^2} = O_p\left(\sqrt{\frac{K_s}{|\mathcal{D}_n|}} + \sqrt{\frac{K_s K_t}{|\mathcal{D}_n|M_n L_n}} + \sqrt{\frac{K_s K_t^2}{|\mathcal{D}_n|M_n^2 L_n^2}}\right).$$

By similar calculations,

$$\left\|\boldsymbol{B}_{[3]}^{\mathrm{T}}\boldsymbol{G}_n^{-1}\boldsymbol{\eta}_n\right\|_{L^2}^2 = \int_T \int_T \int_{[0,\Delta]} \left\{\boldsymbol{B}_{[3]}^{\mathrm{T}}(u, t_1, t_2)\boldsymbol{G}_n^{-1}\boldsymbol{\eta}_n\right\}^2 du\, dt_1\, dt_2$$
$$\leq \frac{C_3}{K_s K_t^2}\|\boldsymbol{G}_n^{-1}\boldsymbol{\eta}_n\|_2^2 \leq \frac{C_3}{K_s K_t^2}\left\{\lambda_{\min}(\boldsymbol{G}_n)\right\}^{-2}\|\boldsymbol{\eta}_n\|_2^2.$$

We derive the upper bound of $\|\boldsymbol{\eta}_n\|_2^2$ as follows,

$$\|\boldsymbol{\eta}_n\|_2^2 \leq \frac{1}{|\mathcal{D}_n|^2 M_n^4 L_n^4}\|R - R^*\|_\infty^2 \cdot \left\|\int_{\mathcal{D}_n^{\otimes 2}} \int_T \int_T \boldsymbol{B}_{[3]}(\|\boldsymbol{s}_1 - \boldsymbol{s}_2\|, t_1, t_2) I\left(\|\boldsymbol{s}_1 - \boldsymbol{s}_2\| \leq \Delta\right)\right.$$
$$\left. \mathcal{N}_{s,2}(d\boldsymbol{s}_1, d\boldsymbol{s}_2)\mathcal{N}_t(dt_1|\boldsymbol{s}_1)\mathcal{N}_t(dt_2|\boldsymbol{s}_2)\right\|_2^2$$

$$\leq \frac{C_4}{|\mathcal{D}_n|^2 M_n^4 L_n^4} \cdot \|R - R^*\|_\infty^2 \cdot \frac{|\mathcal{D}_n|^2 M_n^4 L_n^4}{K_s K_t^2} \asymp \frac{1}{K_s K_t^2} \|R - R^*\|_\infty^2.$$

We conclude that
$$\left\| \boldsymbol{B}_{[3]}^{\mathrm{T}} \boldsymbol{G}_n^{-1} \boldsymbol{\eta}_n \right\|_{L^2} = o_p\left(K_s^{-p_s} + K_t^{-p_t}\right).$$

Hence,
$$\begin{aligned}
\|\widehat{R} - R\|_{L^2} &\leq \left\| \boldsymbol{B}_{[3]}^{\mathrm{T}} \boldsymbol{G}_n^{-1} \boldsymbol{\xi}_n \right\|_{L^2} + \left\| \boldsymbol{B}_{[3]}^{\mathrm{T}} \boldsymbol{G}_n^{-1} \boldsymbol{\eta}_n \right\|_{L^2} + C_1 \|R - R^*\|_\infty \\
&= O_p\left(\sqrt{\frac{K_s}{|\mathcal{D}_n|}} + \sqrt{\frac{K_s K_t}{|\mathcal{D}_n| M_n L_n}} + \sqrt{\frac{K_s K_t^2}{|\mathcal{D}_n| M_n^2 L_n^2}} + K_s^{-p_s} + K_t^{-p_t}\right).
\end{aligned}$$

□

### S.3.2  Proof of Theorem 4.2

Analogous to the proof of Theorem 4.1, we bound $\|\widehat{\Omega} - \Omega\|_{L^2}$ by three terms,

$$\begin{aligned}
&\|\widehat{\Omega} - \Omega\|_{L^2} \\
&= \left\| \int_{[0,\Delta]} \left\{ \widehat{R}(u,\cdot,\cdot) - R(u,\cdot,\cdot) \right\} \mathcal{W}(u) du \right\|_{L^2} \\
&\leq \left\| \int_{[0,\Delta]} \boldsymbol{B}_{[3]}^{\mathrm{T}}(u,\cdot,\cdot) \mathcal{W}(u) du \left(\widehat{\boldsymbol{\beta}} - \boldsymbol{\beta}^*\right) \right\|_{L^2} + C_1 \|R - R^*\|_\infty \\
&\leq \left\| \int_{[0,\Delta]} \boldsymbol{B}_{[3]}^{\mathrm{T}}(u,\cdot,\cdot) \mathcal{W}(u) du \cdot \boldsymbol{G}_n^{-1} \boldsymbol{\xi}_n \right\|_{L^2} + \left\| \int_{[0,\Delta]} \boldsymbol{B}_{[3]}^{\mathrm{T}}(u,\cdot,\cdot) \mathcal{W}(u) du \cdot \boldsymbol{G}_n^{-1} \boldsymbol{\eta}_n \right\|_{L^2} \\
&\quad + C_1 \|R - R^*\|_\infty.
\end{aligned}$$

Applying (S.2) in Lemma 2, we have

$$\begin{aligned}
&\left\| \int_{[0,\Delta]} \boldsymbol{B}_{[3]}^{\mathrm{T}}(u,\cdot,\cdot) \mathcal{W}(u) du \cdot \boldsymbol{G}_n^{-1} \boldsymbol{\xi}_n \right\|_{L^2}^2 \\
&= \int_{T^2} \left( \int_{[0,\Delta]} \boldsymbol{B}_{[3]}^{\mathrm{T}}(u,t_1,t_2) \mathcal{W}(u) du \cdot \boldsymbol{G}_n^{-1} \boldsymbol{\xi}_n \right)^2 dt_1 dt_2 \\
&\leq \frac{C_2}{K_t^2} \sum_{i_2,i_3} \left\{ \sum_{i_1} \left( \sum_{i_1' i_2' i_3'} g^*_{i_1 i_2 i_3, i_1' i_2' i_3'} \cdot \xi_{i_1' i_2' i_3'} \right) \cdot \int_0^\Delta B_{i_1}(u) \mathcal{W}(u) du \right\}^2 \\
&= \frac{C_2}{K_t^2} \sum_{i_2,i_3} \sum_{i_1,j_1} \sum_{i_1' i_2' i_3'} \sum_{j_1' j_2' j_3'} g^*_{i_1 i_2 i_3, i_1' i_2' i_3'} g^*_{j_1 i_2 i_3, j_1' j_2' j_3'} \xi_{i_1' i_2' i_3'} \xi_{j_1' j_2' j_3'} \int_0^\Delta B_{i_1}(u) \mathcal{W}(u) du \int_0^\Delta B_{j_1}(u) \mathcal{W}(u) du.
\end{aligned}$$

50

By taking the conditional expectation $\mathrm{E}(\cdot|\mathcal{G})$, we have

$$\mathrm{E}\left\{\left\|\int_{[0,\Delta]}\boldsymbol{B}_{[3]}^{\mathrm{T}}(u,\cdot,\cdot)\mathcal{W}(u)du\cdot\boldsymbol{G}_n^{-1}\boldsymbol{\xi}_n\right\|_{L^2}^2\Big|\mathcal{G}\right\}$$

$$\leq \frac{C_2}{K_t^2}\sum_{i_2,i_3}\sum_{i_1,j_1}\sum_{i'_1 i'_2 i'_3}\sum_{j'_1 j'_2 j'_3} g^*_{i_1 i_2 i_3, i'_1 i'_2 i'_3} g^*_{j_1 i_2 i_3, j'_1 j'_2 j'_3} \cdot \mathrm{E}\left(\xi_{i'_1 i'_2 i'_3}\xi_{j'_1 j'_2 j'_3}\big|\mathcal{G}\right)$$
$$\times \int_0^\Delta B_{i_1}(u)\mathcal{W}(u)du \int_0^\Delta B_{j_1}(u)\mathcal{W}(u)du$$

$$\lesssim \frac{1}{K_t^2 K_s^2}\sum_{i_2,i_3}\sum_{i_1,j_1}\sum_{i'_1 i'_2 i'_3}\sum_{j'_1 j'_2 j'_3} \left|g^*_{i_1 i_2 i_3, i'_1 i'_2 i'_3}\right|\left|g^*_{j_1 i_2 i_3, j'_1 j'_2 j'_3}\right| \cdot \left|\mathrm{E}\left(\xi_{i'_1 i'_2 i'_3}\xi_{j'_1 j'_2 j'_3}\big|\mathcal{G}\right)\right|.$$

By Lemma 6, when $n$ is sufficiently large, with probability 1,

$$\mathrm{E}\left\{\left\|\int_{[0,\Delta]}\boldsymbol{B}_{[3]}^{\mathrm{T}}(u,\cdot,\cdot)\mathcal{W}(u)du\cdot\boldsymbol{G}_n^{-1}\boldsymbol{\xi}_n\right\|_{L^2}^2\Big|\mathcal{G}\right\}$$

$$\lesssim K_t^2\sum_{i_2,i_3}\sum_{i_1,j_1}\sum_{i'_1 i'_2 i'_3}\sum_{j'_1 j'_2 j'_3} \tau^{|i_1-i'_1|+|j_1-j'_1|}\tau^{|i_2-i'_2|+|i_2-j'_2|}\tau^{|i_3-i'_3|+|i_3-j'_3|} \cdot \left|\mathrm{E}\left(\xi_{i'_1 i'_2 i'_3}\xi_{j'_1 j'_2 j'_3}\big|\mathcal{G}\right)\right|.$$

By Lemma 3, there exists a constant $C_2 > 0$ such that,

- for $|i_1-i'_1|\leq p_s$, $\mathrm{E}\left\{\left|\mathrm{E}\left(\xi_{i'_1 i'_2 i'_3}\xi_{j'_1 j'_2 j'_3}\big|\mathcal{G}\right)\right|\right\} \leq C_2 \left(\frac{1}{|\mathcal{D}_n|K_s K_t^4} + \frac{1}{|\mathcal{D}_n|M_n L_n K_s K_t^3} + \frac{1}{|\mathcal{D}_n|M_n^2 L_n^2 K_s K_t^2}\right)$;

- for $|i_1-i'_1|>p_s$, $\mathrm{E}\left\{\left|\mathrm{E}\left(\xi_{i'_1 i'_2 i'_3}\xi_{j'_1 j'_2 j'_3}\big|\mathcal{G}\right)\right|\right\} \leq C_2 \left(\frac{1}{|\mathcal{D}_n|K_s^2 K_t^4} + \frac{1}{|\mathcal{D}_n|M_n L_n K_s^2 K_t^3}\right).$

Thus, as $n \to \infty$,

$$\mathrm{E}\left\{\left\|\int_{[0,\Delta]}\boldsymbol{B}_{[3]}^{\mathrm{T}}(u,\cdot,\cdot)\mathcal{W}(u)du\cdot\boldsymbol{G}_n^{-1}\boldsymbol{\xi}_n\right\|_{L^2}^2\right\}$$

$$\lesssim K_t^2\sum_{i_2,i_3}\sum_{i_1,j_1}\sum_{i'_1 i'_2 i'_3}\sum_{j'_1 j'_2 j'_3} \tau^{|i_1-i'_1|+|j_1-j'_1|}\tau^{|i_2-i'_2|+|i_2-j'_2|}\tau^{|i_3-i'_3|+|i_3-j'_3|} \cdot \mathrm{E}\left\{\left|\mathrm{E}\left(\xi_{i'_1 i'_2 i'_3}\xi_{j'_1 j'_2 j'_3}\big|\mathcal{G}\right)\right|\right\}$$

$$= K_t^2\sum_{i_2,i_3}\sum_{i_1,j_1}\sum_{i'_1 i'_2 i'_3}\sum_{j'_1 j'_2 j'_3} \tau^{|i_1-i'_1|+|j_1-j'_1|}\tau^{|i_2-i'_2|+|i_2-j'_2|}\tau^{|i_3-i'_3|+|i_3-j'_3|} \cdot \mathrm{E}\left\{\left|\mathrm{E}\left(\xi_{i'_1 i'_2 i'_3}\xi_{j'_1 j'_2 j'_3}\big|\mathcal{G}\right)\right|\right\}$$
$$\times \{I(|i'_1-j'_1|\leq p_s) + I(|i'_1-j'_1|> p_s)\}$$

$$\lesssim K_s K_t^4 \left(\frac{1}{|\mathcal{D}_n|K_s K_t^4} + \frac{1}{|\mathcal{D}_n|M_n L_n K_s K_t^3} + \frac{1}{|\mathcal{D}_n|M_n^2 L_n^2 K_s K_t^2}\right)$$
$$+ K_s^2 K_t^4 \left(\frac{1}{|\mathcal{D}_n|K_s^2 K_t^4} + \frac{1}{|\mathcal{D}_n|M_n L_n K_s^2 K_t^3}\right)$$

$$\asymp \frac{1}{|\mathcal{D}_n|K_t} + \frac{K_t}{|\mathcal{D}_n|M_n L_n} + \frac{K_t^2}{|\mathcal{D}_n|M_n^2 L_n^2}.$$

Consequently, as $n \to \infty$,

$$\left\|\int_{[0,\Delta]} \boldsymbol{B}_{[3]}^{\mathrm{T}}(u,\cdot,\cdot)\mathcal{W}(u)du \cdot \boldsymbol{G}_n^{-1}\boldsymbol{\xi}_n\right\|_{L^2} = O_p\left(\sqrt{\frac{1}{|\mathcal{D}_n|}} + \sqrt{\frac{K_t}{|\mathcal{D}_n|M_nL_n}} + \sqrt{\frac{K_t^2}{|\mathcal{D}_n|M_n^2L_n^2}}\right).$$

By Jensen's Inequality,

$$\begin{aligned}
\left\|\int_{[0,\Delta]} \boldsymbol{B}_{[3]}^{\mathrm{T}}(u,\cdot,\cdot)\mathcal{W}(u)du \cdot \boldsymbol{G}_n^{-1}\boldsymbol{\eta}_n\right\|_{L^2}^2 &= \int_{T^2}\left(\int_{[0,\Delta]} \boldsymbol{B}_{[3]}^{\mathrm{T}}(u,t_1,t_2)\mathcal{W}(u)du \cdot \boldsymbol{G}_n^{-1}\boldsymbol{\eta}_n\right)^2 dt_1 dt_2 \\
&\lesssim \left\|\boldsymbol{B}_{[3]}^{\mathrm{T}} \cdot \boldsymbol{G}_n^{-1}\boldsymbol{\eta}_n\right\|_{L^2}^2 \lesssim \|R - R^*\|_\infty^2.
\end{aligned}$$

Hence, we conclude that, as $n \to \infty$,

$$\|\widehat{\Omega} - \Omega\|_{L^2} = O_p\left(\sqrt{\frac{1}{|\mathcal{D}_n|}} + \sqrt{\frac{K_t}{|\mathcal{D}_n|M_nL_n}} + \sqrt{\frac{K_t^2}{|\mathcal{D}_n|M_n^2L_n^2}} + K_s^{-p_s} + K_t^{-p_t}\right).$$

□

### S.3.3 Proof of Theorem 4.3

Since $\psi_j$'s are eigenfunctions of the covariance function $\Omega$, by the asymptotic expansion of Hall and Hosseini-Nasab (2006),

$$\widehat{\psi}_j - \psi_j = \sum_{k \neq j}(\omega_k - \omega_j)^{-1}\left\langle\int\left(\widehat{\Omega} - \Omega\right)(t_1,t_2)\psi_j(t_1)dt_1, \psi_k\right\rangle\psi_k + O_p\left(\|\widehat{\Omega} - \Omega\|_{L^2}^2\right),$$

for any fixed order $j$. By Bessel's inequality, the above expression leads to

$$\left\|\widehat{\psi}_j - \psi_j\right\|_{L^2} \leq C_1 \cdot \left\|\int\left(\widehat{\Omega} - \Omega\right)(t_1,t_2)\psi_j(t_1)dt_1\right\|_{L^2} + O_p\left(\|\widehat{\Omega} - \Omega\|_{L^2}^2\right).$$

The proof of Theorem 4.3 is complete, if we could show

$$\left\|\int\left(\widehat{\Omega} - \Omega\right)(t_1,t_2)\psi_j(t_1)dt_1\right\|_{L^2} = O_p\left(\sqrt{\frac{1}{|\mathcal{D}|_n}} + \sqrt{\frac{K_t}{|\mathcal{D}|_nM_nL_n}} + K_s^{-p_s} + K_t^{-p_t}\right).$$

Analogous to the proof of Theorem 4.2, we bound the integral by three terms

$$\left\|\int\left(\widehat{\Omega} - \Omega\right)(t_1,t_2)\psi_j(t_1)dt_1\right\|_{L^2}$$

$$\leq \left\|\int_T\int_{[0,\Delta]}\left\{\widehat{R}(u,t_1,\cdot) - R(u,t_1,\cdot)\right\}\mathcal{W}(u)\psi_j(t_1)dudt_1\right\|_{L^2}$$

$$\leq \left\| \int_T \int_{[0,\Delta]} \boldsymbol{B}_{[3]}^{\mathrm{T}}(u,t_1,\cdot)\mathcal{W}(u)\psi_j(t_1)dudt_1 \left(\widehat{\boldsymbol{\beta}} - \boldsymbol{\beta}^*\right)\right\|_{L^2} + C_2\|R - R^*\|_\infty$$

$$\leq \left\| \int_T \int_{[0,\Delta]} \boldsymbol{B}_{[3]}^{\mathrm{T}}(u,t_1,\cdot)\mathcal{W}(u)\psi_j(t_1)dudt_1 \cdot \boldsymbol{G}_n^{-1}\boldsymbol{\xi}_n\right\|_{L^2}$$

$$+ \left\| \int_T \int_{[0,\Delta]} \boldsymbol{B}_{[3]}^{\mathrm{T}}(u,t_1,\cdot)\mathcal{W}(u)\psi_j(t_1)dudt_1 \cdot \boldsymbol{G}_n^{-1}\boldsymbol{\eta}_n\right\|_{L^2} + C_2\|R - R^*\|_\infty.$$

Because

$$\left\| \int_T \int_{[0,\Delta]} \boldsymbol{B}_{[3]}^{\mathrm{T}}(u,t_1,\cdot)\mathcal{W}(u)\psi_j(t_1)dudt_1 \cdot \boldsymbol{G}_n^{-1}\boldsymbol{\eta}_n\right\|_{L^2} \leq C_3 \|R - R^*\|_\infty = O_p\left(K_s^{-p_s} + K_t^{-p_t}\right),$$

we only need to show the bound of

$$\left\| \int_T \int_{[0,\Delta]} \boldsymbol{B}_{[3]}^{\mathrm{T}}(u,t_1,\cdot)\mathcal{W}(u)\psi_j(t_1)dudt_1 \cdot \boldsymbol{G}_n^{-1}\boldsymbol{\xi}_n\right\|_{L^2}.$$

By (S.3) in Lemma 2,

$$\left\| \int_T \int_{[0,\Delta]} \boldsymbol{B}_{[3]}^{\mathrm{T}}(u,t_1,\cdot)\mathcal{W}(u)\psi_j(t_1)dudt_1 \cdot \boldsymbol{G}_n^{-1}\boldsymbol{\xi}_n\right\|_{L^2}^2$$

$$= \int_T \left(\int_T \int_{[0,\Delta]} \boldsymbol{B}_{[3]}^{\mathrm{T}}\mathcal{W}(u)\psi_j(t_1)dudt_1 \cdot \boldsymbol{G}_n^{-1}\boldsymbol{\xi}_n\right)^2 dt_2$$

$$\leq \frac{C_4}{K_t}\sum_{i_3}\left\{\sum_{i_1,i_2}\left(\sum_{i'_1 i'_2 i'_3} g^*_{i_1 i_2 i_3, i'_1 i'_2 i'_3}\xi_{i'_1 i'_2 i'_3}\right) \cdot \int_0^\Delta B_{i_1}(u)\mathcal{W}(u)du \int_0^\Delta B_{i_2}(t_1)\psi_j(t_1)dt_1\right\}^2.$$

By taking the conditional expectation $\mathrm{E}(\cdot|\mathcal{G})$,

$$\mathrm{E}\left\{\left\|\int_T \int_{[0,\Delta]} \boldsymbol{B}_{[3]}^{\mathrm{T}}(u,t_1,\cdot)\mathcal{W}(u)\psi_j(t_1)dudt_1 \cdot \boldsymbol{G}_n^{-1}\boldsymbol{\xi}_n\right\|_{L^2}^2 \bigg| \mathcal{G}\right\}$$

$$\lesssim \frac{1}{K_t^3 K_s^2}\sum_{i_3}\sum_{i_1,j_1}\sum_{i_2,j_2}\sum_{i'_1 i'_2 i'_3}\sum_{j'_1 j'_2 j'_3} \left|g^*_{i_1 i_2 i_3, i'_1 i'_2 i'_3}\right| \cdot \left|g^*_{j_1 j_2 i_3, j'_1 j'_2 j'_3}\right| \cdot \left|\mathrm{E}\left(\xi_{i'_1 i'_2 i'_3}\xi_{j'_1 j'_2 j'_3}|\mathcal{G}\right)\right|.$$

By Lemma 6, when $n$ is sufficiently large,

$$\mathrm{E}\left\{\left\|\int_T \int_{[0,\Delta]} \boldsymbol{B}_{[3]}^{\mathrm{T}}(u,t_1,\cdot)\mathcal{W}(u)\psi_j(t_1)dudt_1 \cdot \boldsymbol{G}_n^{-1}\boldsymbol{\xi}_n\right\|_{L^2}^2\right\}$$

$$\lesssim K_t \sum_{i_3}\sum_{i_1,j_1}\sum_{i_2,j_2}\sum_{i'_1 i'_2 i'_3}\sum_{j'_1 j'_2 j'_3} \tau^{|i_1-i'_1|+|j_1-j'_1|}\tau^{|i_2-i'_2|+|j_2-j'_2|}\tau^{|i_3-i'_3|+|i_3-j'_3|} \cdot \mathrm{E}\left\{\left|\mathrm{E}\left(\xi_{i'_1 i'_2 i'_3}\xi_{j'_1 j'_2 j'_3}|\mathcal{G}\right)\right|\right\}.$$

By the results of Lemma 3, there exists a constant $C_5 > 0$ such that,

- for $|i_1 - i_1'| \leq p_s$ and $|i_2 - i_2'| \leq p_t$,

$$\mathrm{E}\left|\mathrm{E}\left(\xi_{i_1 i_2 i_3} \xi_{i_1' i_2' i_3'} | \mathcal{G}\right)\right| \leq C_5 \left(\frac{1}{|\mathcal{D}_n| K_s K_t^4} + \frac{1}{|\mathcal{D}_n| M_n L_n K_s K_t^3} + \frac{1}{|\mathcal{D}_n| M_n^2 L_n^2 K_s K_t^2}\right);$$

- for $|i_1 - i_1'| \leq p_s$ and $|i_2 - i_2'| > p_t$,

$$\mathrm{E}\left|\mathrm{E}\left(\xi_{i_1 i_2 i_3} \xi_{i_1' i_2' i_3'} | \mathcal{G}\right)\right| \leq C_5 \left(\frac{1}{|\mathcal{D}_n| K_s K_t^4} + \frac{1}{|\mathcal{D}_n| M_n L_n K_s K_t^3}\right);$$

- for $|i_1 - i_1'| > p_s$,

$$\mathrm{E}\left|\mathrm{E}\left(\xi_{i_1 i_2 i_3} \xi_{i_1' i_2' i_3'} | \mathcal{G}\right)\right| \leq C_5 \left(\frac{1}{|\mathcal{D}_n| K_s^2 K_t^4} + \frac{1}{|\mathcal{D}_n| M_n L_n K_s^2 K_t^3}\right).$$

Thus, as $n \to \infty$,

$$\mathrm{E}\left\{\left\|\int_T \int_{[0,\Delta]} \boldsymbol{B}_{[3]}^{\mathrm{T}}(u, t_1, \cdot) \mathcal{W}(u) \psi_j(t_1) du dt_1 \cdot \boldsymbol{G}_n^{-1} \boldsymbol{\xi}_n\right\|_{L^2}^2\right\}$$

$$\lesssim K_t \sum_{i_3} \sum_{i_1, j_1} \sum_{i_2, j_2} \sum_{i_1' i_2' i_3'} \sum_{j_1' j_2' j_3'} \tau^{|i_1 - i_1'| + |j_1 - j_1'|} \tau^{|i_2 - i_2'| + |j_2 - j_2'|} \tau^{|i_3 - i_3'| + |i_3 - j_3'|} \cdot \mathrm{E}\left|\mathrm{E}\left(\xi_{i_1 i_2 i_3} \xi_{i_1' i_2' i_3'} | \mathcal{G}\right)\right|$$

$$\leq C_5 K_t \sum_{i_3} \sum_{i_1, j_1} \sum_{i_2, j_2} \sum_{i_1' i_2' i_3'} \sum_{j_1' j_2' j_3'} \tau^{|i_1 - i_1'| + |j_1 - j_1'|} \tau^{|i_2 - i_2'| + |j_2 - j_2'|} \tau^{|i_3 - i_3'| + |i_3 - j_3'|}$$

$$\times \left\{\left(\frac{1}{|\mathcal{D}_n| K_s K_t^4} + \frac{1}{|\mathcal{D}_n| M_n L_n K_s K_t^3} + \frac{1}{|\mathcal{D}_n| M_n^2 L_n^2 K_s K_t^2}\right) \cdot I(|i_1' - j_1'| \leq p_s, |i_2' - j_2'| \leq p_t)\right.$$

$$+ \left(\frac{1}{|\mathcal{D}_n| K_s K_t^4} + \frac{1}{|\mathcal{D}_n| M_n L_n K_s K_t^3}\right) \cdot I(|i_1' - j_1'| \leq p_s, |i_2' - j_2'| > p_t)$$

$$\left.+ \left(\frac{1}{|\mathcal{D}_n| K_s^2 K_t^4} + \frac{1}{|\mathcal{D}_n| M_n L_n K_s^2 K_t^3}\right) \cdot I(|i_1' - j_1'| > p_s)\right\}$$

$$\lesssim K_s K_t^3 \left(\frac{1}{|\mathcal{D}_n| K_s K_t^4} + \frac{1}{|\mathcal{D}_n| M_n L_n K_s K_t^3} + \frac{1}{|\mathcal{D}_n| M_n^2 L_n^2 K_s K_t^2}\right)$$

$$+ K_s K_t^4 \left(\frac{1}{|\mathcal{D}_n| K_s K_t^4} + \frac{1}{|\mathcal{D}_n| M_n L_n K_s K_t^3}\right)$$

$$+ K_s^2 K_t^4 \left(\frac{1}{|\mathcal{D}_n| K_s^2 K_t^4} + \frac{1}{|\mathcal{D}_n| M_n L_n K_s^2 K_t^3}\right)$$

$$\asymp \frac{1}{|\mathcal{D}_n|} + \frac{K_t}{|\mathcal{D}_n| M_n L_n}.$$

Thus, as $n \to \infty$,

$$\left\| \int_T \int_{[0,\Delta]} \boldsymbol{B}_{[3]}^{\mathrm{T}} \mathcal{W}(u)\psi_j(t_1)dudt_1 \cdot \boldsymbol{G}_n^{-1} \boldsymbol{\xi}_n \right\|_{L^2} = O_p\left(\sqrt{\frac{1}{|\mathcal{D}|_n}} + \sqrt{\frac{K_t}{|\mathcal{D}|_n M_n L_n}}\right).$$

□

### S.3.4  Proof of Theorem 4.4

The expression of $\widehat{\mathcal{C}}_j(u) - \mathcal{C}_j(u)$ can be rewritten into three integrals,

$$\begin{aligned}
\widehat{\mathcal{C}}_j(u) - \mathcal{C}_j(u) &= \int_{T^2} \widehat{R}(u,t_1,t_2)\widehat{\psi}_j(t_1)\widehat{\psi}_j(t_2)dt_1dt_2 - \int_{T^2} R(u,t_1,t_2)\psi_j(t_1)\psi_j(t_2)dt_1dt_2 \\
&= \int_{T^2} \left\{ \widehat{R}(u,t_1,t_2) - R(u,t_1,t_2) \right\} \psi_j(t_1)\psi_j(t_2)dt_1dt_2 \\
&\quad + \int_{T^2} \widehat{R}(u,t_1,t_2) \left\{ \widehat{\psi}_j(t_1) - \psi_j(t_1) \right\} \psi_j(t_2)dt_1dt_2 \\
&\quad + \int_{T^2} \widehat{R}(u,t_1,t_2)\widehat{\psi}_j(t_1) \left\{ \widehat{\psi}_j(t_2) - \psi_j(t_2) \right\} dt_1dt_2.
\end{aligned}$$

Using similar calculations as in Theorems 4.2 and 4.3, we can show that,

$$\int_{T^2} \left\{ \widehat{R}(u,t_1,t_2) - R(u,t_1,t_2) \right\} \psi_j(t_1)\psi_j(t_2)dt_1dt_2 = O_p\left(\sqrt{\frac{K_s}{|\mathcal{D}_n|}} + K_s^{-p_s} + K_t^{-p_t}\right),$$

the detail of which is omitted for brevity. By Hölder's inequality and Theorem 4.3,

$$\begin{aligned}
&\left| \int_{T^2} \widehat{R}(u,t_1,t_2) \left\{ \widehat{\psi}_j(t_1) - \psi_j(t_1) \right\} \psi_j(t_2)dt_1dt_2 \right| \\
&\leq \left[ \int_{T^2} \left\{ \widehat{R}(u,t_1,t_2)\psi_j(t_2) \right\}^2 dt_1dt_2 \right]^{1/2} \cdot |T| \cdot \left\| \widehat{\psi}_j - \psi_j \right\|_{L^2} \\
&= O_p\left( \sqrt{\frac{1}{|\mathcal{D}|_n}} + \sqrt{\frac{K_t}{|\mathcal{D}|_n M_n L_n}} + K_s^{-p_s} + K_t^{-p_t} \right).
\end{aligned}$$

Following the same reasoning,

$$\left| \int_{T^2} \widehat{R}(u,t_1,t_2)\widehat{\psi}_j(t_1) \left\{ \widehat{\psi}_j(t_2) - \psi_j(t_2) \right\} dt_1dt_2 \right| = O_p\left( \sqrt{\frac{1}{|\mathcal{D}|_n}} + \sqrt{\frac{K_t}{|\mathcal{D}|_n M_n L_n}} + K_s^{-p_s} + K_t^{-p_t} \right).$$

Consequently,

$$\left\|\widehat{\mathcal{C}}_j(u) - \mathcal{C}_j(u)\right\|_{L^2}$$
$$= O_p\left(\sqrt{\frac{K_s}{|\mathcal{D}_n|}} + K_s^{-p_s} + K_t^{-p_t}\right) + O_p\left(\sqrt{\frac{1}{|\mathcal{D}|_n}} + \sqrt{\frac{K_t}{|\mathcal{D}|_n M_n L_n}} + K_s^{-p_s} + K_t^{-p_t}\right)$$
$$= O_p\left(\sqrt{\frac{K_s}{|\mathcal{D}_n|}} + \sqrt{\frac{K_t}{|\mathcal{D}|_n M_n L_n}} + K_s^{-p_s} + K_t^{-p_t}\right).$$

□

### S.3.5 Proof of Theorem 4.5

It is easy to see $\left\|\widehat{\Lambda} - \Lambda\right\|_{L^2} \leq \left\|\widehat{\Gamma} - \Gamma\right\|_{L^2} + \left\|\widehat{R}(0,\cdot,\cdot) - R(0,\cdot,\cdot)\right\|_{L^2}$, we derive the rates of the two terms separately and the results of Theorem 4.5 follow immediately.

**Part I**: Convergence rate of $\|\widehat{\Gamma} - \Gamma\|_{L^2}$.

Since $\widehat{\Gamma}$ is the spline estimator of a 2-dim covariance function, derivation of its convergence rate is a simplified version of Theorem 4.1, which provides the convergence rate of the 3-dim covariance estimator $\widehat{R}$. Therefore, we only provide a sketch of the proof.

Define

$$\boldsymbol{H}_n := \frac{1}{|\mathcal{D}_n|M_n^2 L_n} \int_{\mathcal{D}_n} \int_{T^{\otimes 2}} \boldsymbol{B}_{[2]}(t_1, t_2) \cdot \boldsymbol{B}_{[2]}^T(t_1, t_2) I(t_1 \neq t_2) \mathcal{N}_t(dt_1|\boldsymbol{s}) \mathcal{N}_t(dt_2|\boldsymbol{s}) \mathcal{N}_s(d\boldsymbol{s}),$$

and

$$\boldsymbol{\zeta}_n := \frac{1}{|\mathcal{D}_n|M_n^2 L_n} \int_{\mathcal{D}_n} \int_{T^{\otimes 2}} \boldsymbol{B}_{[2]}(t_1, t_2) \{Y(\boldsymbol{s}, t_1) Y(\boldsymbol{s}, t_2) - \Gamma(t_1, t_2)\}$$
$$\times I(t_1 \neq t_2) \mathcal{N}_t(dt_1|\boldsymbol{s}) \mathcal{N}_t(dt_2|\boldsymbol{s}) \mathcal{N}_s(d\boldsymbol{s}).$$

Similar results as Lemmas 3 and 5 exist for bivariate tensor product splines: when $n$ is sufficiently large, with probability 1,

$$\frac{C_1}{K_\Gamma^2} \leq \lambda_{\min}(\boldsymbol{H}_n) \leq \lambda_{\max}(\boldsymbol{H}_n) \leq \frac{C_2}{K_\Gamma^2},$$

for some positive constants $C_1$ and $C_2$, and

$$\|\boldsymbol{\zeta}_n\|_2^2 = O_p\left(\frac{1}{|\mathcal{D}_n|K_\Gamma^2} + \frac{1}{|\mathcal{D}_n|M_n L_n K_\Gamma} + \frac{1}{|\mathcal{D}_n|M_n^2 L_n}\right).$$

By spline approximation theory (analogous to Lemma 1), there exists an $\Gamma^* \in \mathcal{S}_{[2]}^{\Gamma}$ such that $\|\Gamma - \Gamma^*\|_\infty = O(K_\Gamma^{-p_\Gamma})$, as $K_\Gamma \to \infty$. Write $\Gamma^*(t_1, t_2) = \boldsymbol{B}_{[2]}^{\mathrm{T}}(t_1, t_2)\boldsymbol{\gamma}^*$ for some spline coefficient vector $\boldsymbol{\gamma}^* \in \mathbb{R}^{(K_\Gamma + p_\Gamma)^2}$. Then,

$$\begin{aligned}
\left\|\widehat{\Gamma} - \Gamma\right\|_{L^2}^2 &\leq \left\|\widehat{\Gamma} - \Gamma^*\right\|_{L^2}^2 + \|\Gamma^* - \Gamma\|_{L^2}^2 \leq \left\|\boldsymbol{B}_{[2]}^{\mathrm{T}}(\widehat{\boldsymbol{\gamma}} - \boldsymbol{\gamma}^*)\right\|_{L^2}^2 + C_3\|R - R^*\|_\infty^2 \\
&\leq \left\|\boldsymbol{B}_{[2]}^{\mathrm{T}} \boldsymbol{H}_n^{-1} \boldsymbol{\zeta}_n\right\|_{L^2}^2 + C_4\|R - R^*\|_\infty^2 \\
&\leq \frac{1}{K_\Gamma^2}\{\lambda_{\min}(\boldsymbol{H}_n)\}^{-2} \|\boldsymbol{\zeta}_n\|_2^2 + C_4\|R - R^*\|_\infty^2 \\
&= O_p\left(\frac{1}{|\mathcal{D}_n|} + \frac{K_\Gamma}{|\mathcal{D}_n|M_n L_n} + \frac{K_\Gamma^2}{|\mathcal{D}_n|M_n^2 L_n} + K_\Gamma^{-p_\Gamma}\right).
\end{aligned}$$

**Part II**: Convergence rate of $\|\widehat{R}(0, \cdot, \cdot) - R(0, \cdot, \cdot)\|_{L^2}$.

Analogous to the proofs of Theorem 4.1 and Theorem 4.2, we bound $\left\|\widehat{R}(0, \cdot, \cdot) - R(0, \cdot, \cdot)\right\|_{L^2}$ by three terms:

$$\left\|\widehat{R}(0, \cdot, \cdot) - R(0, \cdot, \cdot)\right\|_{L^2} \leq \left\|\boldsymbol{B}_{[3]}^{\mathrm{T}}(0, \cdot, \cdot)\boldsymbol{G}_n^{-1}\boldsymbol{\xi}_n\right\|_{L^2} + \left\|\boldsymbol{B}_{[3]}^{\mathrm{T}}(0, \cdot, \cdot)\boldsymbol{G}_n^{-1}\boldsymbol{\eta}_n\right\|_{L^2} + C_5\|R - R^*\|_\infty.$$

Applying (S.4) in Lemma 2, we have

$$\left\|\boldsymbol{B}_{[3]}^{\mathrm{T}}(0, \cdot, \cdot)\boldsymbol{G}_n^{-1}\boldsymbol{\xi}_n\right\|_{L^2}^2 \leq \frac{C_6}{K_t^2} \sum_{i_2, i_3} \left\{\sum_{1 \leq i_1 \leq p_s} \left(\sum_{i_1' i_2' i_3'} g^*_{i_1 i_2 i_3, i_1' i_2' i_3'} \xi_{i_1' i_2' i_3'}\right) \cdot B_{i_1}(0)\right\}^2.$$

Taking conditional expectation $\mathrm{E}(\cdot|\mathcal{G})$ on both sides of the inequality above, we have

$$\mathrm{E}\left\{\left\|\boldsymbol{B}_{[3]}^{\mathrm{T}}(0, \cdot, \cdot)\boldsymbol{G}_n^{-1}\boldsymbol{\xi}_n\right\|_{L^2}^2 \bigg| \mathcal{G}\right\}$$
$$\lesssim \frac{1}{K_t^2} \sum_{i_2, i_3} \sum_{1 \leq i_1, j_1 \leq p_s} \sum_{i_1' i_2' i_3'} \sum_{j_1' j_2' j_3'} \left|g^*_{i_1 i_2 i_3, i_1' i_2' i_3'}\right| \cdot \left|g^*_{j_1 i_2 i_3, j_1' j_2' j_3'}\right| \cdot \mathrm{E}\left(\left|\xi_{i_1' i_2' i_3'} \xi_{j_1' j_2' j_3'}\right| \bigg| \mathcal{G}\right).$$

By Lemma 6, when $n$ is sufficiently large,

$$\mathrm{E}\left\{\left\|\boldsymbol{B}_{[3]}^{\mathrm{T}}(0, \cdot, \cdot)\boldsymbol{G}_n^{-1}\boldsymbol{\xi}_n\right\|_{L^2}^2 \bigg| \mathcal{G}\right\}$$
$$\lesssim K_s^2 K_t^2 \sum_{i_2, i_3} \sum_{1 \leq i_1, j_1 \leq p_s} \sum_{i_1' i_2' i_3'} \sum_{j_1' j_2' j_3'} \tau^{|i_1 - i_1'| + |j_1 - j_1'|} \tau^{|i_2 - i_2'| + |i_2 - j_2'|} \tau^{|i_3 - i_3'| + |i_3 - j_3'|} \cdot \mathrm{E}\left(\left|\xi_{i_1' i_2' i_3'} \xi_{j_1' j_2' j_3'}\right| \bigg| \mathcal{G}\right).$$

By Lemma 3, there exists a constant $C_7 > 0$ such that,

- for $|i_1 - i_1'| \leq p_s$,

$$\mathrm{E}\left|\mathrm{E}\left(\xi_{i_1 i_2 i_3}\xi_{i_1' i_2' i_3'}|\mathcal{G}\right)\right| \leq C_7 \left(\frac{1}{|\mathcal{D}_n|K_s K_t^4} + \frac{1}{|\mathcal{D}_n|M_n L_n K_s K_t^3} + \frac{1}{|\mathcal{D}_n|M_n^2 L_n^2 K_s K_t^2}\right);$$

- for $|i_1 - i_1'| > p_s$,

$$\mathrm{E}\left|\mathrm{E}\left(\xi_{i_1 i_2 i_3}\xi_{i_1' i_2' i_3'}|\mathcal{G}\right)\right| \leq C_7 \left(\frac{1}{|\mathcal{D}_n|K_s^2 K_t^4} + \frac{1}{|\mathcal{D}_n|M_n L_n K_s^2 K_t^3}\right).$$

Thus,

$$
\begin{aligned}
&\mathrm{E}\left\{\left\|\boldsymbol{B}_{[3]}^{\mathrm{T}}(0,\cdot,\cdot)\boldsymbol{G}_n^{-1}\boldsymbol{\xi}_n\right\|_{L^2}^2\right\} \\
&\leq C_7 K_s^2 K_t^2 \sum_{i_2,i_3} \sum_{1\leq i_1,j_1 \leq p_s} \sum_{i_1' i_2' i_3'} \sum_{j_1' j_2' j_3'} \tau^{|i_1-i_1'|+|j_1-j_1'|} \tau^{|i_2-i_2'|+|i_2-j_2'|} \tau^{|i_3-i_3'|+|i_3-j_3'|} \\
&\quad \times \Bigg\{\left(\frac{1}{|\mathcal{D}_n|K_s K_t^4} + \frac{1}{|\mathcal{D}_n|M_n L_n K_s K_t^3} + \frac{1}{|\mathcal{D}_n|M_n^2 L_n^2 K_s K_t^2}\right) I(|i_1'-j_1'|\leq p_s) \\
&\quad + \left(\frac{1}{|\mathcal{D}_n|K_s^2 K_t^4} + \frac{1}{|\mathcal{D}_n|M_n L_n K_s^2 K_t^3}\right) I(|i_1'-j_1'|>p_s)\Bigg\} \\
&\lesssim K_s^2 K_t^4 \left(\frac{1}{|\mathcal{D}_n|K_s K_t^4} + \frac{1}{|\mathcal{D}_n|M_n L_n K_s K_t^3} + \frac{1}{|\mathcal{D}_n|M_n^2 L_n^2 K_s K_t^2}\right) \\
&\quad + K_s^2 K_t^4 \left(\frac{1}{|\mathcal{D}_n|K_s^2 K_t^4} + \frac{1}{|\mathcal{D}_n|M_n L_n K_s^2 K_t^3}\right) \\
&\asymp \frac{K_s}{|\mathcal{D}_n|} + \frac{K_s K_t}{|\mathcal{D}_n|M_n L_n} + \frac{K_s K_t^2}{|\mathcal{D}_n|M_n^2 L_n^2},
\end{aligned}
$$

which implies

$$\left\|\boldsymbol{B}_{[3]}^{\mathrm{T}}(0,\cdot,\cdot)\boldsymbol{G}_n^{-1}\boldsymbol{\xi}_n\right\|_{L^2} = O_p\left(\sqrt{\frac{K_s}{|\mathcal{D}_n|}} + \sqrt{\frac{K_s K_t}{|\mathcal{D}_n|M_n L_n}} + \sqrt{\frac{K_s K_t^2}{|\mathcal{D}_n|M_n^2 L_n^2}}\right).$$

Now we show the convergence rate of $\left\|\boldsymbol{B}_{[3]}^{\mathrm{T}}(0,\cdot,\cdot)\boldsymbol{G}_n^{-1}\boldsymbol{\eta}_n\right\|_{L^2}$. Define

$$
\begin{aligned}
\eta_{j_1 j_2 j_3} &:= \frac{1}{|\mathcal{D}_n|M_n^2 L_n^2} \int_{\mathcal{D}_n^{\otimes 2}} \int_T \int_T B_{j_1}(\|\boldsymbol{s}_1-\boldsymbol{s}_2\|) B_{j_2}(t_1) B_{j_3}(t_2) I(\|\boldsymbol{s}_1-\boldsymbol{s}_2\|\leq \Delta) \\
&\quad \times \{R(\|\boldsymbol{s}_1-\boldsymbol{s}_2\|,t_1,t_2) - R^*(\|\boldsymbol{s}_1-\boldsymbol{s}_2\|,t_1,t_2)\}\mathcal{N}_{s,2}(d\boldsymbol{s}_1,d\boldsymbol{s}_2)\mathcal{N}_t(dt_1|\boldsymbol{s}_1)\mathcal{N}_t(dt_2|\boldsymbol{s}_2) \\
&\leq \frac{C_8}{K_s K_t^2}\|R-R^*\|_\infty.
\end{aligned}
$$

Then $\boldsymbol{\eta}_n = (\eta_{j_1 j_2 j_3})$, for $j_1 \in \{1,\ldots,K_s+p_s\}$, $j_2, j_3 \in \{1,\ldots,K_t+p_t\}$. Following similar

arguments as the proof of $\left\|\boldsymbol{B}_{[3]}^{\mathrm{T}}(0,\cdot,\cdot)\boldsymbol{G}_n^{-1}\boldsymbol{\xi}_n\right\|_{L^2}$, we have

$$\left\|\boldsymbol{B}_{[3]}^{\mathrm{T}}(0,\cdot,\cdot)\boldsymbol{G}_n^{-1}\boldsymbol{\eta}_n\right\|_{L^2}^2 \leq C_9 K_t^2 \sum_{i_2,i_3}\left\{\sum_{1\leq i_1\leq p_s}\left(\sum_{i_1' i_2' i_3'}\tau^{|i_1-i_1'|+|i_2-i_2'|+|i_3-i_3'|}\cdot|\eta_{i_1' i_2' i_3'}|\right)\right\}^2$$
$$\lesssim \|R-R^*\|_\infty^2.$$

Thus,

$$\left\|\widehat{R}(0,\cdot,\cdot)-R(0,\cdot,\cdot)\right\|_{L^2} = O_p\left(\sqrt{\frac{K_s}{|\mathcal{D}_n|}}+\sqrt{\frac{K_s K_t}{|\mathcal{D}_n| M_n L_n}}+\sqrt{\frac{K_s K_t^2}{|\mathcal{D}_n| M_n^2 L_n^2}}+K_s^{-p_s}+K_t^{-p_t}\right).$$

□

## S.3.6 Proof of Theorem 4.6

Since $\widehat{\sigma}_\epsilon^2 - \sigma_\epsilon^2 = \frac{1}{|T|}\int_T \{\widehat{\sigma}_Y^2(t) - \sigma_Y^2(t)\}\,dt - \frac{1}{|T|}\int_T \{\widehat{\Gamma}(t,t) - \Gamma(t,t)\}\,dt$, we derive the convergence rates for the two terms separately, and the result of Theorem 4.6 follows immediately.

**Part I:** Convergence rate of $\frac{1}{|T|}\int_T \{\widehat{\sigma}_Y^2(t) - \sigma_Y^2(t)\}\,dt$.

Define
$$\boldsymbol{V}_n := \frac{1}{|\mathcal{D}_n| M_n L_n}\int_{\mathcal{D}_n}\int_T \boldsymbol{B}_{[1]}(t)\cdot\boldsymbol{B}_{[1]}^T(t)\mathcal{N}_t(dt|\boldsymbol{s})\mathcal{N}_s(d\boldsymbol{s}),$$

and
$$\varsigma_n := \frac{1}{|\mathcal{D}_n| M_n L_n}\int_{\mathcal{D}_n}\int_T \boldsymbol{B}_{[1]}(t)\cdot\{Y^2(\boldsymbol{s},t)-\sigma_Y^2(t)\}\,\mathcal{N}_t(dt|\boldsymbol{s})\mathcal{N}_s(d\boldsymbol{s}),$$

where
$$\boldsymbol{B}_{[1]}(t) = \{B_{1,K_\epsilon}^{p_\epsilon}(t), B_{2,K_\epsilon}^{p_\epsilon}(t),\ldots,B_{K_\epsilon+p_\epsilon,K_\epsilon}^{p_\epsilon}(t)\}^{\mathrm{T}}$$

is a vector of normalized B-spline functions of order $p_\epsilon$, defined on time domain $T$ with equally spaced interior knots $\kappa_j = j/(K_\epsilon+1)$, $j=1,\ldots,K_\epsilon$.

We write $\boldsymbol{V}_n^{-1} = (V_{j,j'}^*)$ and $\varsigma_n = (\varsigma_j)$, where $V_{j,j'}^*$ is the $(j,j')$th entry of $\boldsymbol{V}_n^{-1}$, and $\varsigma_j$ is the $j$th entry of $\varsigma_n$. By similar arguments as Lemmas 3 and 6, with probability 1, when $n$ is sufficiently large,
$$\sup_{j,j'}\left|\frac{V_{j,j'}^*}{\tau^{|j-j'|}}\right| \leq C_1 K_\epsilon,$$

for some $\tau \in (0,1)$ and some positive constant $C_1$, and for $j,j' \in \{1,\cdots,K_\epsilon+p_\epsilon\}$,

- for $|j-j'| \leq p_\epsilon$, $\mathrm{E}\{|\mathrm{E}(\varsigma_j\varsigma_{j'}|\boldsymbol{\mathcal{G}})|\} \leq C_2\left(\frac{1}{|\mathcal{D}_n|K_\epsilon^2}+\frac{1}{|\mathcal{D}_n|M_n L_n K_\epsilon}\right)$;

- for $|j-j'| > p_\epsilon$, $\mathrm{E}\{|\mathrm{E}(\varsigma_j\varsigma_{j'}|\boldsymbol{\mathcal{G}})|\} \leq \frac{C_2}{|\mathcal{D}_n|K_\epsilon^2}$.

Hence,

$$\mathrm{E}\left\{\int_T \boldsymbol{B}_T(t)\cdot \boldsymbol{V}_n^{-1}\boldsymbol{\varsigma}_n\right\}^2 \lesssim \mathrm{E}\left\{\frac{1}{K_\epsilon^2}\sum_{i,i',j,j'}|V_{i,i'}^*|\cdot|V_{j,j'}^*|\cdot|\mathrm{E}(\varsigma_{i'}\varsigma_{j'}|\mathcal{G})|\right\}$$

$$\lesssim \sum_{i,i',j,j'}\tau^{|i-i'|+|j-j'|}\cdot\mathrm{E}\,|\mathrm{E}(\varsigma_{i'}\varsigma_{j'}|\mathcal{G})| \lesssim \frac{1}{|\mathcal{D}_n|}.$$

Thus,

$$\frac{1}{|T|}\left|\int_T\{\widehat{\sigma}_Y^2(t)-\sigma_Y^2(t)\}\,dt\right| \le \left|\int_T \boldsymbol{B}_T(t)\cdot\boldsymbol{V}_n^{-1}\boldsymbol{\varsigma}_n\right| + O_p\left(K_\epsilon^{-p_\epsilon}\right) = O_p\left(\sqrt{\frac{1}{|\mathcal{D}_n|}}+K_\epsilon^{-p_\epsilon}\right).$$

**Part II:** Convergence rate of $\frac{1}{|T|}\int_T\{\widehat{\Gamma}(t,t)-\Gamma(t,t)\}\,dt$.

Let $\boldsymbol{H}_n$ and $\boldsymbol{\zeta}_n$ be as defined in Section S.3.5. We write $\boldsymbol{H}_n^{-1}=\left(H_{j_1j_2,j_1'j_2'}^*\right)$ and $\boldsymbol{\zeta}_n=(\zeta_{j_1j_2})$, where $H_{j_1j_2,j_1'j_2'}^*$ is the $(j_1j_2,j_1'j_2')$th entry of $\boldsymbol{H}_n^{-1}$, and $\zeta_{j_1j_2}$ is the $(j_1j_2)$th entry of $\boldsymbol{\zeta}_n$, for $j_1,j_2,j_1',j_2'\in\{1,\cdots,K_\Gamma+p_t\}$. By similar arguments as Lemmas 3 and 6, with probability 1, when $n$ is sufficiently large,

$$\sup_{j_1,j_2,j_1',j_2'}\left|\frac{H_{j_1j_2,j_1'j_2'}^*}{\tau^{|j_1-j_1'|+|j_2-j_2'|}}\right| \le C_3 K_\Gamma^2,$$

for some $\tau\in(0,1)$ and some constant $C_3>0$, and for $j_1,j_2,j_1',j_2'\in\{1,\cdots,K_\Gamma+p_\Gamma\}$,

- for $\max(|j_1-j_1'|,|j_2-j_2'|,|j_1-j_2'|,|j_2-j_1'|)\le p_\Gamma$,

$$\mathrm{E}\left\{\left|\mathrm{E}(\zeta_{j_1j_2}\zeta_{j_1'j_2'}|\mathcal{G})\right|\right\} \le C_5\left(\frac{1}{|\mathcal{D}_n|K_\Gamma^4}+\frac{1}{|\mathcal{D}_n|M_nL_nK_\Gamma^3}+\frac{1}{|\mathcal{D}_n|M_n^2L_n^2K_\Gamma^2}\right);$$

- for $\max(|j_1-j_1'|,|j_2-j_2'|,|j_1-j_2'|,|j_2-j_1'|)>p_\Gamma$,

$$\mathrm{E}\left\{\left|\mathrm{E}(\zeta_{j_1j_2}\zeta_{j_1'j_2'}|\mathcal{G})\right|\right\} \le C_5\left(\frac{1}{|\mathcal{D}_n|K_\Gamma^4}+\frac{1}{|\mathcal{D}_n|M_nL_nK_\Gamma^3}\right);$$

- for $\min(|j_1-j_1'|,|j_2-j_2'|,|j_1-j_2'|,|j_2-j_1'|)>p_\Gamma$,

$$\mathrm{E}\left\{\left|\mathrm{E}(\zeta_{j_1j_2}\zeta_{j_1'j_2'}|\mathcal{G})\right|\right\} \le \frac{C_5}{|\mathcal{D}_n|K_\Gamma^4}.$$

Hence,

$$
\begin{aligned}
&\mathrm{E}\left\{\int_T \boldsymbol{B}_{[2]}(t,t)dt \cdot \boldsymbol{H}^{-1}\boldsymbol{\zeta}_n\right\}^2 \\
&= \mathrm{E}\left\{\sum_{j_1 j_2 j_1' j_2'} \int_T B_{j_1}(t)B_{j_2}(t)dt \cdot H^*_{j_1 j_2, j_1' j_2'} \zeta_{j_1' j_2'}\right\}^2 \\
&\leq \mathrm{E}\left\{\sum_{j_1 j_2 j_1' j_2'} \sum_{i_1 i_2 i_1' i_2'} \int_T B_{j_1}(t)B_{j_2}(t)dt \cdot \int_T B_{i_1}(t)B_{i_2}(t)dt \cdot H^*_{j_1 j_2, j_1' j_2'} H^*_{i_1 i_2, i_1' i_2'} \cdot \left|\mathrm{E}\left(\zeta_{j_1' j_2'}\zeta_{i_1' i_2'}|\mathcal{G}\right)\right|\right\} \\
&\lesssim \sum_{|j_1-j_2|<p_\Gamma} \sum_{|i_1-i_2|<p_\Gamma} \sum_{j_1' j_2' i_1' i_2'} K_\Gamma^2 \cdot \tau^{|j_1-j_1'|+|j_2-j_2'|+|i_1-i_1'|+|i_2-i_2'|} \cdot \mathrm{E}\left|\mathrm{E}\left(\zeta_{j_1' j_2'}\zeta_{i_1' i_2'}|\mathcal{G}\right)\right| \\
&\lesssim \frac{1}{|\mathcal{D}_n|} + \frac{K_\Gamma}{|\mathcal{D}_n|M_n L_n}.
\end{aligned}
$$

Thus,

$$
\begin{aligned}
\frac{1}{|T|}\left|\int_T \left\{\widehat{\Gamma}(t,t) - \Gamma(t,t)\right\}dt\right| &\leq C_6 \int_T \boldsymbol{B}_{[2]}(t,t)dt \cdot \boldsymbol{H}^{-1}\boldsymbol{\zeta}_n + O_p\left(K_\Gamma^{-p_\Gamma}\right) \\
&= O_p\left(\sqrt{\frac{1}{|\mathcal{D}_n|}} + \sqrt{\frac{K_\Gamma}{|\mathcal{D}_n|M_n L_n}} + K_\Gamma^{-p_\Gamma}\right).
\end{aligned}
$$

Therefore,

$$
\begin{aligned}
\widehat{\sigma}_\epsilon^2 - \sigma_\epsilon^2 &= \frac{1}{|T|}\int_T \left\{\widehat{\sigma}_Y^2(t) - \sigma_Y^2(t)\right\}dt - \frac{1}{|T|}\int_T \left\{\widehat{\Gamma}(t,t) - \Gamma(t,t)\right\}dt \\
&= O_p\left(\sqrt{\frac{1}{|\mathcal{D}_n|}} + \sqrt{\frac{K_\Gamma}{|\mathcal{D}_n|M_n L_n}} + K_\Gamma^{-p_\Gamma} + K_\epsilon^{-p_\epsilon}\right).
\end{aligned}
$$

□

## S.4 Additional results for the simulation studies

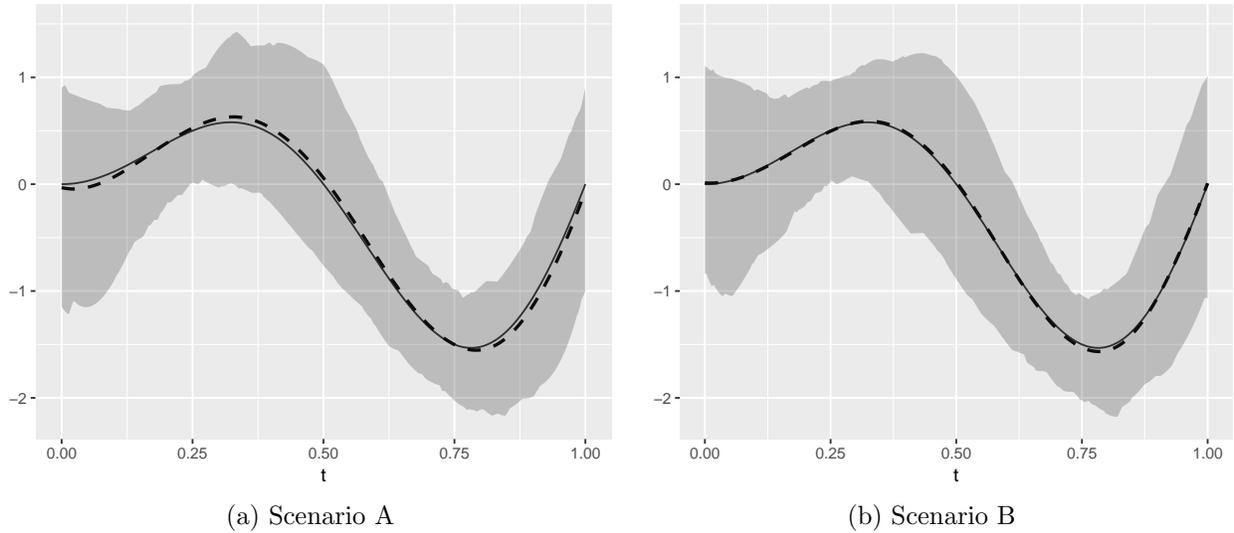

(a) Scenario A

(b) Scenario B

Figure S.1: Mean estimation results for the simulation studies. In each panel, the solid line is the true mean function, the dashed curve is the mean of $\widehat{\mu}(t)$, and the shaded area illustrates the confidence bands formed by the pointwise 5% and 95% percentiles.

Table S.1: Simulation results on the sensitivity of $sFPCA$ to different choices of $\Delta$. The data are generated under the Scenario A described in Section 7, and the results are the mean and standard deviation of integrated square errors (ISE) for functional principal components and the running time (in seconds) of the algorithm.

| Method | ISE($\psi_1$) | ISE($\psi_2$) | ISE($\psi_3$) | Running time (s) |
|---|---|---|---|---|
| $sFPCA$ ($\Delta = 1.5$) | 0.075(0.095) | 0.110(0.109) | 0.075(0.084) | 159.5(11.4) |
| $sFPCA$ ($\Delta = 2.5$) | 0.076(0.104) | 0.104(0.119) | 0.077(0.071) | 384.0(31.4) |
| $sFPCA$ ($\Delta = 3.5$) | 0.069(0.094) | 0.099(0.103) | 0.063(0.044) | 633.7(55.0) |
| $iFPCA$ | 0.411(0.376) | 0.367(0.369) | 1.494(0.311) | 45.2(3.7) |

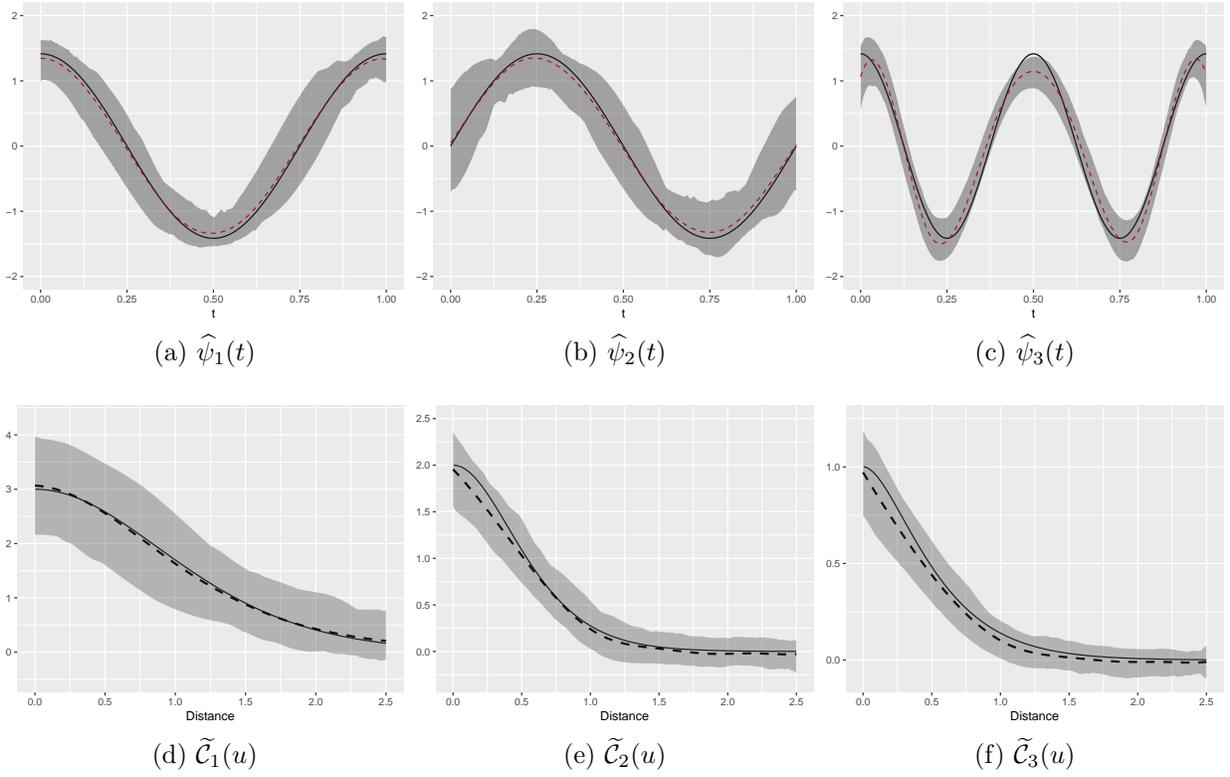

Figure S.2: Estimation results of *sFPCA* under Scenario B. In each panel, the solid line is the true function; the dashed line is the mean of the functional estimator; and the shaded area illustrates the bands of pointwise 5% and 95% percentiles.

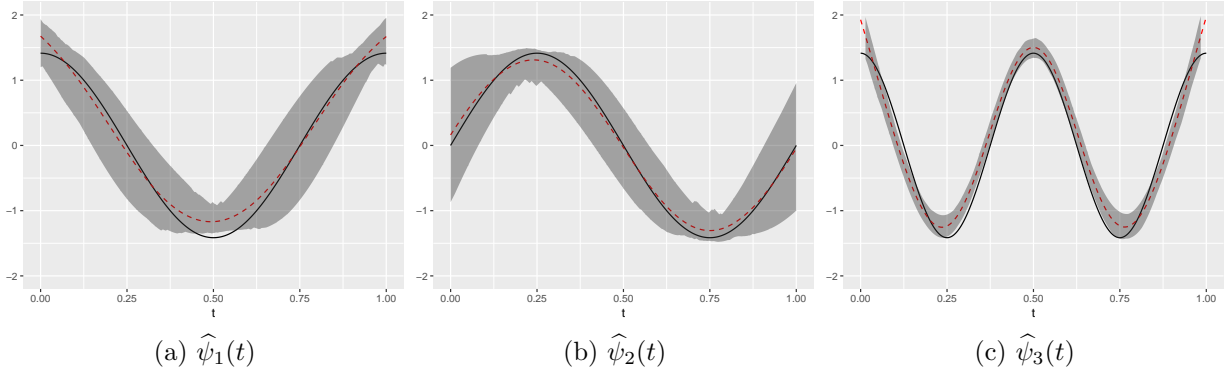

Figure S.3: Estimation results of *iFPCA* under Scenario B.

# S.5 Additional Figures for Real Data Analysis

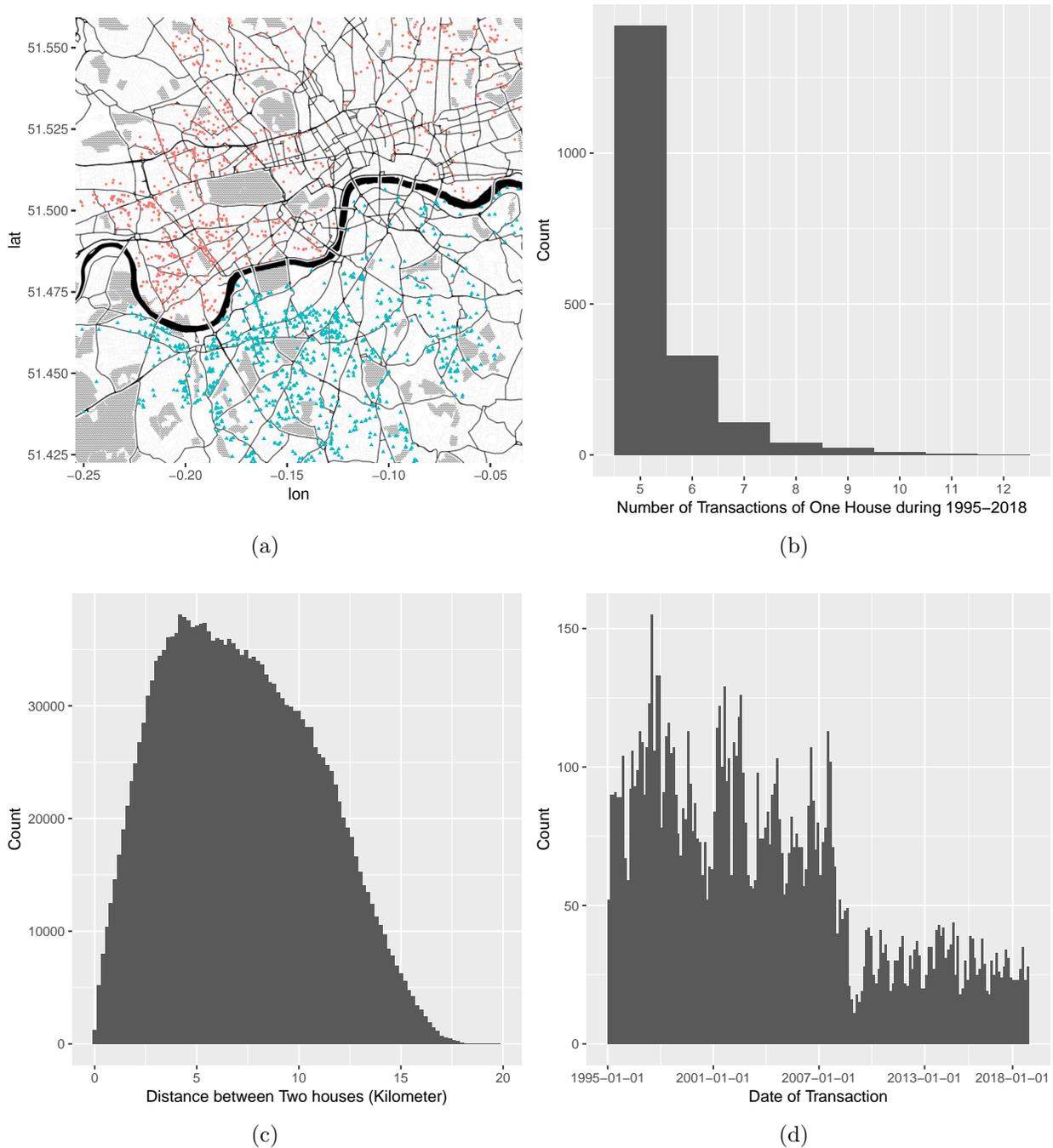

Figure S.4: Spatial-temporal pattern of London house price data: (a) locations of homes (dot points represent homes on the north side of River Thames, and triangular points represent homes on the south side of River Thames); (b) histogram of the number of transactions of the homes in the dataset; (c) histogram of the distance between two homes in the dataset; (d) histogram of transaction dates.

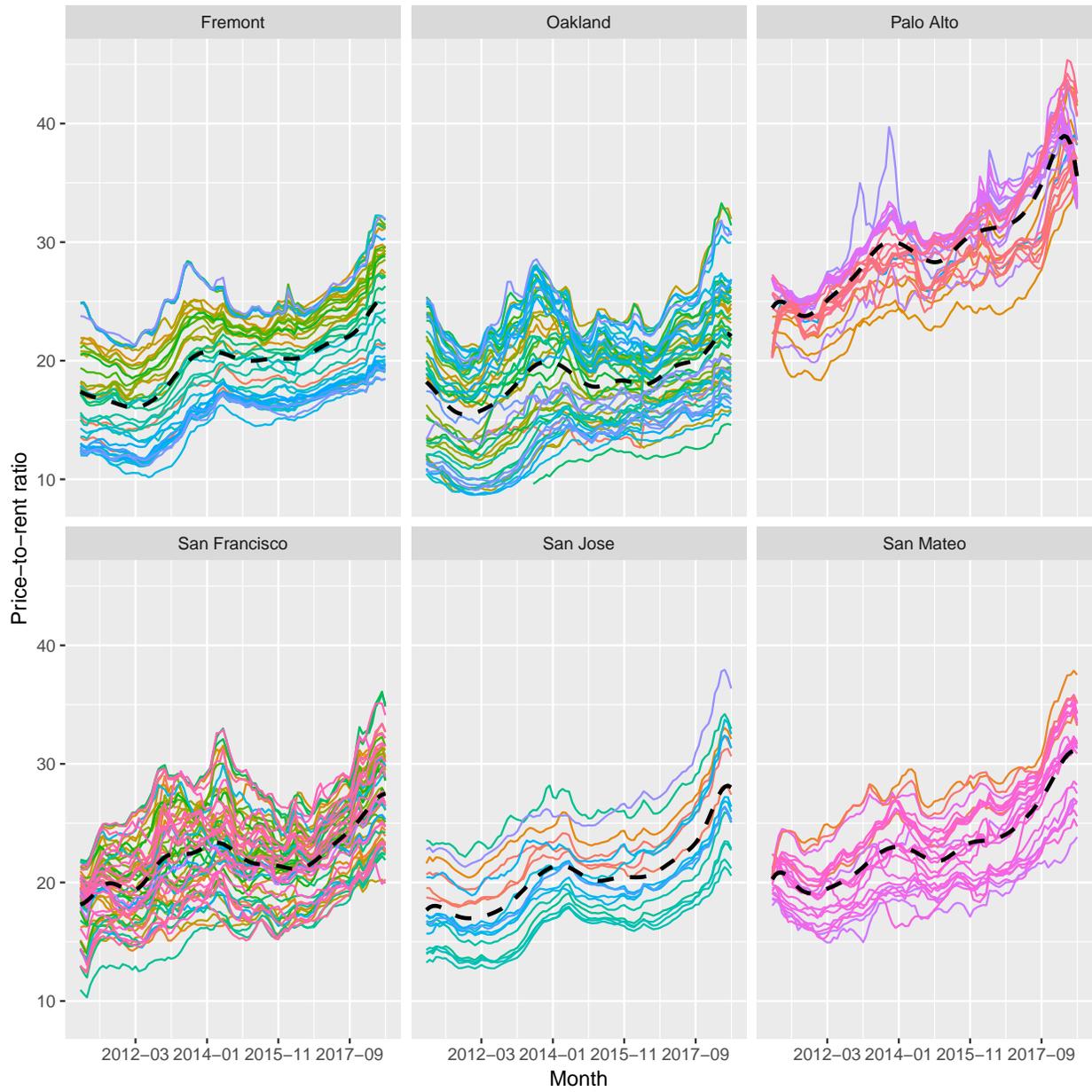

Figure S.5: Zillow price-to-rent ratio trajectories in the six regions of the San Francisco Bay Area and the region-specific mean functions (the dark dashed curve in each panel).

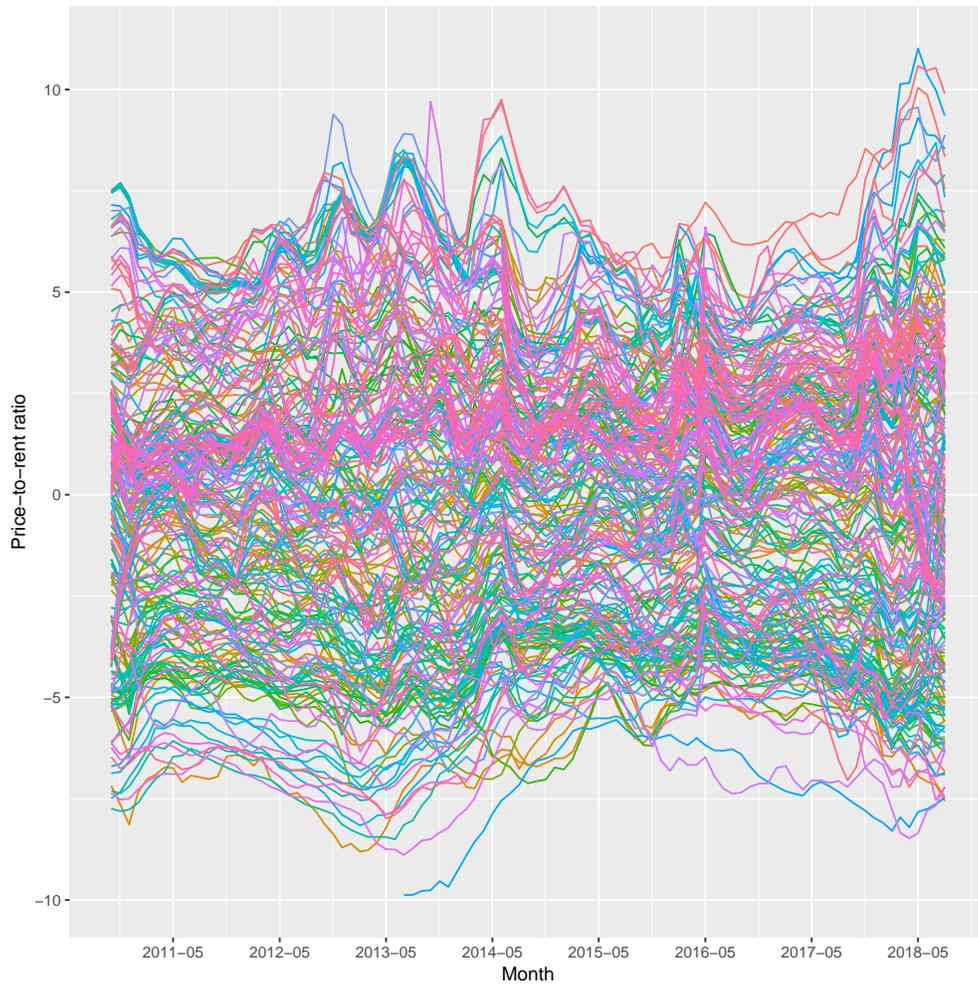

Figure S.6: Zillow price-to-rent ratio trajectories centered by region-specific mean functions.

## S.6 Sensitivity Analysis for the Real Data Analysis

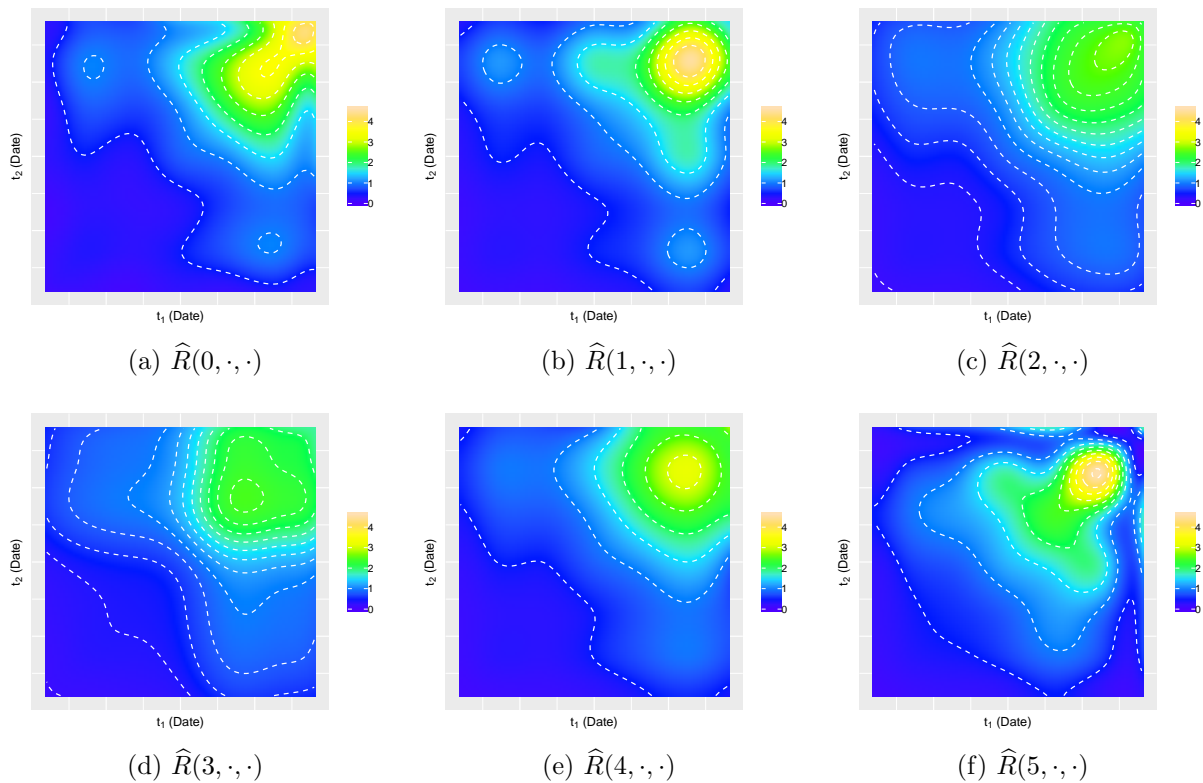

Figure S.7: Spatio-temporal covariance function for London house price Data: contour plots $\widehat{R}(u,\cdot,\cdot)$ standardized by $\|R(u,\cdot,\cdot)\|_1 = \int\int |\widehat{R}(u,t_1,t_2)|dt_1 dt_2/|T|^2$, at $u = 0, 1, 2, 3, 4$, and 5.

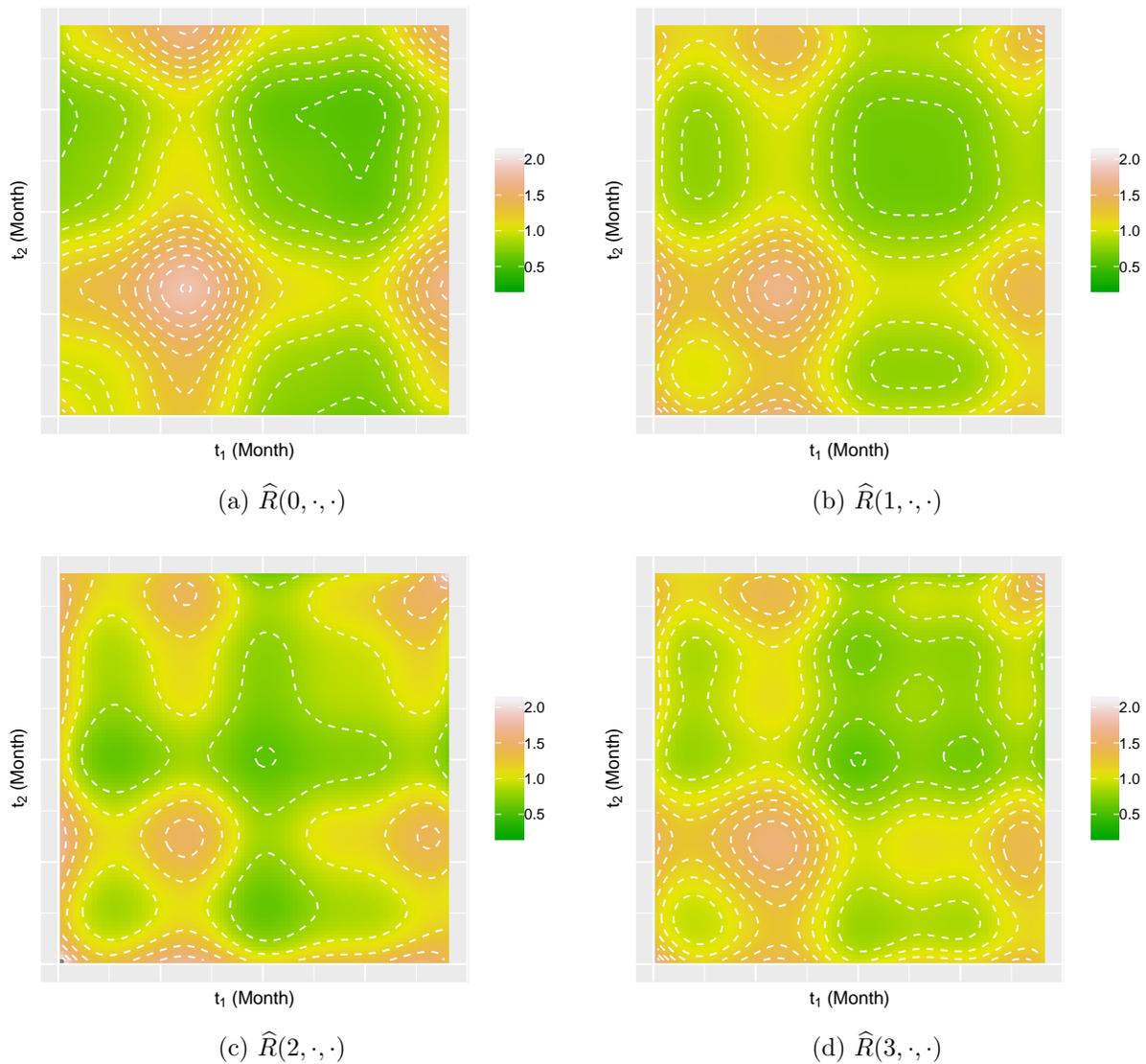

Figure S.8: Spatio-temporal covariance function for Zillow price-to-rent ratio data analysis: contour plots $\widehat{R}(u,\cdot,\cdot)$ standardized by $\|R(u,\cdot,\cdot)\|_1 = \int\int |\widehat{R}(u,t_1,t_2)|dt_1 dt_2/|T|^2$, at $u = 0, 1, 2$, and $3$.

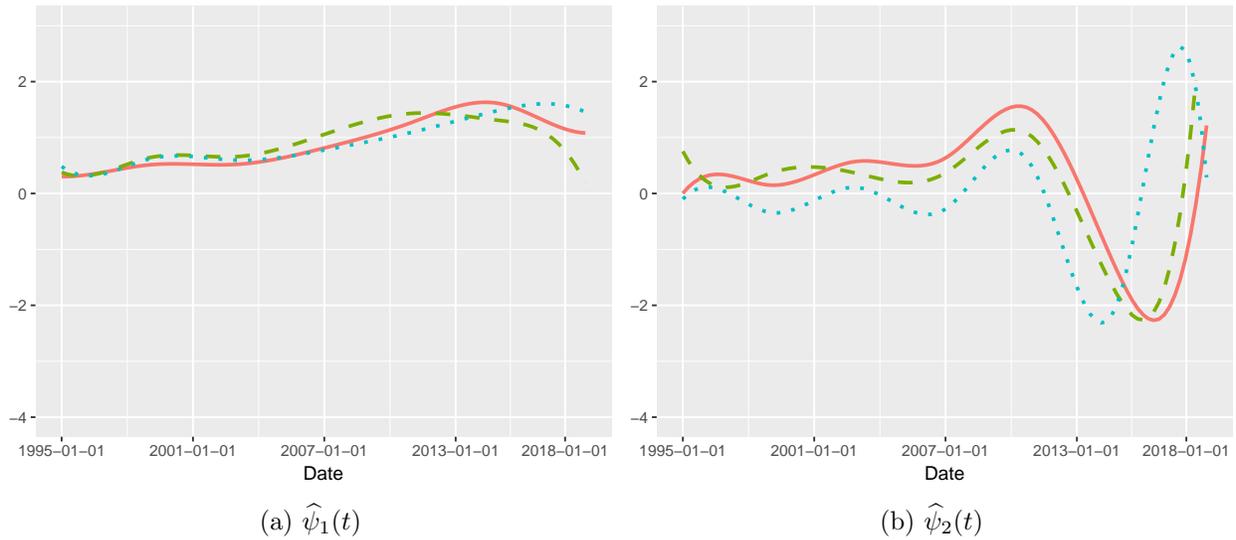

(a) $\widehat{\psi}_1(t)$  (b) $\widehat{\psi}_2(t)$

Figure S.9: Sensitivity analysis on the London house price data. The red lines are the estimated eigenfunctions using the whole dataset, while the green dashed lines and blue dotted lines are the estimated eigenfunctions using subsets of the data to the north and south of River Thames respectively.

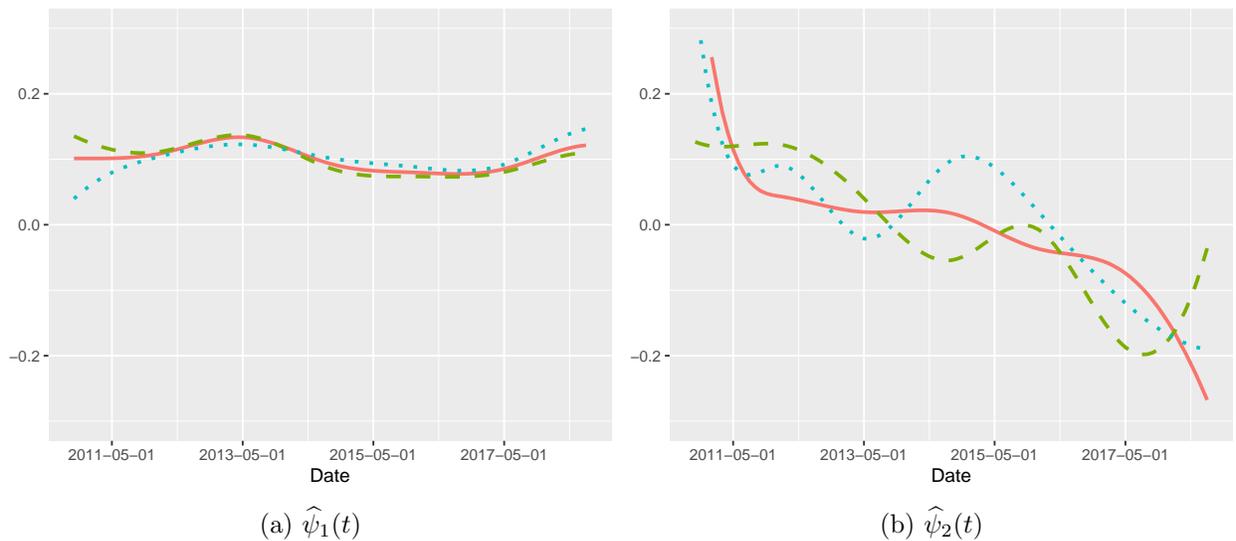

(a) $\widehat{\psi}_1(t)$  (b) $\widehat{\psi}_2(t)$

Figure S.10: Sensitivity analysis on the Zillow price-to-rent ratio data. The red lines are the estimated eigenfunctions using the whole dataset; the green dashed lines are the estimated eigenfunctions by using the data outside the peninsula including neighborhoods in *Fremont, Oakland* and *San Jose*; the blue dotted lines are the estimated eigenfunctions by using the data on the peninsula including the cities of *San Francisco, San Mateo* and *Palo Alto*.

# References

Bosq, D. (2012). *Nonparametric Statistics for Stochastic Processes: Estimation and Prediction*. Springer Science & Business Media, New York.

Demko, S. (1977). Inverses of band matrices and local convergence of spline projections. *SIAM Journal on Numerical Analysis*, 14(4):616–619.

Demko, S., Moss, W. F., and Smith, P. W. (1984). Decay rates for inverses of band matrices. *Mathematics of Computation*, 43(168):491–499.

Guan, Y., Sherman, M., and Calvin, J. A. (2004). A nonparametric test for spatial isotropy using subsampling. *Journal of the American Statistical Association*, 99(467):810–821.

Hall, P. and Hosseini-Nasab, M. (2006). On properties of functional principal components analysis. *Journal of the Royal Statistical Society: Series B*, 68(1):109–126.

Karr, A. F. (1986). Inference for stationary random fields given poisson samples. *Advances in Applied Probability*, 18(2):406–422.

Liang, D., Zhang, H., Chang, X., and Huang, H. (2021). Modeling and regionalization of China's $PM_{2.5}$ using spatial-functional mixture models. *Journal of the American Statistical Association*, 116(533):116–132.

Mastronardi, N., Ng, M., and Tyrtyshnikov, E. E. (2010). Decay in functions of multiband matrices. *SIAM Journal on Matrix Analysis and Applications*, 31(5):2721–2737.

Rudin, W. (1991). *Functional Analysis (International Series in Pure and Applied Mathematics)*. McGraw-Hill, Inc., New York.

Schumaker, L. (1981). *Spline Functions: Basic Theory*. Cambridge University Press, Cambridge.

Wang, L. and Yang, L. (2009). Spline estimation of single-index models. *Statistica Sinica*, 19(2):765–783.

Zhang, H. (2019). *Topics in functional data analysis and machine learning predictive inference*. PhD thesis, Iowa State University.

Zhang, H., Zhu, Z., and Yin, S. (2016). Identifying precipitation regimes in China using model-based clustering of spatial functional data. In *Proceedings of the Sixth International Workshop on Climate Informatics*, pages 117–120.

Zhou, S., Shen, X., and Wolfe, D. (1998). Local asymptotics for regression splines and confidence regions. *The Annals of Statistics*, 26(5):1760–1782.


\end{document}